\documentclass[a4paper,11pt]{article}
\pdfoutput=1 % if your are submitting a pdflatex (i.e. if you have
\usepackage{jcappub} % for details on the use of the package, please
 \graphicspath{{figures/}}

\newcommand{\syn}{{({\rm syn}) }}
\newcommand{\curv}{{({\rm curv}) }}
\newcommand{\coll}{{({\rm coll}) }}
\newcommand{\nocoll}{{({\rm no~coll}) }}

\newcommand{\X}{{{\rm x} }}
\newcommand{\Y}{{{\rm y} }}
\newcommand{\Z}{{{\rm z} }}

\definecolor{darkgreen}{RGB}{0,200,0}

\title{\bf Simulating the universe(s): from cosmic bubble collisions to cosmological observables with numerical relativity}

\author[a]{Carroll L. Wainwright,}
\author[b,c]{Matthew C. Johnson,}
\author[d]{Hiranya V. Peiris,}
\author[a]{Anthony Aguirre,}
\author[c]{Luis Lehner,}
\author[e]{Steven L. Liebling}

\affiliation[a]{SCIPP and Department of Physics, University of California \\ Santa Cruz, CA, 95064, USA}
\affiliation[b]{Department of Physics and Astronomy, York University \\ Toronto, On, M3J 1P3, Canada}
\affiliation[c]{Perimeter Institute for Theoretical Physics \\ Waterloo, Ontario N2L 2Y5, Canada}
\affiliation[d]{Department of Physics and Astronomy, University College London \\ London WC1E 6BT, U.K.}
\affiliation[e]{Department of Physics, Long Island University \\ Brookville, NY, 11548, USA}

\emailAdd{cwainwri@ucsc.edu}
\emailAdd{mjohnson@perimeterinstitute.ca}
\emailAdd{h.peiris@ucl.ac.uk}
\emailAdd{aguirre@scipp.ucsc.edu}
\emailAdd{llehner@perimeterinstitute.ca}
\emailAdd{steve.liebling@liu.edu}

\abstract{The theory of eternal inflation in an inflaton potential with multiple vacua predicts that our universe is one of many bubble universes nucleating and growing inside an ever-expanding false vacuum. The collision of our bubble with another could provide an important observational signature to test this scenario. We develop and implement an algorithm for accurately computing the cosmological observables arising from bubble collisions directly from the Lagrangian of a single scalar field. We first simulate the collision spacetime by solving Einstein's equations, starting from nucleation and ending at reheating.  Taking advantage of the collision's hyperbolic symmetry, the simulations are performed with a 1$+$1-dimensional fully relativistic code that uses adaptive mesh refinement. We then calculate the comoving curvature perturbation in an open Friedmann-Robertson-Walker universe, which is used to determine the temperature anisotropies of the cosmic microwave background radiation. For a fiducial Lagrangian, the anisotropies are well described by a power law in the cosine of the angular distance from the center of the collision signature. For a given form of the Lagrangian, the resulting observational predictions are inherently statistical due to stochastic elements of the bubble nucleation process. Further uncertainties arise due to our imperfect knowledge about inflationary and pre-recombination physics. We characterize observational predictions by computing the probability distributions over four phenomenological parameters which capture these intrinsic and model uncertainties. This represents the first fully-relativistic set of predictions from an ensemble of scalar field models giving rise to eternal inflation, yielding significant differences from previous non-relativistic approximations. Thus, our results provide a basis for a rigorous confrontation of these theories with cosmological data.}

\begin{document} 
\maketitle
\flushbottom

\section{Introduction}
\label{sec:intro}

Do we live in a bubble? In a picture of eternal inflation driven by an inflaton field with multiple potential minima, our universe is predicted to lie inside one of many bubble ``universes" nucleated out of an inflating ``parent" vacuum (see Refs.~\cite{Aguirre:2007gy,Guth:2007ng} for a review of eternal inflation). An active area of inquiry is to search for observable signatures of eternal inflation, which might allow us to confirm this radical picture of the universe on the largest of scales. One unambiguous signature arises from the collision between bubble universes~\cite{Aguirre:2007an,Garriga:2006hw}, which leaves distinctive cosmological signatures in the cosmic microwave background (CMB) radiation~\cite{Aguirre:2007wm,Chang_Kleban_Levi:2009,Czech:2010rg,Kleban_Levi_Sigurdson:2011,Kozaczuk:2012sx,Feeney_etal:2010dd,Feeney_etal:2010jj} and the distribution of large scale structure~\cite{Larjo:2009mt}.

There has been enormous progress in understanding whether or not bubble collisions are a feasible observable consequence of the eternal inflation scenario. Previous work exploring the detailed outcome of bubble collisions~\cite{Turner:1992tz,Hawking:1982ga,Wu:1984eda,Moss:1994pi,Freivogel:2007fx,Aguirre:2009ug,Chang:2007eq,Johnson:2010bn,Giblin:2010bd,Amin:2013dqa,Amin:2013eqa,Easther:2009ft,Aguirre:2008wy,Giblin:2010bd,Deskins:2012tj,Johnson:2011wt,Ahlqvist:2013whn} has firmly established that collisions can be a minor perturbation on standard cosmology, and also addressed the probability with which we can expect to see bubble collisions under various model assumptions~\cite{Garriga:2006hw,Aguirre:2007an,Freivogel_etal:2009it,Salem:2012}. A review of much of this previous work can be found in Refs.~\cite{Aguirre:2009ug,Kleban:2011pg}. However, to date there has been no {\em quantitative} connection between a particular scalar field model giving rise to eternal inflation and the detailed signature of bubble collisions imprinted on the CMB. 

This paper aims to fill this gap, constructing for the first time a quantitative prediction for the signatures of bubble collisions in an ensemble of scalar field models. To this end, fully relativistic numerical simulations are required because the collision between bubbles is a highly non-linear process. We employ a powerful new set of numerical relativity codes,\footnote{See Ref.~\cite{Parry:1998pn,Blanco-Pillado:2003hq,Carone:1989nj,Goldwirth:1989pr,KurkiSuonio:1993fg,Laguna:1991zs} for early applications of numerical relativity to the early universe, and Ref.~\cite{Xue:2013bva} for an interesting recent application.} significantly improving upon previous work~\cite{Johnson:2011wt} (see also Ref.~\cite{Hwang:2012pj}). Using a combination of a specialized gauge, adaptive mesh refinement, and a strategic choice of which regions to simulate, we are able to accurately simulate the collision itself, as well as the evolution of induced perturbations over the entire duration of slow-roll inflation inside each bubble. By tracing geodesics through the simulation, we directly extract the perturbed cosmological metric inside each bubble, allowing us to determine the observational signatures. We show that the signature in the CMB is well described by a set of four phenomenological parameters. The values of these parameters are only probabilistically determined, a consequence of the intrinsic variation in the relative position of the collision and the observer, the center of mass energy of the collision, and uncertainties in the underlying model. To provide a complete case study, we compute the probability distribution over these phenomenological parameters. 

Observational searches for the signature of bubble collisions -- and the interpretation of their findings -- are only as good as the assumptions going into the theoretical predictions. Searches for the predicted signatures of bubble collisions have been performed~\cite{Feeney:2012hj,McEwen:2012uk,Feeney_etal:2010dd,Feeney_etal:2010jj,Osborne:2013hea,Osborne:2013jea} on CMB data from the Wilkinson Microwave Anisotropy Probe (WMAP)~\cite{Bennett:2003ba}. Feeney et. al.~\cite{Feeney:2012hj} performed a Bayesian analysis, which constrained the number of observable collisions on the sky to be $N < 4.0$ at $95 \%$ confidence. Osbourne et. al.~\cite{Osborne:2013hea} took a different approach, constraining a combination of the amplitude $a$ of the initial perturbation and the angular size $\theta_{\rm crit}$ of a bubble collision on the sky to lie in the range $-4.66 \times 10^{-8} < a \left(\sin \theta_{\rm crit} \right)^{4/3} < 4.73 \times 10^{-8} \ [{\rm Mpc}^{-1}]$ at $95 \%$ confidence. Both analyses made assumptions about the form of the signature and the prior probability distribution over parameter values. The methods outlined in this paper will help guide future observational searches, as well as laying the foundations for translating those searches into constraints on fundamental physics.

\section{Summary of background, methods, and results}
\label{sec:summary}

Before launching into the technical details, let us pause to summarize the necessary background on bubble collisions in eternal inflation, outline the most important features of our methods, and highlight some of the most important results. The reader can use this section as a guide to the rest of the paper.

\begin{figure}
\begin{center}
\includegraphics[width=12 cm]{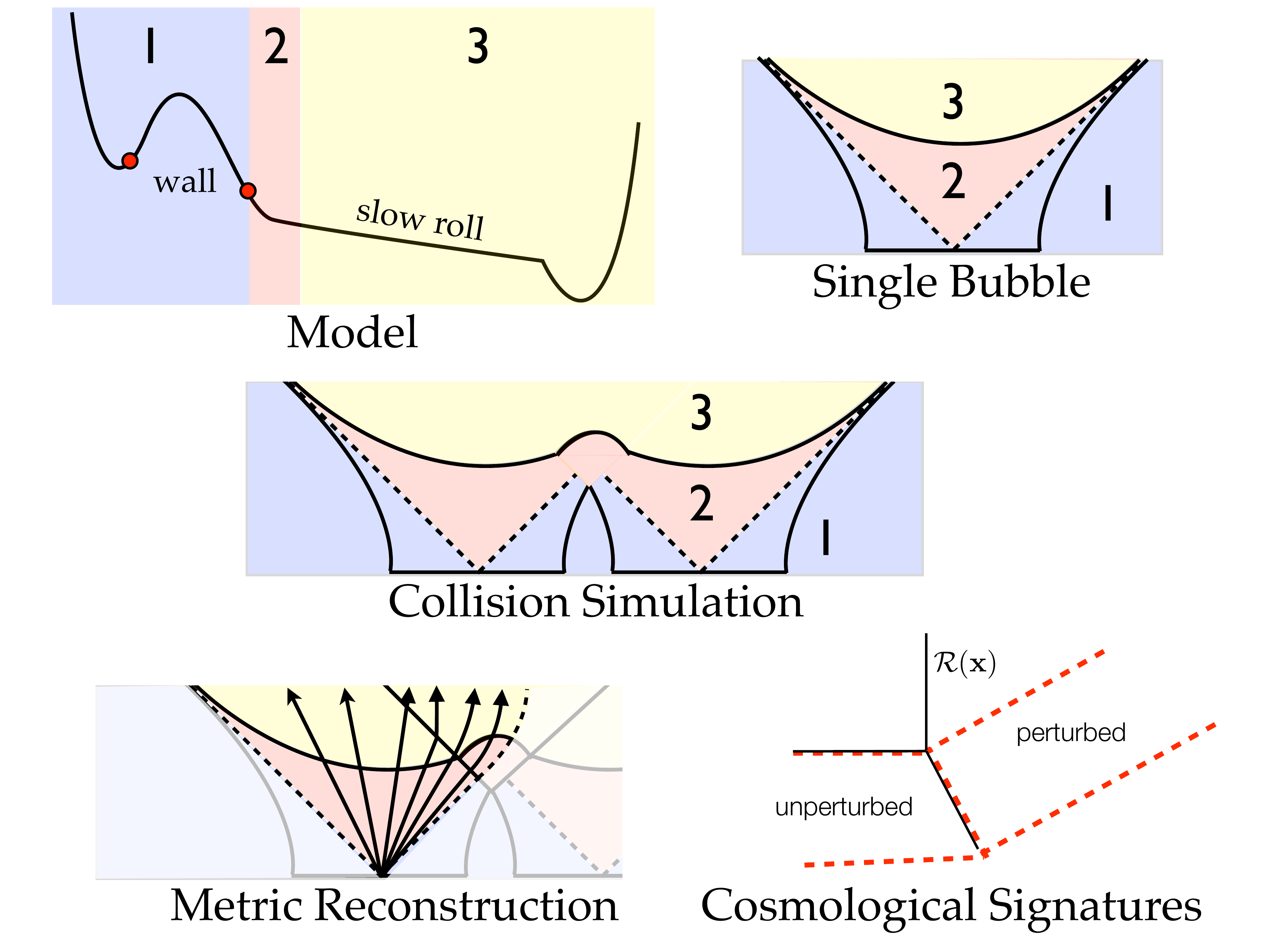}
\end{center}
\caption{{ Mechanics of bubbles, bubble collisions, and reconstruction of observational signatures.} One begins by specifying a scalar-field potential with multiple vacua. The false vacuum is unstable to bubble formation. Each bubble has a wall determined by region 1 of the potential, and contains an open FRW universe, the evolution of which is determined by regions 2 and 3 of the potential. Surfaces of constant density are depicted as solid lines inside the bubble. Bubbles collide, and we can conveniently simulate them in a reference frame, the Collision Frame, where they nucleate at the same cosmological time. Because the field excursion caused by the collision is bounded, the dynamics of the collision are determined entirely by regions 1 and 2 of the potential. The subsequent cosmological evolution is determined by region 3. We reconstruct the perturbed FRW metric inside the observation bubble by evolving a set of geodesics through the simulation. A gauge transformation then allows us to extract the comoving curvature perturbation late in the inflationary epoch. Since this quantity is frozen in, and largely insensitive to the details of reheating, it is possible to calculate cosmological signatures such as the CMB temperature anisotropies.
  \label{fig:summary}
}
\end{figure}

To accurately predict the observational signature of bubble collisions, one must in effect simulate the entire history of the early universe. The steps along the way are sketched in Fig.~\ref{fig:summary}. One begins by specifying a model that drives eternal inflation. In this paper, we will consider a single scalar field with a potential containing a false vacuum and a single true vacuum. Given an initial condition where the field is at rest in the false vacuum over a region of size greater than $H_F^{-1}$ (the false vacuum Hubble radius), bubble nucleation will occur via the Coleman-de Luccia (CDL) instanton~\cite{Coleman:1977py,Coleman:1980aw}. If the rate of bubble formation is somewhat less than one bubble nucleated per Hubble volume per Hubble time, then the phase transition will not complete, and inflation becomes eternal. 

The rate of bubble formation and the profile of the bubble wall is entirely determined by the CDL instanton solution pertaining to the region of the potential labeled ``1". The reader can find a summary of these instantons and the methods for constructing them in Sec.~\ref{sec:CDL_instantons}. The structure of the instanton also implies that the exterior of the null cone emanating from the bubble center contains values of the field between the instanton endpoints, labeled by the red dots in Fig.~\ref{fig:summary}. The symmetry group of the instanton implies that each bubble contains an open Friedmann-Robertson-Walker (FRW) universe. In order to dilute the curvature and to produce phenomenologically-viable density fluctuations, we must invoke an epoch of slow-roll inflation inside the bubble. These models of ``open inflation"~\cite{Bucher:1994gb,Gott:1982zf} have been widely discussed in the literature (see e.g., Ref.~\cite{GarciaBellido:1997uh} for a review). We allow for an initial transient era described by evolution on the section of the potential labeled ``2," before slow-roll occurs in region ``3". This transient region is motivated by the hierarchy in scales necessary to support both slow-roll inflation and tunneling on the same potential\footnote{The potential we use in this paper is shown in Fig.~\ref{fig:L2potential}, where this hierarchy in scales is apparent.}~\cite{Linde:1998iw}. Inside a single bubble, evolution in regions ``2" and ``3" occurs along surfaces of constant density in the open FRW universe, as depicted in Fig.~\ref{fig:summary}. It is important to note that throughout this paper we treat the field evolution in the post-nucleation region entirely classically, neglecting ambiguities about the transition between quantum mechanical and classical behavior after tunneling,\footnote{See Ref.~\cite{turoktunnelling} for some discussion on how one might resolve these ambiguities.} as well as fluctuations during slow-roll inflation and in the bubble wall.

Any given bubble will undergo collisions with an infinite number of other bubbles~\cite{Garriga:2006hw}. The subset of potentially {\em observable} collisions depends on the cosmology inside the bubble and the effect of collisions on it. A discussion can be found in Ref.~\cite{Aguirre:2009ug}. For widely-separated collisions, we can consider each collision with the ``observation bubble" (i.e., the one whose cosmology we are interested in) separately. The potential depicted in Fig.~\ref{fig:summary} allows for only one type of bubble, so all collisions are between identical bubbles. Extensions of this simplest model would allow for the collision between different bubbles, connecting the false vacuum to different basins of attraction. We restrict ourselves in this paper to this simplest class of models, leaving an assessment of the full range of possibilities to subsequent work.

The full collision spacetime for two bubbles possesses an $SO$(2,1) symmetry, allowing us to extract the spacetime (with metric Eq.~\ref{eqn:metric2}) using a $1+1$-dimensional simulation. This symmetry arises from the hyperbolic $SO$(3,1) symmetry of the individual bubbles (rotations and radial boosts): the intersection of two hyperboloids is a hyperboloid of one lower dimension. Rotations about the line separating the bubble centers is retained, as well as boosts transverse to this axis, yielding a residual $SO$(2,1) symmetry~\cite{Hawking:1982ga,Wu:1984eda}. The equations of motion for the scalar field and metric (see Eq.~\ref{eq:equationsofmotion}) respect this symmetry, and unless additional effects are considered, there is no source for breaking it. One such source that violates our symmetry assumption is the effect of fluctuations on the bubble wall induced by tensor modes quantum-mechanically produced in the false vacuum de Sitter space (see e.g., Refs.~\cite{Yamamoto:1995sw,GarciaBellido:1997hy,GarciaBellido:1997uh,Garriga:1998he,Yamamoto:1995tk}). These fluctuations are generally small in amplitude, unless the background false vacuum has an energy density of order $M_{\rm Pl}$. More generally, even though instabilities on domain walls can arise, they become  increasingly length-contracted, and hence unimportant, on expanding bubble walls as they reach relativistic velocities~\cite{Adams:1989su,Aguirre:2005xs}. 

In a previous paper~\cite{Johnson:2011wt}, three of the authors presented the first fully relativistic simulation of bubble collisions. In this paper, we use a significantly improved code which employs adaptive mesh refinement (AMR) and a new gauge which allows us to simulate the entire epoch of inflation inside each bubble. The results (see Fig.~\ref{fig:sim_contour1} and Fig.~\ref{fig:sim_contour2}) are solutions for the metric functions and the scalar field, as functions of the coordinate $x$ separating the bubble centers and a time variable $N$ measuring the elapsed number of $e$-folds of expansion in the false vacuum. The code is convergent (see Fig.~\ref{fig:sim_convergence}) and respects the constraints imposed by Einstein's equations (see Fig.~\ref{fig:sim_constraint}).

Given the structure of the individual bubbles, regions 1 and 2 on the potential are most relevant for determining the immediate outcome of a collision. The immediate aftermath of the collision is quite simple when the bubble walls are relativistic: the field profiles making up the individual bubbles simply superpose~\cite{Giblin:2010bd,Johnson:2011wt}. Therefore, only a finite range in field space is relevant for predicting the immediate outcome of the collision. To evolve further, the full non-linearities of General Relativity become important. Because the bubbles collide in an ambient false vacuum de Sitter space, the center of mass energy for a collision is bounded, and the strength of the disturbance cannot grow arbitrarily large. Since the perturbation can be made small by tailoring the potential, we can ensure the the future of the collision is not a radical deviation from standard cosmological evolution. Region 3 of the potential will be important for determining the evolution of the background inflationary cosmology, and the evolution of the perturbation on that background sourced by the collision.

To simulate the collision and to extract the appropriate information requires multiple coordinate systems, summarized in Table~\ref{tab:coordinate_list}. For the collision itself, we use coordinates that cover the interior of both bubbles, as well as the false vacuum they evolve in. To extract the signature of bubble collisions in the observation bubble, we must first go from the simulation coordinates to a set of perturbed FRW coordinates appropriate for describing the cosmology inside the observation bubble. We accomplish this by tracking a set of geodesics emanating from the nucleation center. Geodesics label coordinate positions $\{\xi, \rho, \varphi \}$, and evolve in proper time $\tau$; this yields a map\footnote{The fact that these maps do not depend on $\rho$ and $\varphi$ is a manifestation of the $SO$(2,1) symmetry of the full collision spacetime.} $\{ N(\xi, \tau), x(\xi, \tau) \}$ between the simulation coordinates and the anisotropic slicing of the open FRW spacetime labeled by the geodesics. Geodesics are evolved numerically; an example of the map is shown in Fig.~\ref{fig:geodesics}. We then find the metric in terms of coordinates $\{\tau, \xi, \rho, \varphi \}$ using the standard tensor transformation law, Eq.~\ref{eq:metric_transform}.

\begin{table}[htbp]
   \centering
   \begin{tabular}{@{} l | l | l @{}} % Column formatting, @{} suppresses leading/trailing space
      Name & Variables & Metric \\
      \hline
      \hline
      \small{\bf Simulation Coordinates}	&	$N,x,\chi,\varphi$	& Eq.~\ref{eqn:metric2} \\
      \small{\emph{Used when running the simulation.}} & & \\
      \hline
      \small{\bf Minkowski Coordinates}	&	$t,x,y,z$	&  $\eta_{\mu\nu}$ \\
      \small{\emph{Describes approximately flat space}} & & \\
      \small{\emph{near the observation bubble's center.}} & & \\
      \hline
      \small{\bf Cartesian Observer Coordinates}	&	$\tau, X,Y,Z$	& Eq.~\ref{eq:unperturbed_cartesian} \\
      \small{\emph{Used to calculate the comoving curvature}} & & \\
      \small{\emph{perturbation as seen by observers.}} & & \\
      \hline
      \small{\bf Anisotropic Hyperbolic Coordinates}	&	$\tau, \xi, \rho, \varphi$	& Eq.~\ref{eq:anisotropic_c} \\
      \small{\emph{Convenient for labeling and integrating}} & & \\
      \small{\emph{geodesics.}} & & \\
      \hline
      \small{\bf False-vacuum Hyperbolic Coordinates}	&	$\Psi, \Upsilon, \theta,\phi$	& Eq.~\ref{eq:outsidehypercoords} \\
	\small{\emph{Used to label the nucleation sites of}} & & \\
	\small{\emph{collision bubbles in the false vacuum.}} & & \\
	\hline
      
   \end{tabular}
   \caption{A list of coordinate systems used in this paper. 
   }
   \label{tab:coordinate_list}
\end{table}

The simulations are performed in a reference frame where the colliding bubbles nucleate at the same time, which we refer to as the Collision Frame. To compute observational signatures, a convenient frame will be one in which the observer is at the center of the bubble. We refer to this as the Observation Frame. The two reference frames are connected by a diffeomorphism that can be enacted by a boost in an embedding-space picture (see Ref.~\cite{Aguirre:2009ug}), which corresponds to a spatial translation in the cosmological spacetime contained inside the observation bubble, and to a time-translation in the spacetime outside the observation bubble.

Because the collision can only affect regions of the observation bubble in its causal future, the universe is split into regions affected and not affected by the collision. Near the causal boundary, the metric must, by continuity, be close to the empty open FRW metric. In this region, we can explicitly find the difference between the metric inside a bubble with a collision and a bubble without a collision, in order to map onto the standard metric fluctuation variables of cosmological perturbation theory. The resulting metric is explicitly in the synchronous gauge, and satisfies to a very good approximation the perturbative expansion of Einstein's equations (see Fig.~\ref{fig:synchronousconst}).
 
The synchronous gauge metric functions can be used to extract cosmological observables. However, a much cleaner variable to calculate is the comoving curvature perturbation $\mathcal{R}$. For the adiabatic perturbations we consider, $\mathcal{R}$ remains constant on superhorizon scales and is insensitive to the details of reheating and subsequent evolution. Fig.~\ref{fig:RvsXY} shows the comoving curvature perturbation near the end of inflation as a function of the Cartesian coordinates $X,Y,Z$ in an open universe. As can be seen in this figure, $\mathcal{R}$ varies mainly along the direction connecting the bubble centers, with deviations from this near-planar symmetry only becoming apparent on scales close to the radius of curvature of the open FRW spacetime. Near the collision boundary, we find that $\mathcal{R}$ is well described by a power law in the coordinate $\xi$, as in Eq.~\ref{eq:Rtemplate}. This allows us to develop a set of phenomenological parameters that describe the collision. The detailed profile depends on the initial separation $\Delta x$; the free parameters in the template Eq.~\ref{eq:Rtemplate} vary with $\Delta x$ as shown in Fig.~\ref{fig:template_plots}. 

The comoving curvature perturbation can be converted into the observed temperature and polarization signatures in the CMB. In this paper, we work in the Sachs-Wolfe~\cite{Sachs:1967er} approximation, where the temperature anisotropy is proportional to the comoving curvature on the surface of last-scattering, evaluated on the past light-cone of an observer, $\delta T / T = \mathcal{R}(x_{\rm ls}) / 5$. 
By symmetry, collisions project onto a disc on the CMB sky. From the detailed form of $\mathcal{R}$, the  temperature profile is a power law in the cosine of the angular distance from the center of the disc, Eq.~\ref{eq:deltat_template}. This fitting function agrees with the full projection of $\mathcal{R}$ over most scales, as shown in Fig.~\ref{fig:projectedR} and Fig.~\ref{fig:projectedR_log}.

\begin{figure}
\begin{center}
\includegraphics[width=12 cm]{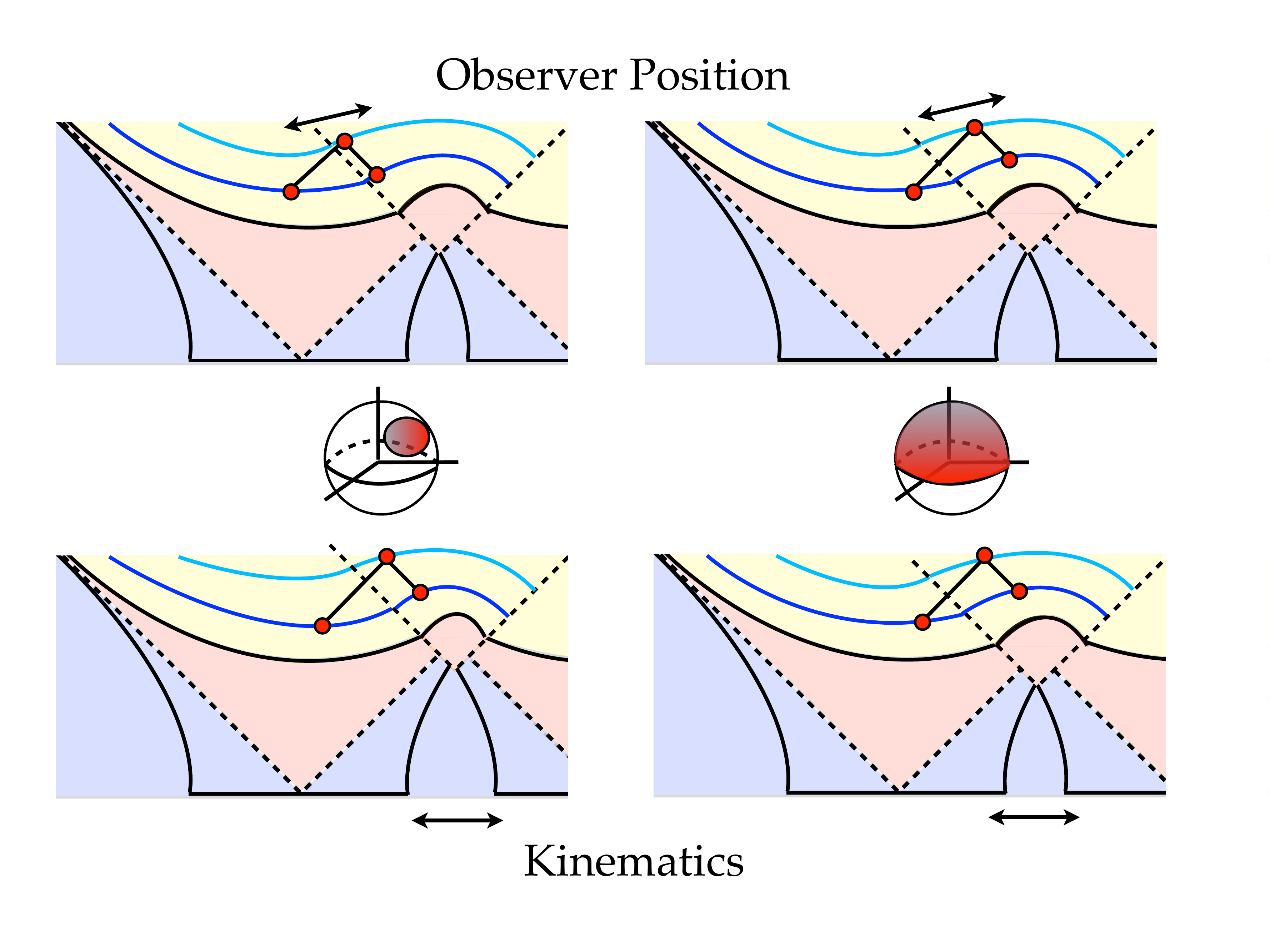}
\end{center}
\caption{{ Observer position and kinematics: the two intrinsic sources of variation in the collision signature in the Collision Frame.} An observer's position today, as well as the intersection of their past light-cone with the surface of last scattering, are depicted using filled dots. Moving the observer along this constant-time hypersurface (left panel to right panel), more of the region affected by the collision is seen. The causal boundary (dashed light-cone emanating from the collision) and perturbed region projects onto a disc of variable size on the observer's CMB sky. Alternatively, fixing the observer's position in its bubble, and decreasing the separation of the colliding bubbles, more of the region affected by the collision is seen, and the strength of the underlying perturbation will change. Both of these effects  project onto a variable size, amplitude, and perturbation profile in the affected region on the sky. 
  \label{fig:position_kinematics}
}
\end{figure}

We have mentioned two sources of variation in the observed signature after the underlying inflationary potential is fixed.  In the Collision Frame, as depicted in Fig.~\ref{fig:position_kinematics}, variations in the observed signature arise from variations in the position of the observer relative to the causal boundary of the collision $\xi_{\rm obs}$ and from variations in the initial separation of the bubbles $\Delta x$. To these, we add two parameters describing variations in the underlying scalar-field potential itself.  First, we can perform an overall rescaling of region 1 of the potential by a factor $\beta$; second, we can alter region 3 so as to effect a change in the observable portion of the surface of last scattering $R_{\rm ls}$. This yields four fundamental parameters $\{\xi_{\rm obs}, \Delta x, \beta, R_{\rm ls} \}$, which we translate into four phenomenological parameters describing the characteristics that determine the observational signatures of this ensemble of scalar-field models: the observed angular scale $\theta_c$, number of collisions $N_{\rm coll}$, power-law index $\kappa$, and $R_{\rm ls}$. These parameters are analogous to, for example, the amplitude and spectral index predicted by an inflationary model --- complicated scalar-field models map onto a few phenomenological parameters. A main result of this paper is a derivation of the probability distribution over these phenomenological parameters, Eq.~\ref{eq:finalprior}. The marginalized probability distribution for a fiducial class of models is plotted in Fig.~\ref{fig:distribution}. 

The probability distribution, along with the phenomenological parameters and fitting function describing the signal, are the complete prediction of this set of models. Similar distributions exist for other models --- or classes of models --- and are the central input for any observational search. In future work, following the procedure outlined above will allow us to identify how features of the underlying model interface with observational constraints.

\section{Simulating Colliding Universes}
\label{sec:simulation_desc}

\subsection{CDL Instantons}
\label{sec:CDL_instantons}

To simulate the collision of two bubbles in an inflating background we first need the initial conditions: a field and metric configuration for each bubble that interpolates between the metastable false vacuum state and a region with lower vacuum energy separated from the false vacuum by a potential barrier. If one of the bubbles is to represent our observable universe, its interior must undergo a period of slow-roll inflation.

Tunneling between vacuum states was first studied in detail by Coleman and de Luccia~\cite{Coleman:1977py,Coleman:1980aw}. Their basic conclusion was that quantum tunneling occurs via bubble nucleation --- a bubble with a lower energy vacuum state fluctuates into existence in a sea of false vacuum --- with an exponentially-suppressed tunneling rate. The rate, as well as the field configuration --- and with gravity, the metric --- of the bubble is found by solving Euclidean field equations to find an ``instanton" solution that extremizes the Euclidean action.

Once nucleated, the bubble's internal pressure will accelerate its bubble wall outwards, quickly approaching the speed of light. The instanton with the smallest action, and thus the dominant contribution to the tunneling rate, comes from the bubble which is $O(4)$-symmetric in the Euclidean metric. 

The general CDL instanton with $O(4)$ symmetry has a Euclidean metric of the form
\begin{equation}
\label{eqn:metric1}
ds^2 = dr^2 + \rho(r)^2 d\Omega_3^2,
\end{equation}
where $r$ is the bubble's (Euclidean) radius and $d\Omega_3$ is the metric on a three-sphere. The function $\rho(r)$ determines the curvature. For example, an instanton in pure de Sitter space has $\rho(r) = R_{\rm dS} \sin(r/R_{\rm dS})$, where $R_{\rm dS}$ is the de Sitter radius. Such an instanton has a maximum radius of $\pi R_{\rm dS}$, corresponding to half the circumference of the de Sitter space. Any instanton that interpolates between regions in a positive-definite potential will likewise be compact with some maximum radius $r_{\rm max}$. The instanton equations take the form:
\begin{gather}
\frac{d^2\phi}{dr^2} + \frac{3}{\rho}\frac{d\rho}{dr}\frac{d\phi}{dr} = \frac{dV}{d\phi}; \\
\frac{d^2\rho}{dr^2} = -\frac{8\pi}{3M_\mathrm{Pl}^2}\rho(r) \left[\left(\frac{d\phi}{dr}\right)^2 + V(\phi)\right].
\end{gather}
The field $\phi$ should be in different basins of attraction (near different minima in the potential) at $r=0$ and $r=r_{\rm max}$ in order to achieve a non-trivial solution to the equations of motion. Conventionally, the field is in the lower energy vacuum at $r=0$. Non-singular solutions must also satisfy $\left.{d\phi}/{dr}\right|_{r=0} = \left.{d\phi}/{dr}\right|_{r=r_{\rm max}} = 0$ and $\rho(r=0) = \rho(r=r_{\rm max}) = 0$. We also specify $\left.{d\rho}/{dr}\right|_{r=0} = 1$, which essentially just sets $\rho$ and $r$ to have the same units. Note that an overall rescaling of $V$ just rescales the radius $r$ (and $\rho$); it does not otherwise affect the instanton's shape.

We solve the instanton equations using an iterative overshoot/undershoot method. 
First, we make an initial guess for the value of the field $\phi$ at $r=0$. The guess should have $V(\phi_{\rm guess}) < V(\phi_{\rm metastable})$. The equations of motion dictate that the field will roll up the potential barrier as $r$ increases, gaining momentum. If the field has enough momentum ($|d\phi/dr|$ is large enough), it will continue down the other side of the barrier and, if the initial guess was correct, stop with $d\phi/dr=0$ exactly at $r=r_{\rm max}$ near the metastable vacuum. If the field does not have enough momentum, it will stop rolling before $r=r_{\rm max}$ and reverse course towards the top of the potential barrier. This is an undershoot, and the initial guess must be modified to have a lower vacuum energy and sit closer to the stable minimum. If the initial guess is very close to a minimum (or an inflection point) such that $dV/d\phi$ is very small, the field will not move much over a large radius and the friction term $\frac{3}{\rho}\frac{d\rho}{dr}\frac{d\phi}{dr}$ will become small or negative. Therefore, the field can always be made to have enough momentum to cross the barrier. If the momentum is too large, the field will either cross the metastable minimum or not stop ($d\phi/dr \neq 0$) by $r=r_{\rm max}$. These are overshoots, and the initial guess must be modified to have a higher vacuum energy and sit closer to the potential barrier. One can converge upon the correct initial condition by making a sequence of initial guesses, integrating the equations of motion, and then modifying the guesses in response to overshoots and undershoots.

We implement the overshoot/undershoot method using the {\tt CosmoTransitions} software package~\cite{Wainwright:2011kj} with several modifications to account for gravity. Including the dynamics for $\rho$ (whereas $\rho(r) = r$ in flat space) is relatively straightforward; the criterion for an undershoot remains unchanged, but we had to add an additional criterion for an overshoot. The field overshoots the desired end point if $d\phi/dr \neq 0$ when $\rho(r)=0$, even if the field has not passed the metastable vacuum. Note that if the bubble wall radius is of similar size to, or larger than, the de Sitter radius, then the only solution may be the Hawking-Moss instanton~\cite{Hawking:1981fz}.  In this case, the field is constant and resides at the top of the potential barrier separating the two vacua. The modified {\tt CosmoTransitions} code will converge to this solution when appropriate.

\begin{figure}[t] 
   \centering
   \includegraphics[width=6in]{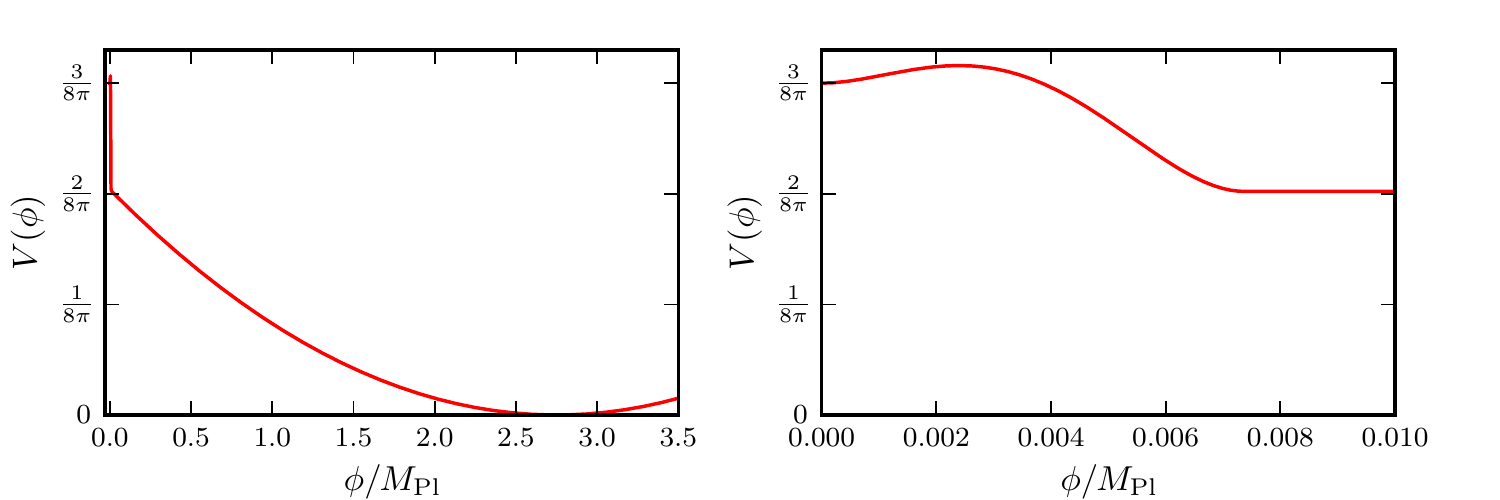} 
   \caption{The inflationary potential used throughout this study, rescaled such that $V(\phi_F) = \tfrac{3}{8\pi}$ (its original scale only affects the conversion between dimensionless simulation coordinates and physical scales). In order to achieve $\sim 60$ $e$-foldings of inflation, the potential requires a large slow-roll region (left), while the requirement that the instanton's size be much smaller than the de Sitter radius forces the potential barrier to be very narrow (magnified on the right).
   }
   \label{fig:L2potential}
\end{figure}

\begin{figure}[t] 
   \centering
   \includegraphics[width=2.8in]{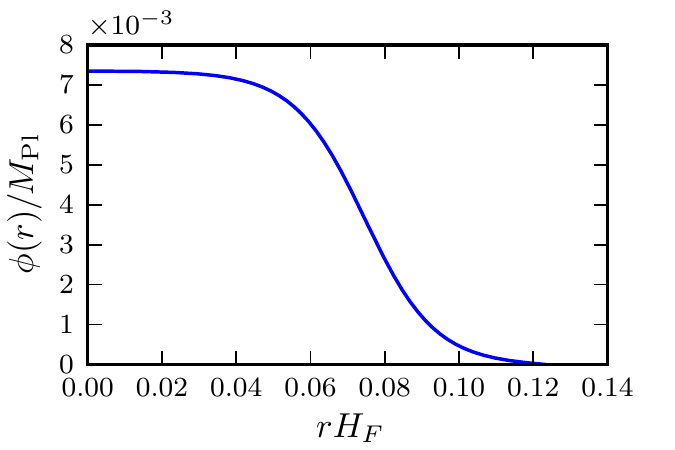} 
   \caption{ The instanton's radial profile, measured in units of the false vacuum de Sitter radius $H_F^{-1}$. It is fairly thick-walled, as we expect given the relatively small barrier separating the phases.  }
   \label{fig:instanton}
\end{figure}

For the purposes of this paper, we consider just a single potential with both bubbles coming from the same type of instanton. The potential is identical to the `L2' potential in Ref.~\cite{Johnson:2011wt}, replotted here in Fig.~\ref{fig:L2potential}. The metastable false vacuum is at $\phi_F = 0$; there is an inflection point at $\phi_I = 0.0074\, M_\mathrm{Pl}$; and the true vacuum (with zero cosmological constant) is at $\phi_T = 2.75\, M_\mathrm{Pl}$. Fig.~\ref{fig:instanton} shows the shape of the associated instanton. A broad survey of different scalar field potentials will appear in future work.

\subsection{Simulation equations and initial conditions}

Each bubble by itself is invariant under rotations and boosts about its origin. When there are two bubbles, the $SO(3,1)$ symmetry is broken down to $SO(2,1)$: rotation about, and boosts transverse to, the vector separating the two bubbles are still symmetries of the field configuration. Therefore, we can fully model the collision with a $(1+1)$-dimensional simulation. We choose coordinates with metric 
\begin{equation}
\label{eqn:metric2}
H_F^2 ds^2 = -\alpha(N, x)^2 \ dN^2 +  a(N,x)^2 \ \cosh^2 N \ dx^2 + \sinh^2N \ (d\chi^2 + \sinh^2\chi d\varphi^2),
\end{equation}
where $H_F$ is the false vacuum Hubble constant (and $H_F^{-1}$ is the false vacuum de Sitter radius), and $\alpha(N,x)$ and $a(N,x)$ are unknown metric functions. In the pure de Sitter case, $\alpha = a = 1$ and $N$ measures the number of false vacuum $e$-foldings. 
Inside the bubbles, where $\alpha, a \neq 1$, the variable $N$ does not correspond to the number of slow-roll $e$-foldings in a straightforward way. Instead, the number of $e$-foldings can be found by taking $N$ in concert with the metric functions $\alpha$ and $a$.
Distances and times are measured in units of $H_F^{-1}$; this is a natural set of units to describe the collision, since all dimensionless quantities will be order one. For the potential we simulate in this paper, the inflationary energy scale is comparable to the false vacuum Hubble scale, and we expect $a$ and $\alpha$ to be order one until the simulation reaches the end of slow-roll inflation inside the bubbles. However, this will not necessarily be true for other potentials which have a large hierarchy between the energy scale of the false vacuum and slow-roll inflationary epoch inside the bubbles. 
The coordinate $x$ is measured in radians such that the point at $(N=0,x=0)$ is antipodal to the point at $(N=0,x=\pi)$. The $SO(2,1)$ symmetry is manifest in the coordinates $\chi$ and $\varphi$: a boost with rapidity $\eta$ transverse to the $x$-axis will transform a point at $\chi=0$ to $\chi=\eta$, and a rotation about the $x$-axis simply shifts the coordinate $\varphi$.

These coordinates are much better suited to the simulation of collisions than those used in Ref.~\cite{Johnson:2011wt}. The relations between the coordinates and metric functions used by Ref.~\cite{Johnson:2011wt} and in this work are given by: $N = \sinh^{-1} \tilde{z}$, $\alpha_\mathrm{new} = \alpha_\mathrm{old} \cosh N$, and $a_\mathrm{new} = a_\mathrm{old}/\cosh N$. It can be seen that we avoid exponential changes in the lapse and shift, and can follow the simulation through the end of inflation ($N \sim 60$).

The equation of motion for the field $\phi$ is given by $\square \phi = \partial_\phi V$, and the equations for the metric functions $\alpha$ and $a$ come from the Einstein equations. These were worked out in detail in Ref.~\cite{Johnson:2011wt}. We restate the results here in terms of our new gauge choice. First, let us define
\begin{align}
\Pi &\equiv \frac{a}{\alpha}\frac{d\phi}{dN} \\
A &\equiv  \tanh (N) + \frac{1}{2\tanh N} - \frac{\alpha^2}{2}\left(\frac{1}{\cosh (N) \ \sinh (N)} + 8\pi\tanh (N)\; V(\phi)\right) \\
B &\equiv  2\pi \ \tanh (N) \ \frac{\alpha^2}{a^2} \left(\frac{\phi'^2}{\cosh^2(N)}+\Pi^2\right) ,
\end{align}
where $\phi$ is measured in units of the Planck mass, $\phi' = d\phi/dx$, and $V(\phi)$ has been rescaled such that $V(\phi_F) = \tfrac{3}{8\pi}$. 
Then the field and metric functions evolve as
\begin{align}\label{eq:equationsofmotion}
\frac{d\alpha}{dN} &= \alpha(A+B) \\
\frac{da}{dN} &= a(-A+B) \\
\frac{d\Pi}{dN} &= -\left(\tanh (N) + \frac{2}{\tanh (N)}\right)\Pi + \frac{d}{dx} \left(\frac{\alpha \phi'}{a\cosh^2 (N)}\right) - \alpha a \partial_\phi V.
\end{align}
The solutions to these equations should satisfy the constraint
\begin{equation}
\label{eq:sim_constraint}
\frac{d\alpha}{dx} = \frac{4\pi \tanh (N)\; \alpha^2 \phi' \Pi}{a}.
\end{equation}

The initial field configuration comes almost directly from the instanton calculation. In the limit where $N \rightarrow 0$, the field-dependent terms in the equations of motion Eq.~\ref{eq:equationsofmotion} are sub-dominant. At $N=0$, the metric functions are $\alpha(N=0, x) = a(N=0, x) = 1$. Comparing metric equations \ref{eqn:metric1} and \ref{eqn:metric2} near $N=0$, we can see that for a single bubble, after appropriately rescaling $\phi$ and $r$, we simply have $\phi(x, N=0) = \phi(r=|x|)$. Since the instanton is symmetric in time, $\Pi(N=0) = 0$. A second instanton can be added by translating the first instanton along the $x$-axis and then summing the fields. Technically, a single instanton modifies the field and the metric across the entire compact Euclidean space. Its effect upon a second instanton's field configuration is generally non-negligible. 
However, we shall assume that for cases of interest, the two instantons can be considered to have finite size, and as long as their separation significantly exceeds the sum of their radii, they can be considered independently.

We start the simulation at some small value of $N$, where the metric Eq.~\ref{eqn:metric2} becomes arbitrarily close to a slicing of Minkowski space with $SO$(2,1) symmetry. The spatial dependence of the metric functions will be negligible in this limit, where anisotropic curvature dominates the Einstein equations. For further discussion of this point, see Ref.~\cite{Johnson:2011wt}. Performing a Taylor expansion about $N=0$, the lowest order surviving terms are
\begin{align}
\alpha &= 1 - \alpha_2(x) N^2 \\
a &= 1 + a_2(x) N^2 \\
\phi &= 1 + \phi_2(x) N^2 \\
\Pi &= 2\phi_2 N.
\end{align}
Inserting these into the equations of motion Eq.~\ref{eq:equationsofmotion}, we obtain
\begin{align}
\alpha_2 &= -\frac{1}{2} + \frac{2\pi}{3}\left(2V(\phi_0) - \phi_0'^2\right) \\
a_2 &= -\frac{1}{2} + \frac{4\pi}{3}\left(V(\phi_0) + \phi_0'^2\right) \\
\phi_2 &= \frac{1}{6}\left(\phi_0'' - \left.\partial_\phi V\right|_{\phi_0}\right),
\end{align}
where a prime denotes differentiation with respect to $x$. We then use this as the input for the simulation.

\subsection{Integration and adaptive mesh refinement}
\label{sec:simulation_integration}

We integrate the above equations of motion using the method of lines (see e.g., Ref.~\cite{nrbook} for a general discussion of the numerical methods incorporated below). In the simplest possible implementation, one would discretize the field in the $x$-direction on a uniform grid and find the $x$-derivatives using finite differences. One would then integrate the field in time using any standard numerical integrator, with the field's value at each discrete point being a separate integration variable. This was the technique used in Ref.~\cite{Johnson:2011wt}, and with sufficient resolution it should work here as well. Unfortunately, as the bubbles expand, their walls quickly approach the speed of light and become severely length-contracted. The shape of the wall is extremely important for accurately reproducing the perturbed metric, as we will show later, so we need extremely high resolution in its surrounding region. This is not feasible with a uniform grid.

So instead, we switch to a method involving adaptive mesh refinement (AMR).  AMR describes computational algorithms which introduce resolution dynamically where and when such resolution is needed. A number of variations appear in the literature, and the one developed here can be considered a variant of that introduced by Berger and Oliger~\cite{Berger1984484}. In common with that work, this scheme adapts both the spatial resolution $\Delta x$ and the time-step $\Delta N$. However, it is important to note the following points, which are unique to this approach and appropriate for just a single spatial dimension: (i) spatial resolution is continuously adaptable instead of discretely, (ii) only a single set of points covers any particular region of the spatial domain, (iii) AMR boundaries are handled in the spirit of the `tapered boundary' approach of Ref.~\cite{0264-9381-23-16-S08}, and (iv) regridding occurs only globally when all regions are time-aligned.  Our algorithm proceeds as follows:

\begin{enumerate}
\item The first step, after setting the initial conditions for the simulation, is to create the initial grid.
To do this, the algorithm calculates a desired grid spacing at each of the input points. The grid density should be high near the bubble walls where the field and metric functions change rapidly, and relatively low elsewhere. We set the grid density $m \equiv dn_{\rm grid}/dx$ to be directly proportional to $d\phi/dx$, such that the number of grid points along the bubble wall is $n_{\rm wall} = \int_{\rm wall} m\; dx$. We also enforce a minimum grid density $m_{\rm min}$. In this paper, we use $n_{\rm wall} = 70$ and $m_{\rm min} = 500$. It is helpful if the grid density does not vary too quickly: otherwise, sharp features in the simulation could migrate from high resolution regions to low resolution regions before the algorithm recomputes the grid. We set an additional minimum of 
\begin{equation}
m(x_i) \geq \left[ m^{-1}(x_j) + \frac{\log 2}{n_2} \left| x_i - x_j \right| \right]^{-1}
\end{equation}
for all points $i$ and $j$. This guarantees that there will be at least $n_2$ points along the grid before the density drops by a factor of 2. We take $n_2 = 40$. With the grid density set, the code integrates $m$ to create the new grid points.

\item Next, the values of the field and metric functions need to be computed along the new grid. The algorithm creates a cubic spline using the { SciPy} \cite{scipy} routine {\tt interpolate.splrep()}, which  matches each of the input points along the old grid and is $C_2$ continuous. The new values are then interpolated from the spline at the new grid points.

\item The equations of motion depend on first and second spatial derivatives, which one can calculate using the differences between neighboring grid values. For a uniform grid, the differentiation stencil is the same for all grid points (excluding the boundaries). For example, to fourth order in $\Delta x$,
\begin{equation}
\frac{dy_i}{dx} = \frac{  y_{i-2} - 8 y_{i-1} + 8 y_{i+1} - y_{i+2} }{12\Delta x} + \mathcal{O}(\Delta x^4).
\end{equation}
For a non-uniform grid, a different stencil must be calculated at each grid point. The code finds these stencils to 4th order in $\Delta x$ for first derivatives and to 3rd order for second derivatives using the recurrence relations defined in Ref.~\cite{fornberg1988}. At the simulation boundaries the stencils are the same order, but they are no longer centered on the grid points whose derivatives they calculate.

\item  The entire computational domain naturally separates into regions based on their local resolution.
 These regions will share a common time-step $\Delta N$ and will be integrated in time together (as per point 6 below), and thus their grid spacing will all be within roughly a factor of two of each other. 
 We additionally require that each region contain at least 40 grid points.\footnote{
Grid spacing variations slightly larger than a factor of 2 are allowed if they keep the number of grid points per region larger than 40.}
 The boundaries of these regions are purely an artifact of the AMR itself and are called AMR boundaries. To enable proper time integration at such boundaries, the regions are extended spatially to overlap with neighboring regions.\footnote{
Regions with finer grid spacing extend 16 grid points into their coarser neighbors, and coarser regions extend 8 grid points into their finer neighbors.}
  This overlap is similar in spirit to the `tapered boundaries' introduced in Ref.~\cite{0264-9381-23-16-S08} and avoids the generally lower-order temporal interpolation of the original Berger and Oliger scheme~\cite{Berger1984484}.

\item Each region evolves independently for at least one time-step using a fourth-order Runge-Kutta integration. The time-step of the most finely-spaced region is set by the Courant-Friedrichs-Lewy (CFL) condition~\cite{CFL_ref}, which demands that the time-step be smaller than the amount of time it takes information to pass between neighboring grid points. That is, $c\Delta N \lesssim \Delta x$ (we use $c\Delta N \approx 0.2 \Delta x$). Time-steps in more coarsely-spaced regions are set to be powers of 2 larger than that in the most finely-spaced region (so that such regions will be periodically time-aligned).  Because of the overlap, we need not worry about boundary effects at the edge of each region: no information can propagate from outside the overlap into a region's interior before a single time-step has completed.
At the end of each time-step in each region, the overlapping points are replaced with their corresponding values in the interior of each regions' neighbors, as long as the neighbors have been integrated up to the same time. 
Since the time-steps come in powers of two, a fine region must integrate two steps for every one step in a coarser neighboring region before the two will match.
We set an upper bound on the time-step of the coarsest region to be no more than $2^6$ times greater than the time-step in the finest region. This, along with the above-defined minimum region width and time-step size, guarantees that features in high resolution regions will not evolve into low density regions before the coarsest-gridded region has taken a single time-step.

\item After each region has made at least one time-step (the finest-spaced region may make as many as $2^6$ steps so that all regions can be matched), the code loops back to step 1 and recreates the non-uniform grid. At specified values of $N$, the entire grid can be saved to file for later retrieval.

\end{enumerate}

In addition to the CFL condition, the integration time-step is limited by the frequency of oscillations about minima in the potential. This will always become the limiting factor near the end of inflation, since $\alpha$ increases roughly exponentially when $N$ is large and $V(\phi)=0$, and $d^2\phi/dN^2 \approx -\alpha^2 \partial_\phi V$ so $\omega_\mathrm{osc} \approx \alpha \sqrt{\partial_\phi^2 V}$ becomes very large. We require that the time-step be no larger than $\frac{1}{30}$ of the oscillatory period about the inflationary minimum.

The entire simulation is written in C with python wrapper functions for easy interactive analysis and for communication with the {\tt CosmoTransitions} package.

\subsection{Universes collide} 

\begin{figure}[t] 
   \centering
   \includegraphics[width=3in]{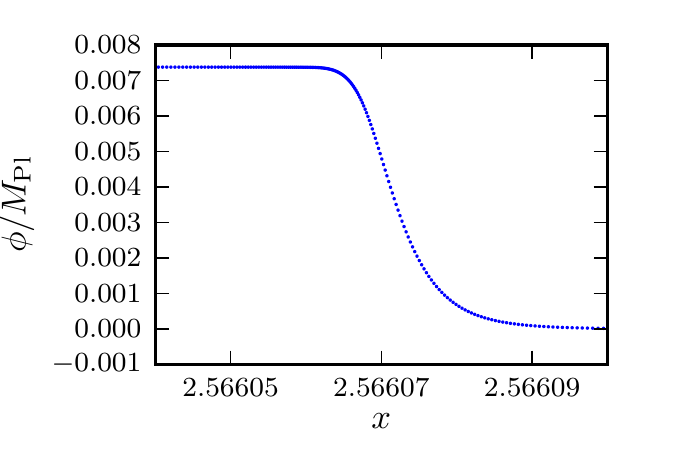} 
   \caption{Bubble wall profile at $N=6$ at the full simulation resolution. Each dot is one of the discrete points in the simulation itself. The wall is highly length-contracted, so the density of points needs to be extremely large in this region. The region shown here is roughly one hundred thousand times smaller than the size of the full simulation.
   }
   \label{fig:sim_wall}
\end{figure}

\begin{figure}[t] 
   \centering
   \includegraphics[width=6in]{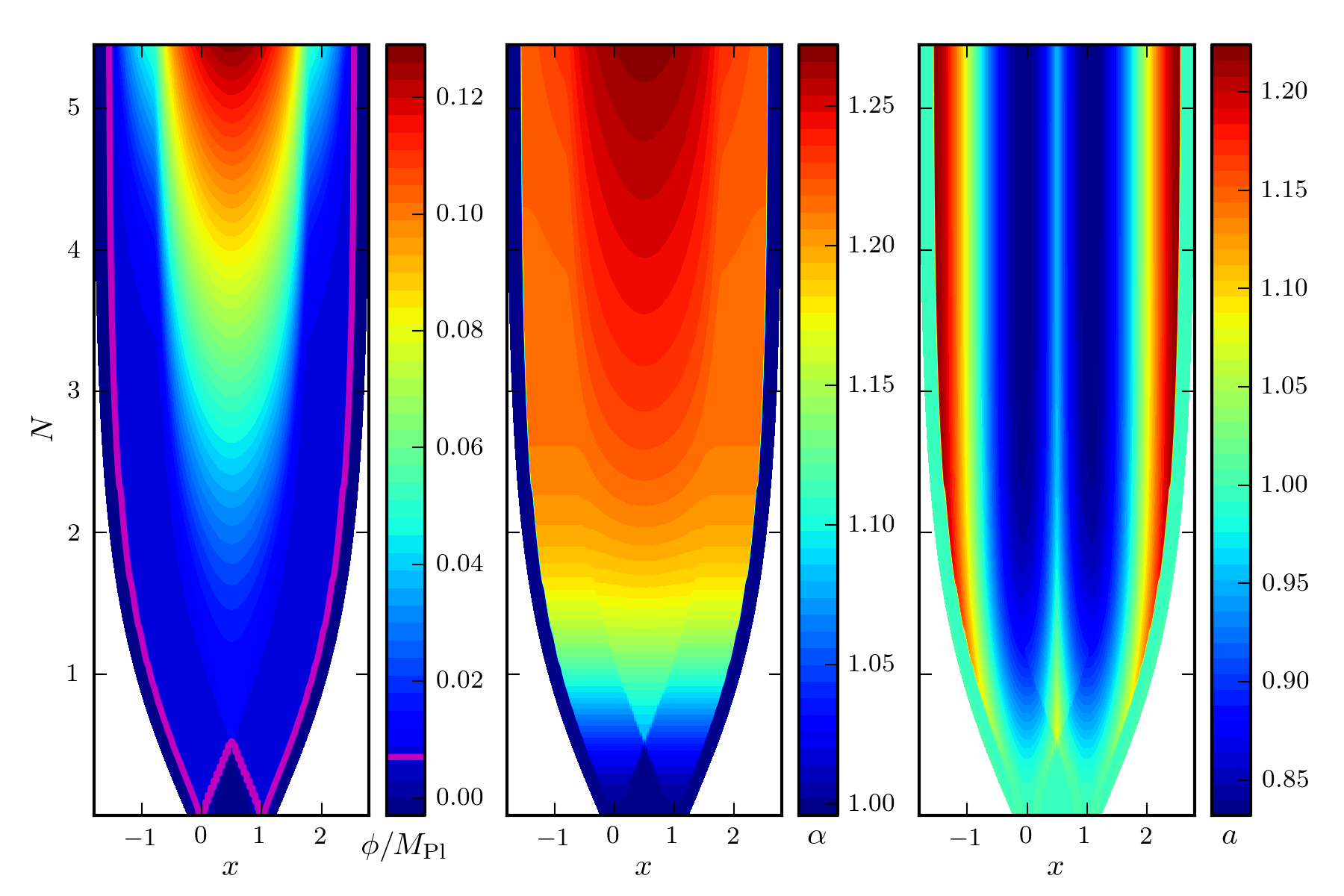} 
   \caption{Contour plots of the field and metric functions up to $N=5.5$. The white regions are outside the simulation boundaries, and the thick magenta line in the left plot shows the location of the bubble wall.}
   \label{fig:sim_contour1}
\end{figure}

Here we present the results for a single simulation of two bubbles colliding in the `L2' potential with an initial separation of $\Delta x_\mathrm{sep} = 1.0$. We run the simulation in two steps. In the first step, the boundaries of the simulation are outside the bubble walls, and at each time-step we enforce the boundary conditions $\phi = \Pi = 0$; $\alpha = a = 1$. The simulation region grows along with the bubbles\footnote{
The boundaries grow during the spline interpolation step. The expansion happens along null geodesics so that the entire simulation region is (just barely) within the causal future of the initial simulation region.
} so that it does not needlessly integrate the field in the false vacuum (see Fig.~\ref{fig:sim_contour1}). As the bubbles grow, their walls get length-contracted, and more points are added by the adaptive grid algorithm to resolve the wall structure (see Fig.~\ref{fig:sim_wall}). The adaptive grid follows the bubble walls very well, but eventually the extra points noticeably slow down the computation.

In the second step, which we choose to start at $N=7$, 
the simulation region is trimmed to exclude the bubble walls and thereafter held constant. We calculate the spatial derivatives near the boundary using non-centered finite differences. Of course, we should not trust the results of the simulation near the boundary, or anywhere in the boundary's causal future, but luckily this region is very small. For constant $\alpha$ and $a$, null geodesics obey
\begin{gather}
x(N) = \frac{\alpha}{a}2\tan^{-1}\left[ \tanh \left(\frac{N}{2} \right)\right] \\
\longrightarrow x(\infty) - x(N) \approx 2e^{-N}.
\end{gather}
Since $\alpha/a \sim 1$ near the region's boundary at $N=7$, the maximum width of the boundary's future light cone is $\delta x \approx 0.002$, which is insignificant in comparison with the total size of the simulation region.
By simulating all the way to end of the inflation, we can explicitly check the assumption that the perturbations are frozen in.

The entire simulation takes about 7 minutes to complete on a single core of a modern 2.66 GHz desktop computer, about 6 minutes of which is spent evolving the bubble walls up to $N=7$.

\begin{figure}[t] 
   \centering
   \includegraphics[width=6in]{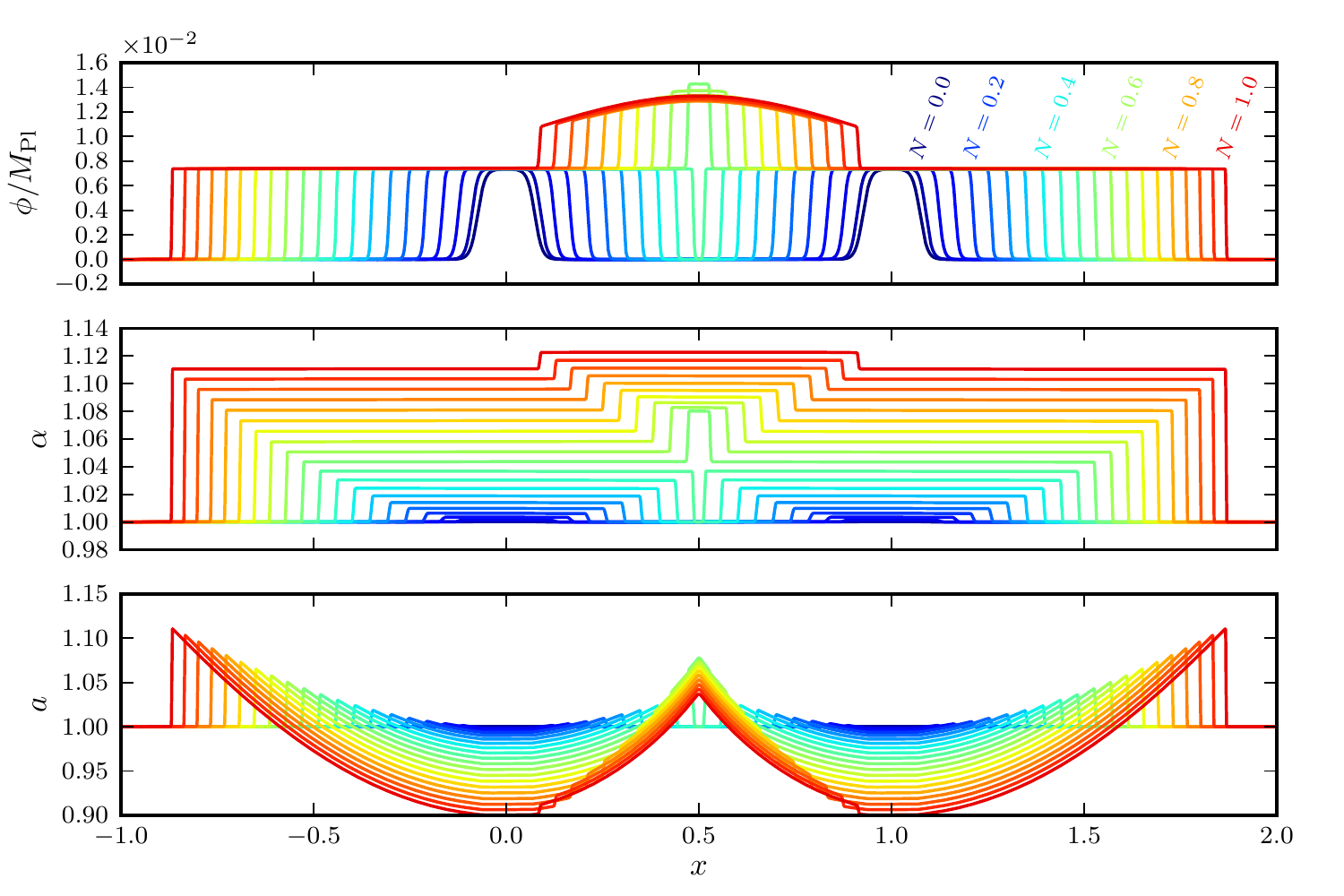} 
   \caption{Evolution of the field and metric functions up to $N=1$. Individual lines represent time slices at constant $N$. 
   }
   \label{fig:sim_lines1}
\end{figure}

Fig.~\ref{fig:sim_lines1} shows the initial field configuration and the immediate aftermath of the collision. At $N=0$, $\alpha=a=1$ and the two instantons are stationary. They then begin to grow, colliding around $N \approx 0.5$. The centers of each bubble evolve very little during this time --- they are very close to the inflection point $\phi_I$, so they do not move much along the potential. After the collision, the two bubbles essentially superimpose --- the `free passage' approximation~\cite{Giblin:2010bd,Easther:2009ft,Amin:2013dqa,Amin:2013eqa}. The sharp edges of the superimposed bubble walls relax somewhat since there is no longer a strong pressure (potential energy) gradient supporting them. Since the field in the collision region is farther away from $\phi_I$, it begins to slowly roll along the inflationary potential. In contrast, the field outside the collision region does not move significantly away from $\phi_I$ for several more $e$-folds.

\begin{figure}[h] 
   \centering
   \includegraphics[width=6in]{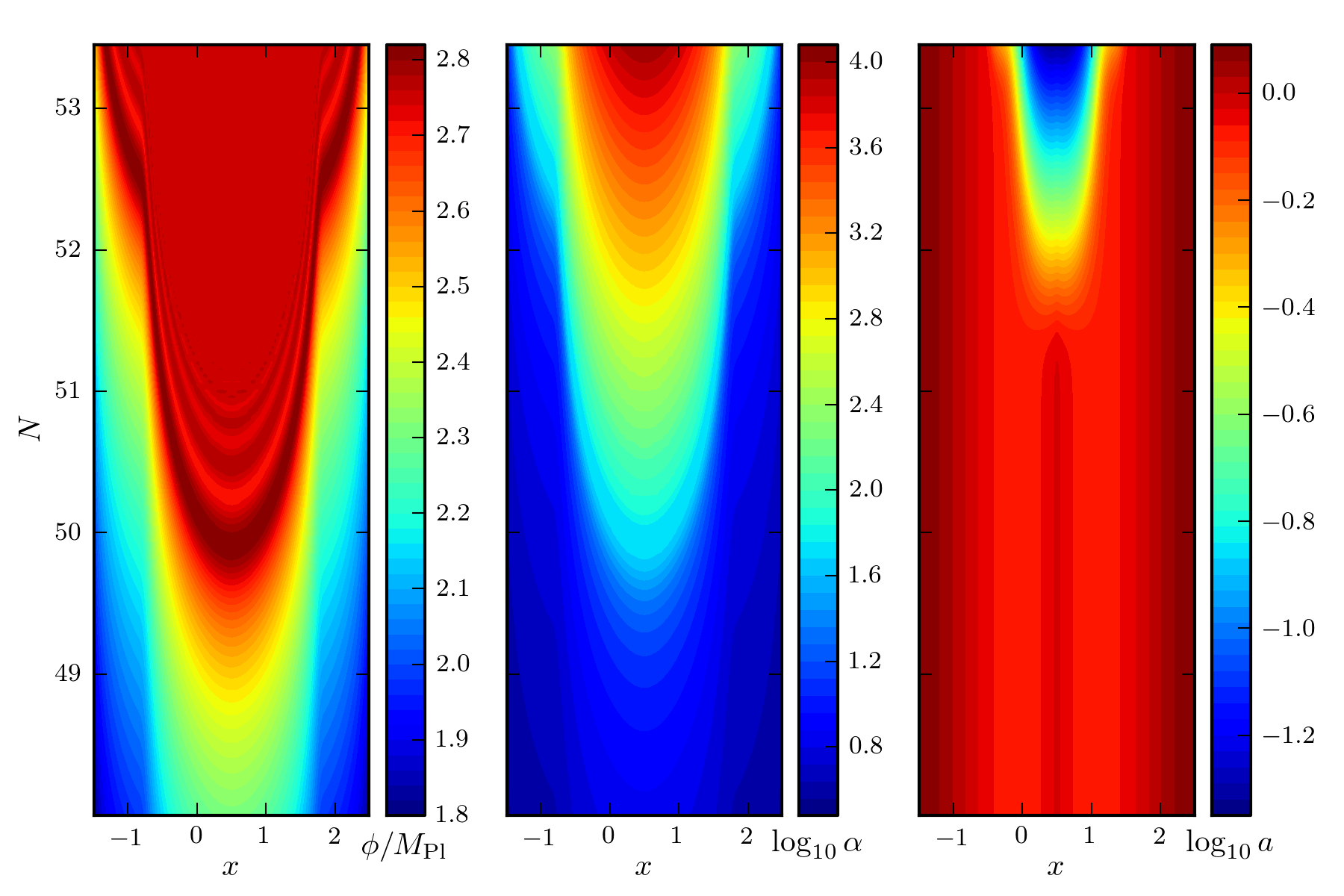} 
   \caption{Contour plots of the field and metric functions near the end of inflation. Note that the metric function contours are on a log scale, and $\alpha$ increases exponentially.}
   \label{fig:sim_contour2}
\end{figure}

Fig.~\ref{fig:sim_contour1} shows contours of the field and the metric functions for $N<5.5$. Again, we see that the field in the collision region starts rolling down the inflationary potential much sooner than the field outside the collision region. The metric function $a$ quickly approaches a stable configuration, while $\alpha$ continues to increase with increasing $N$. This matches the results from Ref.~\cite{Johnson:2011wt}. Fig.~\ref{fig:sim_contour2} shows the behavior near the end of inflation. At $N\approx 50$, the field $\phi$ reaches the inflationary minimum in the collision region and begins oscillating coherently inside each bubble. The metric function $\alpha$ starts to grow exponentially, while $a$ decreases (although $a\cosh N$ continues to increase, so the universe continues to expand). As noted above, the frequency of oscillations about the minimum becomes very large in this region, while the oscillations themselves die down in amplitude. Since lines of constant $\phi$ do not correspond to constant $N$, high frequency temporal oscillations also imply high frequency spatial oscillations. The density of grid points does not significantly increase during this period (the density of points is set to be proportional to $|d\phi/dx|$, which remains relatively small), so eventually the frequency of oscillations becomes smaller than the grid spacing and information about the oscillations is lost. The quality of the solution degrades at this point. Nevertheless, for our purposes we can reconstruct the perturbed metric well within the regime of accurate solutions.

\begin{figure}[h] 
   \centering
   \includegraphics[width=6in]{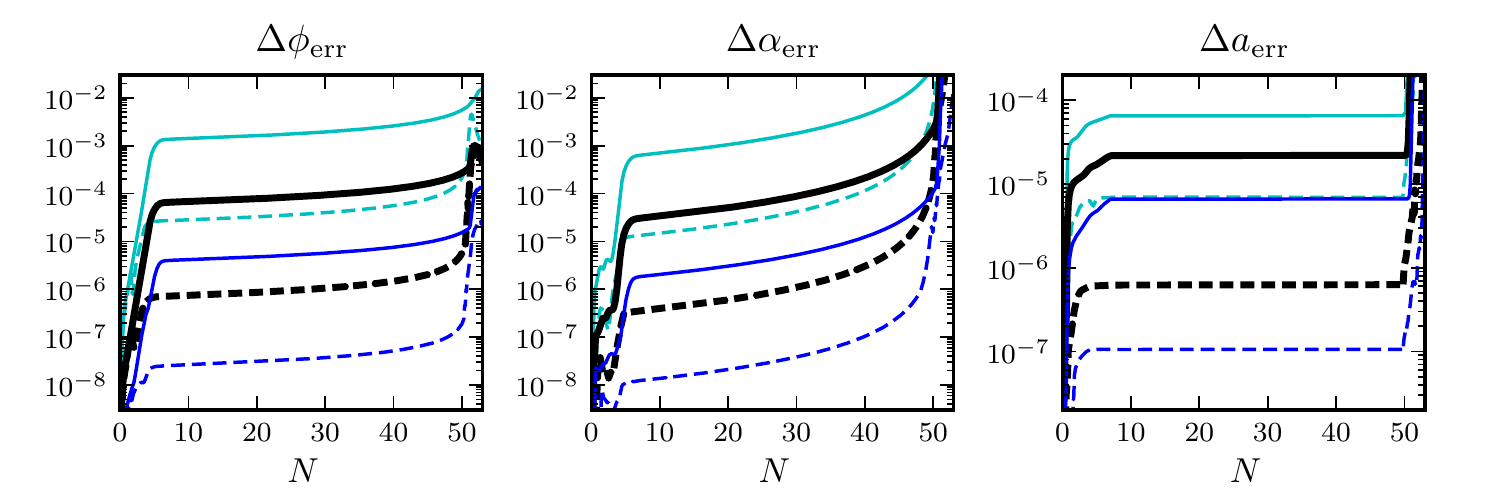} 
   \caption{
   Estimated errors for the field $\phi$ and the metric functions $\alpha$ and $a$ with an initial bubble separation of $\Delta x_{\rm sep} = 1.0$ for a simulation utilising adaptive mesh refinement.  The error at any given grid point is estimated as the value at that point minus the value at the same point in a simulation with half the grid spacing. Solid lines show the root-mean-squared error across individual time slices (excluding the exterior bubble walls), and dashed lines show the median absolute error across time slices. The thick black lines show the error estimates for a simulation with our default input parameters. The thin light-blue (dark-blue) lines show error estimates for simulations with double (half of) the default grid spacing. Note that they tend to plateau before $N=7$ --- truncating the simulation domain at $N=7$ does not noticeably impact the error estimates and is not related to the plateauing behavior.
   }
   \label{fig:sim_convergence}
\end{figure}

\begin{figure}[h] 
   \centering
   \includegraphics[width=6in]{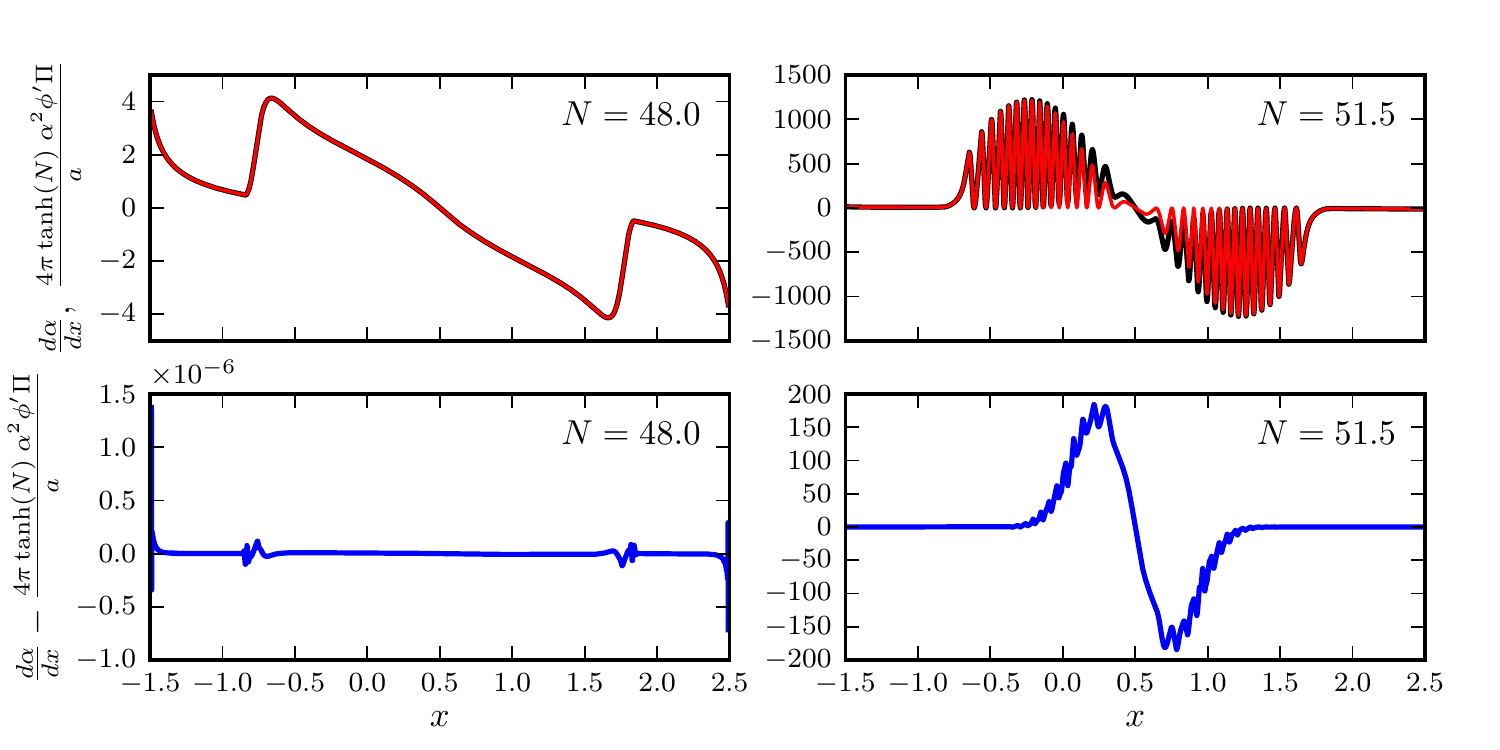} 
   \caption{
   An examination of the residual of the constraint equation before
   (left) and after (right) the end of inflation. In the top plots, the thick black line represents $d\alpha/dx$ (the left-hand side of Eq.~\ref{eq:sim_constraint}) and the thin red line represents $4\pi \tanh(N) \alpha^2 \phi' \Pi/a$ (the right-hand side of Eq.~\ref{eq:sim_constraint}). These would ideally be exactly coincident. The bottom plots 
   show the residuals, the differences between the left and right sides of the constraint equation, at the two times.
   Note that the $y$-axis in the bottom-left plot has a maximum value of $1.5\times10^{-6}$.
   }
   \label{fig:sim_constraint}
\end{figure}

We explicitly check the convergence of the simulation by varying the grid spacing parameters and, implicitly, the integration time-step. To uniformly double the density of grid points across the simulation, we need to double the input parameters $n_{\rm wall}, m_{\rm min}, n_2$, the minimum number of grid points per region, and the minimum number of time-steps per oscillation about the inflationary minimum. The Runge-Kutta integration has an error of order $\Delta N_{\rm step}^4 \propto \Delta x^4$, and the first spatial derivative has an error of order $\Delta x^4$. The second derivative formally has an error of order $\Delta x^3$, but by symmetry this leading term is zero for uniform grids and the error becomes order $\Delta x^4$. Since the grid is increasingly uniform for smaller grid spacing, the error is effectively $\Delta x^4$. Fig.~\ref{fig:sim_convergence} shows the error associated with the field $\phi$ and the metric functions $\alpha$ and $a$ as a function of time $N$. The error at any given grid point is estimated as the value at that point minus the value at the same point in a simulation with half the grid spacing. Solid lines show the root-mean-squared error across individual time slices, and dashed lines show the median absolute error. For our chosen input parameters, the error tends to be smaller inside the collision region, which is better represented by the median error estimate. The error in both $\phi$ and $\alpha$ converges as $\Delta x^4$ (that is, the error increases by a factor of $\sim\!16$ when the grid spacing doubles), as expected. The error in the metric function $a$ only converges as $\Delta x^2$, but with a seemingly subleading contribution. As expected, the error is large after the end of inflation when the spatial oscillation frequency is comparable to the grid spacing.

We also check that the constraint equation (Eq.~\ref{eq:sim_constraint}) holds (see Fig.~\ref{fig:sim_constraint}). The constraint holds extremely well (to about 1 part in $10^7$ in relative terms) until inflation ends and the field begins rapid oscillation,
after which the simulation (expectedly) performs increasingly poorly in satisfying the constraint.
Together with the convergence tests described above, this indicates that our code very accurately models the collision spacetime. Both the convergence properties and the behavior of the constraint residuals in this code represent very significant improvements over the code used in Ref.~\cite{Johnson:2011wt}.

\section{Finding the Observer Metric}
\label{sec:cosmo_metric}

To simulate the collision between two bubbles, we choose coordinates that cover the entire collision spacetime. These are clearly not the coordinates appropriate for describing a nearly homogeneous and isotropic cosmology inside the colliding bubbles. It is possible to describe cosmology in terms of manifestly geometric and covariant quantities~\cite{Ellis:1989jt}. This was the approach in Ref.~\cite{Xue:2013bva}. For extracting signatures of bubble collisions, it is most convenient to describe the bubble interior in terms of perturbed open FRW coordinates. The challenge in this case is to define a map between such a set of coordinates and the coordinates used in the simulation. In general, this is an ill-defined problem --- it is not easy to find a coordinate transformation connecting two generic metrics. 

The key insight is to note that as the proper time goes to zero, an open FRW universe can be very well approximated by the Milne slicing of Minkowski space. In the Milne patch, position is labeled by the initial velocity of geodesics emanating from the origin of spherical coordinates. This suggests a prescription for reconstructing the synchronous gauge metric inside bubbles. First, follow a set of geodesics from the bubble center, each labeled by their initial velocity, and track the proper time along each by evolving the geodesic equations through the simulation. Then identify a new set of coordinates associated with the geodesic labels and the proper time, and explicitly construct the map to the simulation coordinates. For an unperturbed bubble, transforming the metric to the geodesic coordinates yields the metric on an open FRW universe. For a perturbed bubble, this procedure yields a perturbed open FRW metric in the synchronous gauge. The synchronous gauge is in general a continuous family of coordinate systems, leading to an ambiguity. However, in our case this ambiguity is fixed by the geometrical prescription used to construct the metric. In effect, the gauge is anchored by the requirement that geodesics probing the unperturbed region of the bubble must describe precisely an unperturbed open FRW universe.

In the following subsections, we build the metric for arbitrary observers in the observation bubble. In Sec.~\ref{sec:observer_metric}, we describe two coordinate systems in terms of geodesic trajectories: a Cartesian coordinate system ($\tau, X,Y,Z$), and an anisotropic hyperbolic coordinate system ($\tau, \xi, \rho, \varphi$). The former is useful for describing individual observers and perturbations, while the latter is convenient for integrating a grid of geodesics. In both cases, the time coordinate $\tau$ labels the proper time along the geodesics.
In Sec.~\ref{sec:geodesics} we describe the geodesic integration, which supplies the map from simulation coordinates ($N, x, \chi, \varphi$) to anisotropic hyperbolic coordinates, and we show the integrated results. Importantly, $N$ and $x$ are functions of $\tau$ and $\xi$ only, so we only need to vary the initial trajectory $\xi$ to recover all features of all geodesics in the ($3+1$)-dimensional space.
We would then like to find observer-centered Cartesian coordinates --- that is, coordinates in which the observer resides at $X=Y=Z=0$ --- and to do this we must perform a change in frame which takes the observer to the origin, which we describe in Sec.~\ref{sec:frames}. Since the coordinate systems describe an open universe, the frame-changing coordinate transformation is more complicated than the Euclidean translation that one would find in flat space. 
In Sec.~\ref{sec:metric_reconstruction}, we describe how to transform from the simulation metric to the observer metric using this multi-step map between the simulation coordinates and the observer-centered Cartesian coordinates. Table~\ref{tab:coordinate_list} summarizes the coordinate systems discussed below.

\subsection{Coordinate Systems for the Observation Bubble}
\label{sec:observer_metric}

An important first step is to identify a convenient coordinate system
on the bubble interior in the absence of a collision. An unperturbed single bubble contains an infinite open universe, a fact dictated by the $SO(3,1)$ symmetry of the instanton.  We utilize two coordinate systems on the open FRW universe, each of which manifests different properties of the spacetime. 
\begin{enumerate}
\item {\bf Cartesian Coordinates}. The open FRW universe can be covered by a Cartesian set of coordinates with metric
\begin{equation}\label{eq:unperturbed_cartesian}
H_F^2 ds^2 = -d\tau^2 + \left[\frac{a(\tau)}{1-\tfrac{R^2}{4}}\right]^2\left(dX^2 + dY^2 + dZ^2\right),
\end{equation}
where $R^2 \equiv X^2 + Y^2+ Z^2$, $a(\tau)$ is the (dimensionful) scale factor, and the coordinates take values in the ranges $0 < \tau < \infty$, $0 < R < 2$. 
For ease of comparison with the simulation coordinates, the proper time $\tau$ and the scale factor $a(\tau)$ are both measured in terms of the false-vacuum Hubble scale. The coordinates $X, Y$ and $Z$ are all unitless.
If tunneling occurs directly to an inflationary plateau inside the bubble, the scale factor is given by $a(\tau) =  \frac{H_F}{H_I}  \sinh \left( \frac{H_I}{H_F} \tau \right)$, and the comoving curvature radius is $R_0 = \frac{H_F}{H_I}$. For $R \ll 1$, the Cartesian coordinates approach those of a flat universe.  
The Cartesian foliation of an open universe is sometimes referred to as the Poincar\'e disk.

As $\tau \rightarrow 0$, for any non-singular configuration we have $a(\tau) \rightarrow \tau$, and the metric approaches the Milne slicing of Minkowski space:
\begin{equation}\label{eq:unperturbed_cartesian2}
H_F^2 ds^2 = -d\tau^2 + \left[\frac{\tau}{1-\frac{R^2}{4}}\right]^2 \left(dX^2 + dY^2 + dZ^2\right).
\end{equation}
In terms of the coordinates on Minkowski space $t,x,y,z$, we have\footnote{
The $\tau=0$ Milne slice of Minkowski space is the lightcone emanating from the origin. However, the simulated spacetime is approximately flat only at the origin, especially when we start considering colliding bubbles. Therefore, we need to be careful to take the limit $\tau \rightarrow 0$ after first fixing $R<2-\epsilon$ for any finite $\epsilon$. Using this limiting procedure, the spacetime can always be considered flat for small $\tau$.
}
\begin{equation}
t = \tau\frac{4+R^2}{4-R^2}, \ x = \tau\frac{4X}{4-R^2}, \ y = \tau\frac{4Y}{4-R^2}, \ z = \tau\frac{4Z}{4-R^2}.
\end{equation}

Defining 
\begin{equation}
R = 2 \tanh \left[ \frac{\eta}{2} \right] 
\end{equation}
and $X = R \cos \theta$, $Y = R \sin \theta \cos \phi$, $Z = R \sin \theta \sin \phi$, we obtain
\begin{equation}
t = \tau \cosh \eta, \ x = \tau\sinh \eta \cos \theta, \ y = \tau \sinh \eta \sin \theta \cos \phi, \ z = \tau \sinh \eta \sin \theta \sin \phi.
\end{equation}
Therefore, we see that the Cartesian coordinates in an open universe can be assigned a very physical interpretation. The coordinates can be thought of as labels attached to a congruence of geodesics emanating from the origin of Minkowski space: $\tau$ is the proper time along each geodesic, $\eta$ is the initial rapidity (the radial velocity and rapidity are related by $v = \tanh \eta$) of each geodesic, and $\theta$ and $\varphi$ parameterize the initial direction of each geodesic. Therefore, we could build up the metric inside an open FRW universe by tracking a set of labeled geodesics. This is the basic method we employ below; however, there is a slightly more convenient coordinate system, for which the symmetries of the collision spacetime are manifest.

\item {\bf Anisotropic Hyperbolic Coordinates}. To make direct connection with the symmetries of the collision spacetime, we define a new set of coordinates. In terms of the open universe Cartesian coordinates, these are defined by
\begin{align}
\label{eq:xiToX}
X &= \frac{2  \sinh\xi}{1+\cosh\xi \cosh\rho} \\
\label{eq:Y_from_xirho}
Y &= \frac{2 \sinh\rho\cosh\xi}{1+\cosh\xi\cosh\rho} \cos \varphi \\
Z &= \frac{2 \sinh\rho\cosh\xi}{1+\cosh\xi\cosh\rho} \sin \varphi,
\end{align}
or equivalently, 
\begin{align}\label{eq:xirhophi}
\sinh\xi &= \frac{X}{\left( 1-\frac{R^2}{4} \right)},&\;\; 
\tanh\rho &= \frac{\sqrt{Y^2+Z^2}}{ \left( 1+\frac{R^2}{4} \right)},&\;\;
\tan\varphi &= \frac{Z}{Y}.
\end{align}
These coordinates take values in the ranges $-\infty < \xi < \infty$, $-\infty < \rho < \infty$, $0 < \varphi < 2 \pi$, and cover the entirety of the bubble interior. The metric is
\begin{equation}\label{eq:anisotropic_c}
H_F^2 ds^2 = -d\tau^2 + a(\tau)^2 \left[ d\xi^2 + \cosh^2 \xi \left( d\rho^2 + \sinh^2 \rho d\varphi^2 \right) \right] .
\end{equation}
As $\tau \rightarrow 0$, again we have $a(\tau) \rightarrow \tau $,
 and we can see that this metric is an anisotropic slicing of Minkowski space,
\begin{equation}\label{eq:anisotropic_m}
H_F^2 ds^2 = -d\tau^2 + \tau^2 \left[ d\xi^2 + \cosh^2 \xi \left( d\rho^2 + \sinh^2 \rho \ d\varphi^2 \right) \right] . 
\end{equation}
In terms of the Cartesian coordinates on Minkowski space, we have
\begin{equation}\label{eq:minkowski_anihyp}
t = \tau \cosh \xi \cosh \rho, \ \ x= \tau \sinh \xi, \ \ y = \tau \cosh \xi \sinh \rho \cos \varphi, \ \ z = \tau \cosh \xi \sinh \rho \sin \varphi.
\end{equation}
\end{enumerate}

\subsection{Geodesic Evolution}
\label{sec:geodesics}

As discussed, the coordinate systems for the observation bubble defined above have a clear geometric interpretation as a congruence of labeled geodesics emanating from the center of the bubble at $\tau = 0$. Since the collision will never affect the point where the geodesics begin, we can straightforwardly extend any of the coordinate systems defined above to describe the interior of the perturbed observation bubble. The map between the simulation coordinates and the coordinates foliating the observation bubble can then be obtained by evolving the geodesic equation for each of the labeled geodesics. Because they explicitly manifest the symmetry of the collision spacetime, we employ the anisotropic hyperbolic coordinates. The simulation metric functions are independent of $\rho$ and $\varphi$, and therefore geodesic motion in these directions is trivial. Evolving the geodesics produces two maps: $N(\tau, \xi)$ and $x(\tau, \xi)$. 

The metric for the simulation coordinates (Eq.~\ref{eqn:metric2}) in the limit where $N \rightarrow 0$ is given by
\begin{equation}
H_F^2 ds^2 = -dN^2 + dx^2 + N^2 \left( d\rho^2 + \sinh^2 \rho \ d\varphi^2 \right) .
\end{equation}
Comparing this with Eq.~\ref{eq:anisotropic_m}, the coordinate map between the two metrics is
\begin{equation}\label{eq:xNxitau}
N = \tau \cosh \xi, \ \ x = \tau \sinh \xi .
\end{equation}
This provides a map at early times, allowing one to label geodesics by their initial conditions. From Eq.~\ref{eq:xNxitau}, the initial conditions for geodesics are
\begin{eqnarray}
\left.\frac{dN}{d\tau}\right|_{\tau=0} &=& \cosh \left(\xi \right),\;\;     N(\tau=0) = 0, \\
\left.\frac{dx}{d\tau}\right|_{\tau=0} &=& \sinh \left(\xi \right), \;\;    x(\tau=0) = 0, \\
\left.\frac{d \rho}{d\tau}\right|_{\tau=0} &=& 0,\;\;     \rho(\tau=0) = \rho, \\
\left.\frac{d \varphi}{d\tau}\right|_{\tau=0} &=& 0, \;\;    \varphi(\tau=0) = \varphi.
\end{eqnarray}

The relevant geodesic equations are 
\begin{gather}
\frac{d^2 N}{d\tau^2} + \Gamma^N_{NN} \left(\frac{dN}{d\tau}\right)^2 + 2\Gamma^N_{Nx} \frac{dN}{d\tau}\frac{dx}{d\tau} + \Gamma^N_{xx}  \left(\frac{dx}{d\tau}\right)^2 = 0, \\
\frac{d^2 x}{d\tau^2} + \Gamma^x_{NN} \left(\frac{dN}{d\tau}\right)^2 + 2\Gamma^x_{Nx} \frac{dN}{d\tau}\frac{dx}{d\tau} + \Gamma^x_{xx}  \left(\frac{dx}{d\tau}\right)^2 = 0, 
\end{gather}
where the Christoffel symbols in the simulation coordinates are
\begin{align}
\Gamma_{NN}^N &= \partial_N \log \alpha,  & 
\Gamma_{Nx}^N &= \partial_x \log \alpha, & 
\Gamma_{xx}^N &= \frac{ \partial_N (a \cosh N)^2}{2\alpha^2}, \\
\Gamma_{NN}^x &= \frac{ \partial_x \alpha^2 }{ 2a^2 \cosh^2 N}, &
\Gamma_{Nx}^N &= \partial_N \log( a\cosh N),  &
\Gamma_{xx}^N &= \partial_x \log a.
\end{align}
The geodesic equations for $\rho(\tau)$ and $\varphi(\tau)$ are always satisfied for $\rho(\tau) = {\rm const.}$ and $\varphi(\tau) = {\rm const.}$ in both the unperturbed and perturbed regions --- the collision is symmetric in these coordinates. Integrating the geodesics is fairly simple given $\alpha(N,x)$ and $a(N,x)$ --- they are just ordinary differential equations --- but finding $\alpha(N,x)$ and $a(N,x)$ requires some extra work. 
The geodesic integration and simulation integration are logically separate procedures, so we perform them in separate steps.\footnote{
It is possible to convert the geodesic equations such that $N$ is the integration variable rather than $\tau$, which would allow the geodesics to be integrated concurrently with the simulation. We may switch to this method in future studies.}
We save the Christoffel symbols to a table during the simulation, and then interpolate from the table to get each symbol $\Gamma$ at arbitrary $N$ and $x$. We additionally save $\partial_N \Gamma, \partial_x \Gamma$ and $\partial_N \partial_x \Gamma$ to file so that we can use bicubic rather than bilinear interpolation. Bicubic interpolation is not well-defined for an arbitrary non-uniform grid, but the Christoffel symbols are saved in constant-$N$ slices. 
We first interpolate along a single slice $N_i$ using cubic interpolation
which produces functions $\Gamma(N_i, x)$ and $\partial_N \Gamma(N_i, x)$ that are $C_1$-continuous in $x$. We can then interpolate between adjacent slices for any given value of $x$. The resultant function is continuous and smooth.

\begin{figure}[t] 
   \centering
   \includegraphics[width=6in]{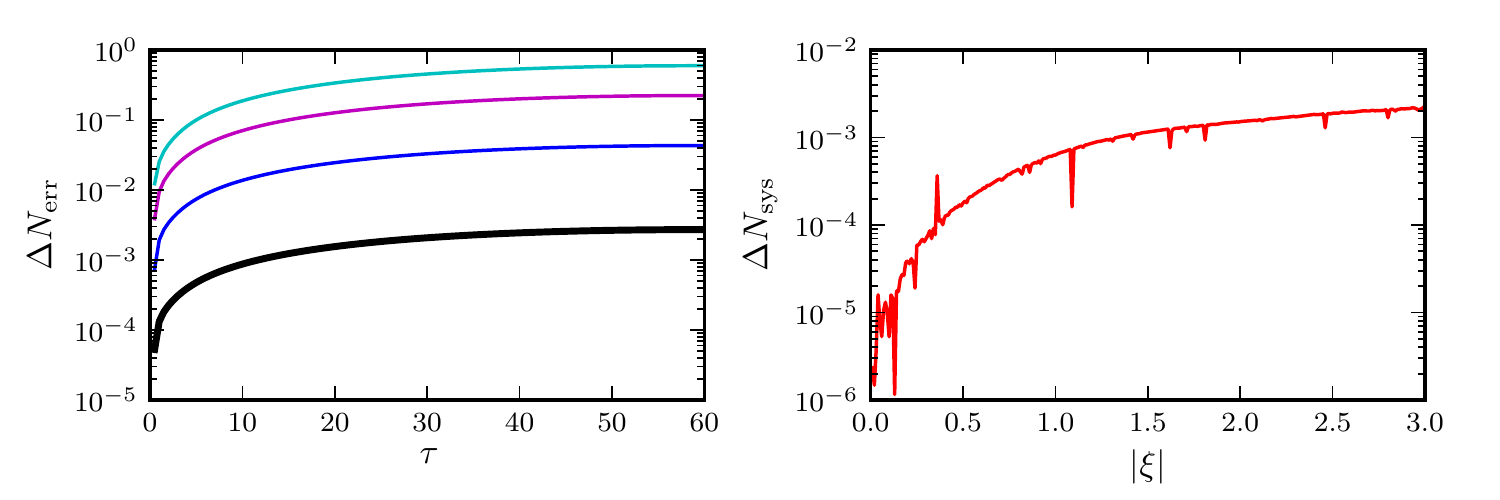} 
   \caption{Testing the geodesic integration at $\Delta x_{\rm sep} = 1.8$. The left plot shows the average error of the geodesic integration for different simulation output time-steps. The error estimate for $N(\xi, \tau)$ for one output time-step is taken to be the difference between its value calculated with that time-step and with a time-step that is half-again smaller. Each line is the root-mean-squared error for $-3 \leq \xi \leq 3$. The bottom (thick black) line is for our default time-step, which is doubled in each line above that. The right plot shows the small systematic bias in the unperturbed region for $\tau=60$ with the default output time-step: $\Delta N_{\rm sys} \equiv N(\xi) - \sinh^{-1}\left[ \sinh(N(\xi=0)) \cosh(\xi) \right]$. Analytically, we expect $\sinh N \propto \cosh\xi$ exactly, and $\Delta N_{\rm sys}= 0$. 
   }
   \label{fig:geoConvergence}
\end{figure}

For straightforward post-processing, it is important that we choose the resolution of output slices carefully: too few slices will not resolve important features of the collision front, while having too many slices results in impractically large files. We choose the recording such that each geodesic crosses approximately 20 saved slices in the time that it takes each to cross the collision front. We check that the geodesic integration converges for higher output resolutions, and that the error associated with the chosen resolution is much smaller than the integrated result (see Fig.~\ref{fig:geoConvergence}). It can be shown analytically that $\sinh N \propto \cosh\xi$ for the single bubble simulation, which matches our numerical results. 

\begin{figure}[t] 
   \centering
   \includegraphics[width=6in]{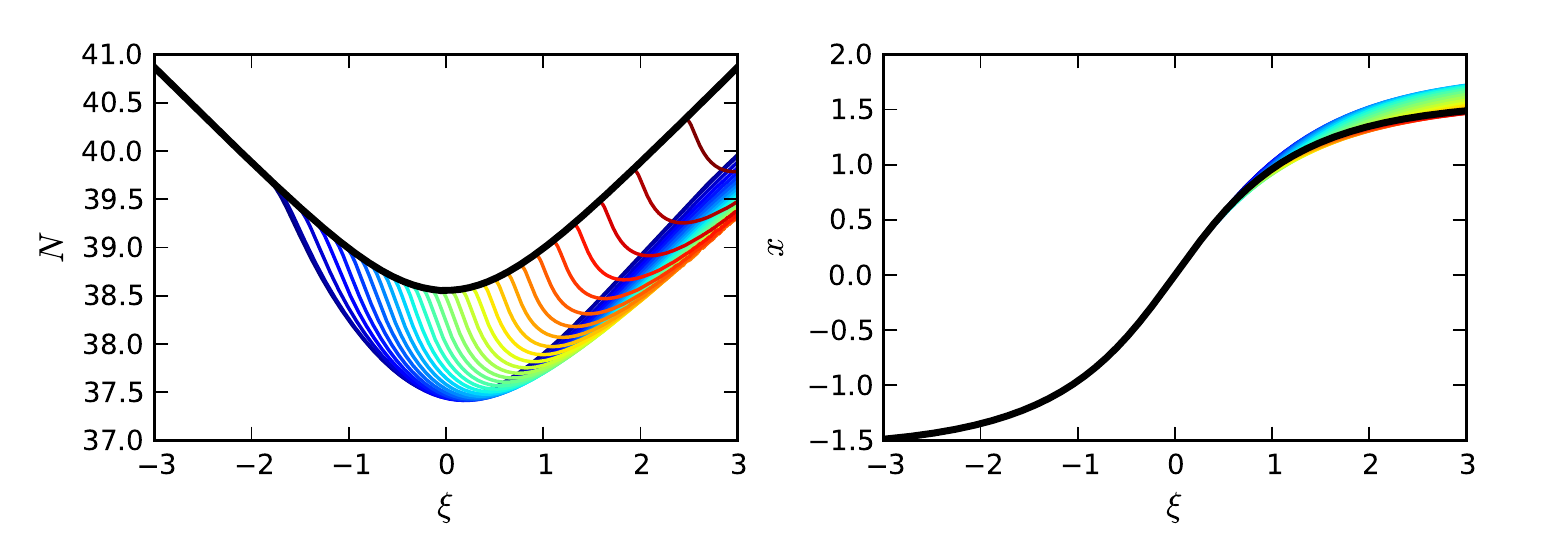} 
   \caption{Simulation coordinates as functions of the geodesic trajectory parameter $\xi$ at constant geodesic time $\tau=60$. Each line shows geodesics in a different simulation: the thick black line is for a simulation with no collision, while the thinner colored lines are for collision simulations with initial separation between bubbles ranging from $\Delta x_{\rm sep} = 0.3$ (bluest line; deviates from the unperturbed case at the lowest value of $\xi$) to $\Delta x_{\rm sep} = 3.0$ (reddest line; deviates from the unperturbed case at the highest value of $\xi$).}
   \label{fig:geodesics}
\end{figure}

For each different bubble collision simulation, we make a grid of geodesics, each starting with a different trajectory $\xi$. By interpolating along this grid (using linear interpolation, for simplicity), we can find $N(\xi, \tau)$, $x(\xi, \tau)$, and their $\xi$ and $\tau$ derivatives at arbitrary points. Fig.~\ref{fig:geodesics} shows $N(\xi)$ and $x(\xi)$ at $\tau=60$ for many different bubble collision simulations. 
Smaller values of the kinematic separation between bubbles yield collision regions which extend farther into the observation bubble, creating perturbations at lower values of $\xi$. 
Near the collision boundary, almost all of the perturbation is in $\delta N$, with very little in $\delta x$. Both perturbations are continuous and smooth --- the collision boundary does not result in a discontinuous jump in $dN/d\xi$.

\subsection{Frames}\label{sec:frames}

In an unperturbed observation bubble, the surfaces of constant $\tau$ inside the bubble are homogeneous --- there is no preferred position. However, the collision induces anisotropies inside the observation bubble, making different positions physically inequivalent. The simulated collisions described above are performed in a reference frame where the bubbles both nucleate simultaneously at $N=0$. This is the so-called Collision Frame. According to the geometric interpretation of the open universe Cartesian or anisotropic hyperbolic coordinates discussed above, at early times observers at different positions can be thought of as living on boosted trajectories in Minkowski space. To make contact with observables, it is convenient to transform to the so-called Observation Frame, where a hypothetical observer is at rest, or equivalently, at the origin of the anisotropic hyperbolic and Cartesian coordinates. This amounts to performing Lorentz transformations on the coordinates at early times, or equivalently, performing a re-labeling of the geodesics running through the simulation.

To see how this works, consider the Minkowski embedding of the anisotropic hyperbolic coordinates, Eq.~\ref{eq:minkowski_anihyp}. A boost along the $y$-direction with rapidity $\rho_{\rm obs}$ has the following action
\begin{eqnarray}
\tau' &=& \tau, \\
\xi' &=& \xi, \\
\cosh \rho' &=& \cosh \rho_{\rm obs} \cosh \rho -  \sinh \rho_{\rm obs} \sinh \rho \cos \varphi, \\
\sin\varphi' &=& \frac{\sinh \rho \sin \varphi}{ \sinh \rho'} .
\end{eqnarray}
Along $\varphi = 0$, using the identity $ \cosh \rho_{\rm obs} \cosh \rho -  \sinh \rho_{\rm obs} \sinh \rho = \cosh \left( \rho- \rho_{\rm obs} \right)$, we see that this boost corresponds to a translation of a point $\rho = \rho_{\rm obs}$ to $\rho = 0$. Importantly, the boost only mixes the $\rho$ and $\varphi$ coordinates in such a way that $d \rho^2 + \sinh^2 \rho\, d\varphi^2 \rightarrow d \rho'^2 + \sinh^2 \rho' d\varphi'^2$.
 Similarly, a boost in the $z$-direction only mixes the $\rho$ and $\varphi$ coordinates. In the collision spacetime, since the metric functions $a$ and $\alpha$  are independent of $\rho$ and $\varphi$, boosts in the $y$- or $z$- direction do nothing to the mappings $N(\tau,\xi)$ and $x(\tau, \xi)$. This is the underlying $SO$(2,1) symmetry of the collision spacetime.

Now, consider a boost along the $x$-direction in the Minkowski embedding with rapidity $\xi_{\rm obs}$. This has the action
\begin{eqnarray}
\tau' &=& \tau, \\
\label{eq:xi_boost}
\sinh \xi' &=& \cosh \xi_{\rm obs} \sinh \xi  - \sinh \xi_{\rm obs} \cosh \xi \cosh \rho, \\
\label{eq:rho_boost}
\cosh \rho' &=& \frac{\cosh \xi_{\rm obs} \cosh \xi \cosh \rho - \sinh \xi_{\rm obs} \sinh \xi}{\cosh \xi'}, \\
\varphi' &=& \varphi .
\end{eqnarray}
Along $\rho = 0$, using the identity mentioned above, we see that this boost corresponds to a translation of a point at $\xi = \xi_{\rm obs}$ to $\xi = 0$. Boosts in the $x$-direction alter the mapping between the simulation and cosmological coordinates, taking $N(\tau,\xi) \rightarrow N(\tau,\xi', \rho')$ and $x(\tau, \xi) \rightarrow x(\tau, \xi', \rho')$. 

\begin{figure}[t] 
   \centering
   \includegraphics[width=6in]{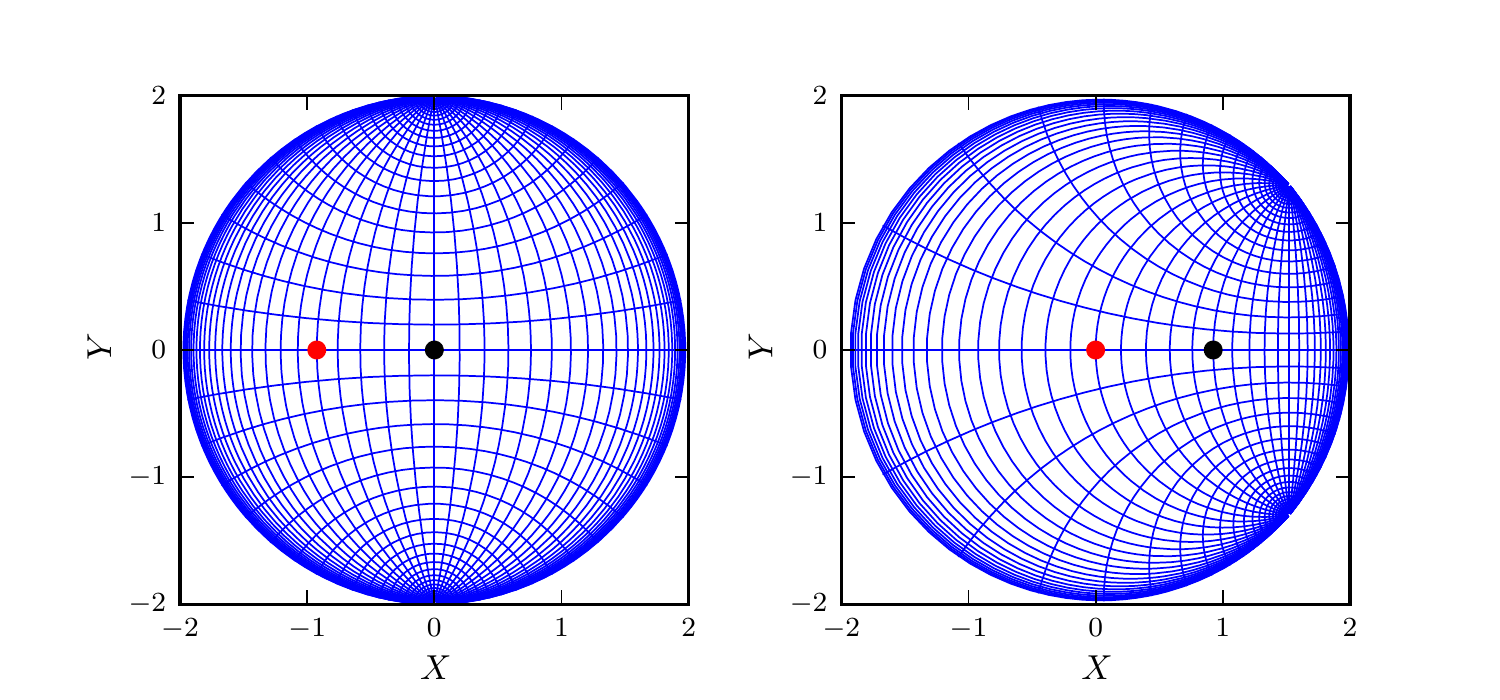} 
   \caption{
   { Mapping between the Collision Frame and Observation Frame.}
   In the left panel, we show lines of constant $\xi$ (vertical light blue) and $\rho$ (horizontal light blue) as seen by an observer (black dot) centered in the Collision Frame. The red dot (towards the left) represents a secondary observer with $\xi_{\rm obs} = -1.0$. By symmetry, the field $\phi$ and metric functions $\alpha$ and $a$ are constant along lines of constant $\xi$. In the right panel, we transform to the Observation Frame such that the secondary observer is at the origin. In this frame, lines of constant field and metric functions become distorted.
   }
   \label{fig:XYxirho}
\end{figure}

To make contact with observables, we ultimately want to go from anisotropic hyperbolic coordinates to Cartesian coordinates, which will allow us to calculate the comoving curvature perturbation. This can be accomplished by first going to the Observation Frame, and then using Eq.~\ref{eq:xirhophi} to find the mapping between the simulation coordinates and the Cartesian coordinates in the Observation Frame. In Fig.~\ref{fig:XYxirho}, we show the mapping from the anisotropic hyperbolic coordinates in the Collision Frame to the Cartesian coordinates in the Observation Frame. The field and metric functions are, by symmetry, constant along lines of constant $\xi$ in the Collision Frame. In the Observation Frame, these surfaces of constant field and metric functions get distorted by the coordinate transformation. Schematically, Fig.~\ref{fig:XYxirho} depicts what the general structure of the collision spacetime will look like in the Observation Frame.

The final result of the steps described above is a map from the simulation coordinates to the Observation Frame in Cartesian coordinates for a specific observer located at $\{\rho_{\rm obs}, \xi_{\rm obs}\}$ in the Collision Frame. Schematically, we have
\begin{equation}
N, x \ \xrightarrow{\rm Geodesics} \ \xi, \rho \ \xrightarrow{\rm Obs. \ Frame} \ \xi'(\xi_{\rm obs}) , \rho'(\xi_{\rm obs}) \ \xrightarrow{\rm Cartesian} \ X(\xi_{\rm obs}), Y(\xi_{\rm obs}), Z(\xi_{\rm obs}) , \nonumber
\end{equation}
where we have made explicit the fact that there is no physical difference between observers at different $\rho_{\rm obs}$. The map $\{N(X,Y,Z), \ x(X,Y,Z) \}$ can be created for arbitrary observers, allowing us to describe the collision spacetime from an arbitrary vantage point.

\subsection{Metric Reconstruction}
\label{sec:metric_reconstruction}

The metric inside the perturbed observation bubble is explicitly obtained from the coordinate transformations discussed above. Using the standard tensor transformation law, we have
\begin{equation}
\label{eq:metric_transform}
g_{\mu \nu} [X] = \frac{dx^{\alpha}}{dX^{\mu}} \frac{dx^{\beta}}{dX^{\nu}} g_{\alpha \beta} [x(X)] ,
\end{equation}
where $x^\alpha$ represent the simulation coordinates, $X^\mu$ represent the Cartesian coordinates in the Observation Frame, and $g_{\alpha \beta} [x(X)]$ is the metric from the simulation. The resulting metric inside the observation bubble, $g_{\mu \nu} [X] $, is by construction in the synchronous gauge: $g_{\tau\tau} = -1$ and $g_{\tau i} = 0$. 

 In practice, we perform the various coordinate transformations in sequence. We first find the metric in terms of the anisotropic hyperbolic coordinates, which are the direct products of the geodesic evolution. The accuracy of the metric reconstruction is dependent on how well the geodesics reproduce the coordinate map and its first derivatives. 
We have explicitly shown convergence in both the simulation and geodesic integration, 
and thus we expect an accurate reconstruction of the metric. 
We then transform the metric in the Collision Frame (anisotropic hyperbolic coordinates) to the Observation Frame (described in Cartesian coordinates) for an ensemble of observers at different values of $\xi_{\rm obs}$.

\section{The Comoving Curvature Perturbation}
\label{sec:comoving_perturbation}

The metric obtained in the last section describes the interior of a bubble in the presence of a collision. However, a full description of the bubble interior is not necessarily relevant for constructing cosmological observables. Because we observe a nearly homogeneous and isotropic universe, any relevant signatures arise from small (linear) deviations from FRW. By mapping the full metric onto a metric describing a linear perturbation of open FRW, we can take advantage of the extensive results of cosmological perturbation theory (see e.g., Ref.~\cite{dodelson}). 

In making this map, we must choose a gauge (coordinate system) for the perturbed FRW metric. The metric we obtain from the geodesics is explicitly in the synchronous gauge. This metric can, in principle, be used to extract any observable quantity of interest. However, transforming to the comoving gauge facilitates a more direct connection with observables, which are determined entirely by the comoving curvature perturbation. In regions where the metric is nearly FRW, this is accomplished by a linear gauge transformation, and, in linear theory, the comoving curvature perturbation is conserved on super-horizon scales. This allows us to directly connect the output of our simulations to cosmological observables without considering the details of reheating.\footnote{
The details of reheating do not affect the time-independence of $\mathcal{R}$, only its projection onto the observer's sky: the reheating temperature helps to determine the observable size of the present-day universe and, importantly, the distance $R_{\rm ls}$ to the surface of last-scattering. However, $R_{\rm ls}$ also depends upon the details of the inflationary cosmology, and we take it to be an independent observational parameter.
}

The rest of this section describes how to calculate the comoving curvature perturbation $\mathcal{R}$ and its behavior for different initial bubble separations $\Delta x_{\rm sep}$.
A summary of the necessary steps is presented below. After reading the summary, the more results-oriented reader may wish to skip directly to Sec.~\ref{sec:perturbation_behavior}.

We start in Sec.~\ref{sec:synchronous_perturbation} by mapping the full metric into scalar perturbations on a background FRW metric in the synchronous gauge.
We write the metric perturbation (the fractional difference between metrics in perturbed and unperturbed bubbles) as $\delta g_{ij} = -2D^\syn \delta_{ij} + E^\syn_{ij}$, where $E_{ij}$ is traceless and symmetric. The $D$-term is a purely scalar conformal perturbation, whereas $E_{ij}$ contains scalar, vector, and tensor components. 
To make the extraction of $E$ from $E_{ij}$ tractable, we assume
that the tensor and vector components are negligable and that the perturbation is approximately planar (a function of the coordinate $X$ only).
We justify these assumptions in Sec.~\ref{sec:synchronous_perturbation}.
For the particular Lagrangian under study, we later find that $E_{ij}$ is small and therefore the exact planar assumption is not necessary to calculate the scalar perturbations. (We have not verified that this is true for other Lagrangians, so we present a way to calculate $E$ here.)
Together, $D$ and $E$ represent scalar perturbations on an open FRW metric. To convert to perturbations on a flat metric, we assume that the region of interest satisfies $R = \sqrt{X^2+Y^2+Z^2} \ll 1$, allowing us to write the overall curvature of the open bubble universe as an additional small scalar perturbation $D^\curv$. 
Note that unlike perturbations in flat FRW, the scalar perturbations have coordinate-dependent contributions: $D^\curv$ goes to zero at the origin of coordinates and can be large far from the origin. 
It is most convenient to place the observer at the origin, so picking a point to be the origin amounts to choosing the Observation Frame. The scalar perturbations $D^\syn$, $D^\curv$, and $E$ are then defined in that frame.

Collectively the perturbations should satisfy the linearized Einstein equations; we check that this is the case in Sec.~\ref{sec:perturbation_check}. 
We then describe how to switch to comoving gauge and retrieve the comoving curvature perturbation $\mathcal{R}$ in Sec.~\ref{sec:sync_to_comoving}.

We describe our numerical results in Sec.~\ref{sec:perturbation_behavior}, verifying that $\mathcal{R}$ is time-independent for all initial bubble separations $\Delta x_{\rm sep}$ when the linear Einstein equations are satisfied. 
We find that $E^\syn_{ij} \ll D^\syn$ for all values of $\Delta x_{\rm sep} \leq 3.0$ and $R \lesssim 1$. If $E^\syn_{ij} = 0$ exactly, one can show that $D^\syn$ depends only upon the map $N(\xi,\tau)$ from anisotropic hyperbolic coordinates to simulation coordinates. 
If we drop the contribution from $D^\curv$ (which is constant on any sphere surrounding the observer, and constant along the surface of last scattering), the perturbation $\mathcal{R}$
then becomes explicitly independent of the choice of coordinate system: transforming $\mathcal{R}$ from the Collision Frame to an Observation Frame amounts to a straightforward change in the underlying coordinates, and we can treat $\mathcal{R}$ as a function of anisotropic hyperbolic coordinates $\xi$ and $\tau$ only. When $\mathcal{R}$ is time-independent (`frozen out'), it is a function of $\xi$ only.
This is an important convenience, allowing us to later determine the CMB signature for arbitrary observers in Sec.~\ref{sec:cmb_signatures} in a straightforward manner.
Our working assumptions in Sec.~\ref{sec:perturbation_behavior} and onwards are then: $E^\syn_{ij} = 0$ and $\delta g_{ij} \propto \delta_{ij}$ (so that $\mathcal{R}$ is explicitly coordinate-independent); and $R \ll 1$ (so that the overall curvature perturbation can be treated linearly). Note that we no longer need to assume that the perturbation is exactly planar, since that was only needed to calculate $E$.

Finally, in Sec.~\ref{sec:perturbation_fits}, 
we find that $\mathcal{R}(\xi)$ is well described by a power law near the collision boundary. Simulating a number of collisions, we show the power law index $\kappa$ to vary from $\kappa \approx 3.5$ for small $\Delta x_{\rm sep}$ to $\kappa \approx 2$ for large $\Delta x_{\rm sep}$.

\subsection{Perturbations in the synchronous gauge}
\label{sec:synchronous_perturbation}

With the metric Eq.~\ref{eq:metric_transform} from the geodesic simulations in hand, we can compare with standard results of cosmological perturbation theory to make contact with cosmological observables. The most general perturbed metric in synchronous gauge can be written as
\begin{equation}\label{eq:sychronous1}
H_F^2 ds_{\rm syn}^2 =  - d\tau^2 +  \frac{a^2(\tau)}{\left( 1- \frac{R^2}{4} \right)^2} \left[  \delta_{ij}  + \delta g_{ij} \right] dX^i dX^j,
\end{equation}
where
\begin{equation}
\delta g_{ij} \equiv - 2 D^\syn (\vec{X},\tau) \delta_{ij} + E_{ij}^\syn (\vec{X},\tau).
\end{equation}
When there is no collision, $D = E = 0$, and we recover the open FRW metric, Eq.~\ref{eq:unperturbed_cartesian}.

We find the components of the perturbed metric Eq.~\ref{eq:sychronous1}  as follows. First, we perform a simulation of a single bubble to define a reference background geometry. We then use the geodesic metric reconstruction method described above to find the components of the metric in terms of Cartesian coordinates, $g_{ij}^{(\rm no \ coll)}$. This matches with Eq.~\ref{eq:unperturbed_cartesian} to high accuracy (we quantify this below), 
providing an important check of our geodesic reconstruction method. Next, we perform a simulation of a collision, and re-run the geodesic reconstruction routine to obtain the metric functions $g_{ij}^{\rm (coll)}$ inside the perturbed observation bubble. 

Comparing with Eq.~\ref{eq:sychronous1}, we can identify
\begin{equation}
\label{eq:delta_gij}
\delta g_{ij} = \left(g_{ij}^{\rm (coll)} - g_{ij}^{(\rm no \ coll)} \right) \frac{\left( 1- \frac{R^2}{4} \right)^2}{a^2(\tau)}.
\end{equation}
Taking the trace, we find
\begin{equation}
D^\syn = - \frac{1}{6} {\rm Tr} \left(\delta g_{ij} \right),
\end{equation}
and
\begin{equation}
\label{eq:Eij_syn}
E^\syn_{ij} = \delta g_{ij} + 2 D^\syn \delta_{ij}.
\end{equation}

The components of $E_{ij}$ can be written as
\begin{equation}\label{eq:Eijdef}
E^\syn_{ij} = \partial_i \partial_j E^\syn - \frac{1}{3} \delta_{ij} \partial^2 E^\syn + \text{(vector and tensor components)}.
\end{equation}
This is traceless and symmetric.
The tensor contribution is zero by a hyperbolic version of Birkhoff's theorem~\cite{Hawking:1982ga}. We find the vector component, which decays during inflation, to be small empirically. Therefore, we consider only the scalar component in what follows. 

We will later need to separate out the various partial derivatives in the scalar part of $E^\syn_{ij}$  given the components of the perturbed metric  Eq.~\ref{eq:delta_gij}. Equating Eq.~\ref{eq:Eij_syn} and Eq.~\ref{eq:Eijdef} (which holds up to the neglected vector contribution),
 we obtain a set of elliptic equations for $E^\syn$,
\begin{equation}
 \delta g_{ij} + 2 D^\syn \delta_{ij} =  \partial_i \partial_j E^\syn - \frac{1}{3} \delta_{ij} \partial^k\partial_k E^\syn .
\end{equation}
This set of equations must be supplemented by boundary conditions. As we will see shortly, an understanding of the correct boundary conditions is not strictly necessary.

Two essential simplifications result when we consider cosmologies where curvature is not dominant today, implying $R \ll 1$ (where $R$ is the Cartesian radius).  First, we can treat the spatial curvature as a small perturbation in the region of interest,
\begin{equation}
H_F^2 ds_{\rm syn}^2 =  - d\tau^2 +  a^2(\tau) \left[ \delta_{ij} + \delta g_{ij} - 2 D^\curv\right] dX^i dX^j ,
\end{equation}
where $D^\curv = - \frac{R^2}{4}$. In linear theory, we can treat this separately from the perturbations sourced by the collision. Second, since the collision is a function of $\xi$ only, and close to $R=0$ we have $X \approx \xi-\xi_{\rm obs}$, we can assume that $E^\syn = E^\syn (X)$. That is, we assume that the perturbation $E$ is exactly planar. We can then write
\begin{equation}\label{eq:Echecks}
E^\syn_{\X\X} = \frac{2}{3}\partial_\X^2 E^\syn, \ \  E^\syn_{\Y\Y} = E^\syn_{\Z\Z} = - \frac{1}{3} \partial_\X^2 E^\syn .
\end{equation}
The elliptical equation for $E$ becomes an ordinary second-order differential equation in $X$, with the most natural initial condition being $E = \frac{dE}{dX} = 0$ in the unperturbed region. However, we need not solve this equation, since only the derivatives of $E$ are necessary to calculate the comoving curvature perturbation.
We will later show that $E_{ij} \ll D$, so the precise calculation of $E$ and its derivatives has little effect upon our results. 

We will find it convenient to define a new variable $\Psi$, the curvature perturbation in synchronous gauge
\begin{equation}
\label{eq:Psi_syn}
 \Psi^\syn = D^\syn + D^\curv + \frac{1}{4} E^\syn_{\X\X}.
\end{equation}
Making this change, the metric is
\begin{equation}\label{eq:sychronous2}
H_F^2 ds_{\rm syn}^2 =  - d\tau^2 +  a^2(\tau) \left[ \left(1 - 2 \Psi^\syn \right) \delta_{ij} + \partial_i \partial_j E^\syn \right] dX^i dX^j.
\end{equation}

\subsection{Checking the perturbative description}
\label{sec:perturbation_check}

With these definitions, we can verify our perturbative description. First, we check that the components of $E_{ij}$ satisfy Eq.~\ref{eq:Echecks}. These equations imply that $E$ is very nearly planar, an important assumption in our analysis below. Indeed, in the vicinity of $R=0$, we find that the planar assumption works well, as described below.

Another important test is to check that the metric in synchronous gauge obeys the Einstein equations. To zeroth order in perturbations, the metric functions should satisfy the Friedmann equations
\begin{equation}
\left( \frac{\partial_\tau a}{a} \right)^2 = \frac{8 \pi }{3} \left( \frac{(\partial_\tau \phi_0)^2}{2} + V(\phi_0) \right) + \frac{1}{a^2},
\end{equation}
where $\phi_0$ is the field with no collision. 

At first order in perturbations, we get a constraint from the $\tau$--$X$ component of the Einstein equation:
\begin{equation}
2 \partial_\tau \partial_\X \Psi^\syn = 8 \pi  \partial_\tau \phi_0 \partial_\X \delta \phi,
\end{equation}
where $\delta\phi = \phi^\syn - \phi_0$, and we work along the $X$-axis such that $Y=Z=0$. Going to second order, we have
\begin{gather}
\label{eq:syn_constraint1}
8 \pi  \left( \partial_\tau \phi_0 +  \partial_\tau \delta \phi  \right)    \partial_\X \delta \phi = 
\left(  2\partial_\tau + 4 \partial_\X \Psi^\syn + 4 \Psi^\syn \partial_\tau - \partial_\tau \partial_\X^2 E \right)   \partial_\X \Psi^\syn.
\end{gather}
Substituting Eq.~\ref{eq:Psi_syn} into Eq.~\ref{eq:syn_constraint1} yields
\begin{align}
\label{eq:syn_constraint2}
8 \pi  \partial_\tau \phi^\syn  \partial_\X \delta \phi &=
2\partial_\tau\partial_\X \left(  D^\syn + \tfrac{1}{4}E_{\X\X}  \right) \nonumber \\
&+ 4 \partial_\X \left[\left(  D^\syn + \tfrac{1}{4} \left( E_{\X\X} - X^2 \right) \right) \partial_\tau \left(  D^\syn + \tfrac{1}{4} E_{\X\X} \right) \right] \nonumber \\
&- \partial_\X \left(  D^\syn + \tfrac{1}{4} \left( E_{\X\X} - X^2 \right) \right) \tfrac{3}{2} \partial_\tau E_{\X\X}.
\end{align}

\begin{figure}[t] 
   \centering
   \includegraphics[width=6in]{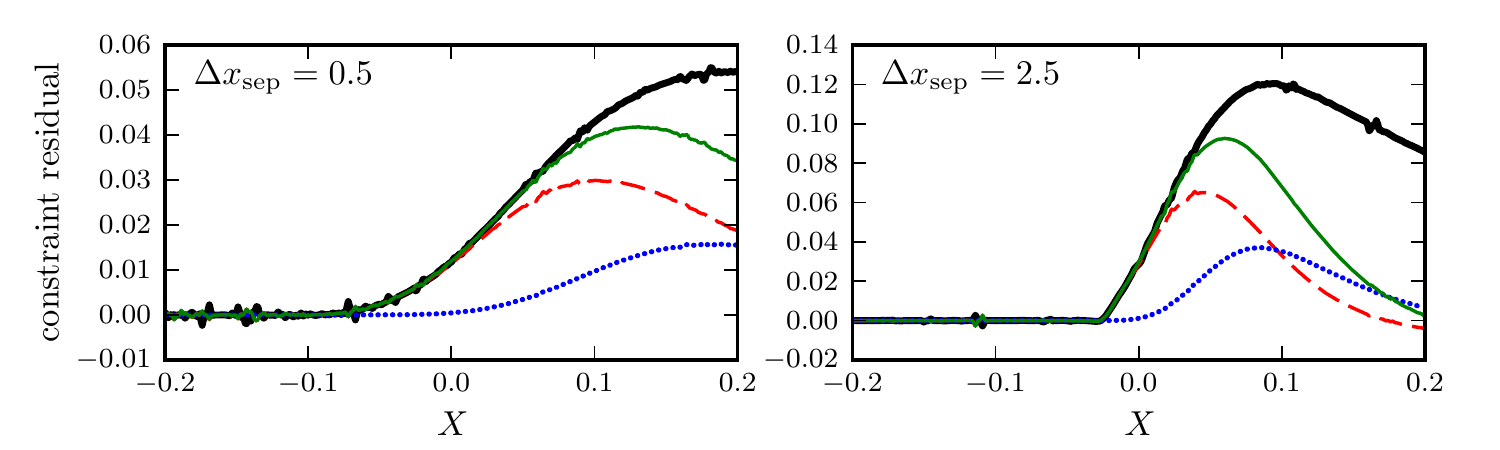} 
   \caption{ Plot of the constraint residual in synchronous gauge for two different observers at $\tau=60$ experiencing collisions with different kinematics. The thick black line shows the left-hand side of Eq.~\ref{eq:syn_constraint2}; the dashed red line shows the first-order piece of its right-hand side ($2\partial_\tau\partial_\X \left(  D^\syn + \tfrac{1}{4}E_{\X\X}  \right)$); the blue-dotted line shows the remaining second-order piece; and the thin solid green line shows the total right-hand side. The linearized Einstein equations hold to first order where the red dashed lines and thick black lines approximately coincide.
}
   \label{fig:synchronousconst}
\end{figure}

We expect this constraint to hold near the collision boundary where the perturbation is relatively small and the use of perturbation theory is valid. We can explicitly check it by computing the perturbation variables using the procedure outlined above. The result is shown in Fig.~\ref{fig:synchronousconst}. The constraint holds quite well at the edge of the collision, but does not match arbitrarily far into the collision region. It can also be seen that the second-order correction yields a result which matches further into the collision region than when including the first order correction only. This is the expected behavior --- perturbation theory is working, but higher order terms become more important as the magnitude of $\mathcal{R}$ increases deep inside the collision region. The collision can be treated as a small perturbation up to $\Delta X \approx 0.15$ beyond the collision boundary for $\Delta x_{\rm sep} = 0.5$, and up to $\Delta X \approx 0.05$ beyond the boundary for $\Delta x_{\rm sep} = 2.5$. This follows a general pattern, which we will explore in more detail below, of smaller kinematic separations between bubbles yielding smoother and wider collision boundaries.

We conclude that --- sufficiently close to the causal boundary of the collision --- the perturbative description, and the assumption of planarity, are accurate. This is also an excellent test of our code, since numerical errors would also lead to violations of the constraint equation.

\subsection{Synchronous gauge to comoving gauge}
\label{sec:sync_to_comoving}

In order to facilitate comparison with observations, we would like to go to the comoving gauge, in which the metric is
\begin{equation}\label{eq:comoving}
H_F^2 ds_{\rm com}^2 =  - (1+ 2 A^{(\rm com)}) d\tau^2 +  a^2(\tau) \left[ \left(1 - 2 \Psi^{(\rm com)} \right) \delta_{ij} +\partial_i \partial_j E^{(\rm com)} \right] dX^i dX^j,
\end{equation}
and the energy momentum tensor has the property that
$T_{i 0}^{(\rm com)} = 0.$
One can show that metric functions in comoving gauge are related to those in synchronous gauge by
\begin{eqnarray}
A^{(\rm com)} &=& -H \frac{\delta \phi}{\partial_\tau \phi_0} - \partial_\tau \frac{\delta \phi}{\partial_\tau \phi_0} \\
\Psi^{(\rm com)} &=& \Psi^\syn + H \frac{\delta \phi}{\partial_\tau \phi_0} \\
E^{(\rm com)} &=& E^\syn - \frac{\epsilon}{2},
\end{eqnarray}
where
$x_{\rm (com)}^\mu = x_{\rm (syn)}^{\mu} + \epsilon^{\mu} (x_{\rm (syn)})$
and $  \epsilon^{\mu} (x_{\rm (syn)}) = \left( \epsilon^0, \partial_i \epsilon \right)$. See e.g., Ref.~\cite{dodelson} for a discussion of linear gauge transformations in cosmological perturbation theory.

The three-curvature on spatial slices in the comoving gauge is given by
\begin{equation}
^{(3)}R^{(\rm com)} = \frac{4}{a^2} \partial^2 \Psi^{(\rm com)} =  \frac{4}{a^2} \partial^2 \left[ \Psi^\syn + H \frac{\delta \phi}{\partial_\tau \phi_0} \right].
\end{equation}
This defines the comoving curvature perturbation
\begin{equation}
\mathcal{R} \equiv  \Psi^{(\rm com)} =  \Psi^\syn + H \frac{\delta \phi}{\partial_\tau \phi_0}.
\end{equation}
Substituting Eq.~\ref{eq:Psi_syn}, we can write this as
\begin{equation}
\label{eq:comoving_perturb_full}
\mathcal{R} = D^\syn + \frac{1}{4} E_{\X\X} + H \frac{\delta \phi}{\partial_\tau \phi_0},
\end{equation}
where we have now dropped the curvature perturbation, which is a small constant on any sphere of fixed distance from the origin (relevant for the projection discussed in the next section).

\subsection{Behavior of the comoving curvature perturbation}
\label{sec:perturbation_behavior}

\begin{figure}[t] 
   \centering
   \includegraphics[width=6in]{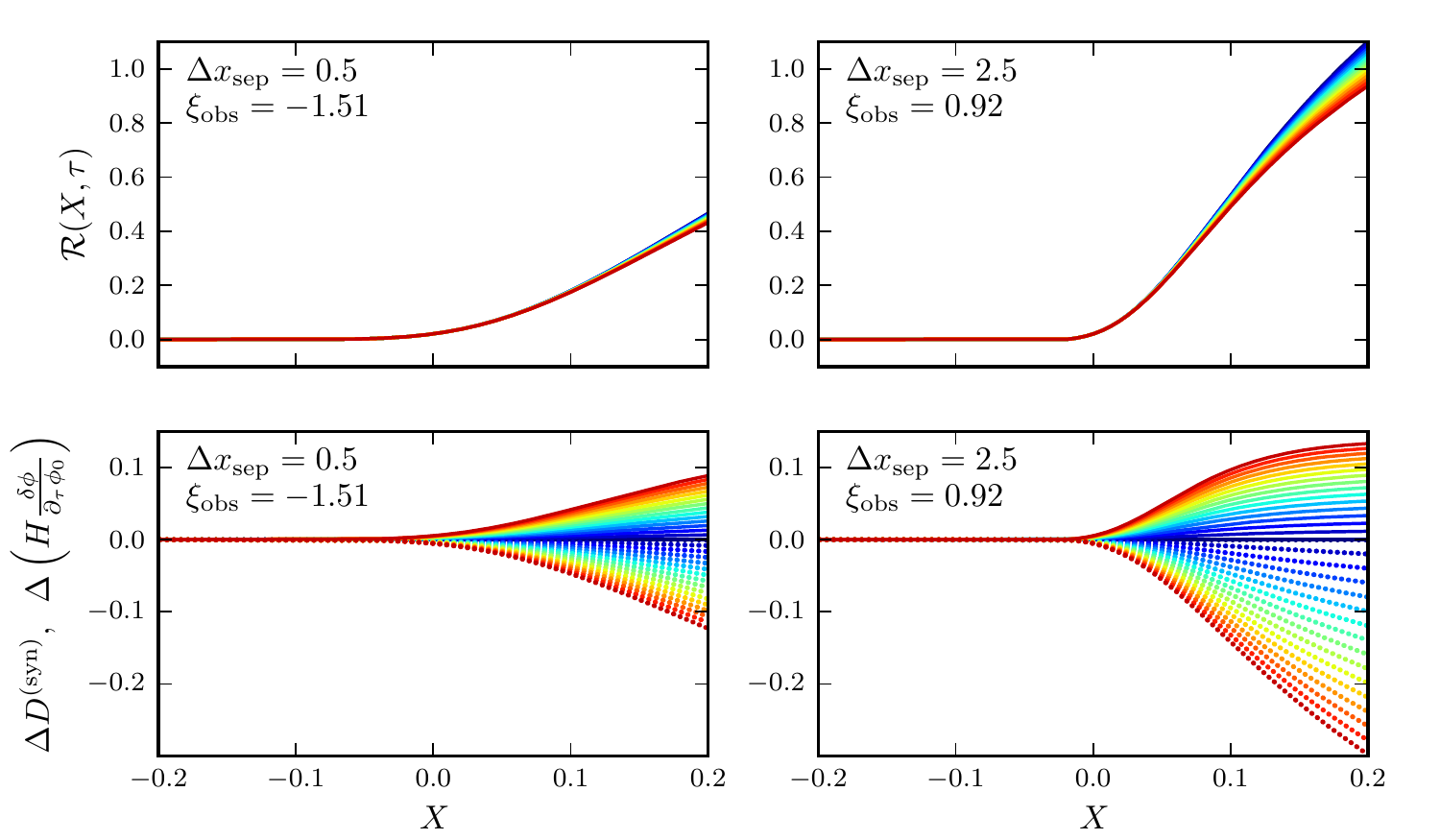} 
   \caption{
   Perturbation evolution as seen by two different observers
   for different values of $\Delta x_{\rm sep}$. 
   We calculate the comoving perturbation $\mathcal{R}$ using Eq.~\ref{eq:comoving_perturb_full} and the full set of coordinate transformations, without any of the further simplifying assumptions described below.
   In the top panels, $\mathcal{R}$ is shown for different time slices $\tau$, ranging from $\tau=30$ to $\tau=60$ (with blue to red denoting increasing time). The perturbation tends to decrease slightly with increasing time. In the bottom panels, the individual components of the perturbation are shown separately (excluding $E_{ij}$): $\Delta D^\syn = D^\syn(\tau) - D^\syn(\tau=30)$ (solid lines), and likewise for $\Delta \left(H \frac{\delta\phi}{\partial_\tau \phi_0}\right)$ (dotted lines). The evolution of the individual components almost exactly cancel everywhere that the linearized Einstein equations hold, indicating that the perturbation has `frozen in.'
   }
   \label{fig:time_dependence}
\end{figure}

 With the above machinery we can calculate the comoving perturbation. Fig.~\ref{fig:time_dependence} shows the shape and evolution of the perturbation for two different collisions as a function of $X$. In both cases, the perturbation smoothly increases from $\mathcal{R} = 0$ --- there is no sharp edge to the collision boundary. Comparing the two collisions, a higher value of $\Delta x_{\rm sep}$ yields a faster and sharper rise in the perturbation. Both the metric perturbation $D^\syn$ and the Hubble rate $H$ evolve in time,\footnote{The field perturbation $\delta \phi$ and $\partial_\tau \phi_0$ evolve as well, but these are subdominant to the evolution in $H$.} but near the collision boundary the two contributions almost exactly cancel and $\mathcal{R}$ is roughly constant. In perturbation theory, for adiabatic fluctuations, the comoving curvature perturbation should freeze in as exhibited in Fig.~\ref{fig:time_dependence}. Away from the collision boundary, the perturbation exhibits a small but significant time-dependence. The magnitude of the dependence is roughly proportional to the error in the first-order piece of the metric constraint equation (see Fig.~\ref{fig:synchronousconst}). The perturbative freeze-in is a result of the linear approximation, so the perturbation `thaws' to the extent that the linear approximation fails.  This is an important consistency check, indicating that we are accurately constructing the curvature perturbation.

\begin{figure}[t]
   \centering
   \includegraphics[width=4in]{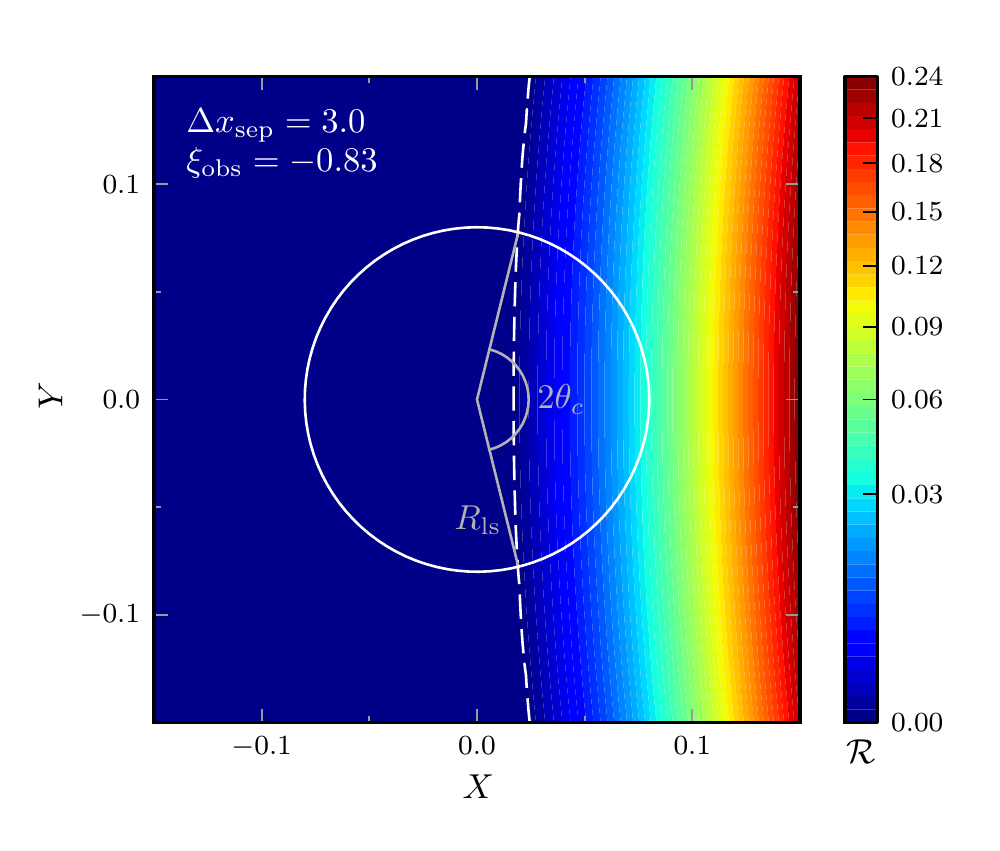} 
   \caption{Contour plot of the comoving perturbation $\mathcal{R}$ in the Observation Frame at $\tau = 60$, calculated using Eq.~\ref{eq:comoving_perturb_full}. The white dashed line shows the edge of the collision region. If the observer's past light-cone intersects the surface of last scattering at $R_{\rm ls} = 0.08$, then the white circle would represent a great circle across the observer's sky, with a visible perturbation everywhere that the circle intersects the collision region. Null geodesics that intersect the origin follow straight lines. The perturbed patch in the sky would have an angular diameter of $2\theta_c$. Note that the contour levels are spaced uniformly in $\sqrt{\mathcal{R}}$ to increase contrast near the collision border.
 }
   \label{fig:RvsXY}
\end{figure}

In Fig.~\ref{fig:RvsXY}, we show the collision in the $X-Y$ plane near the end of inflation. Because the comoving curvature perturbation near the causal boundary of the collision is frozen in, this snapshot of $\mathcal{R}$ will be representative of $\mathcal{R}$ after inflation. Recall that $\mathcal{R}$ is defined in the Observation Frame, where the hypothetical observer is at the origin of coordinates. This observer will have causal access to some portion of this early-time hypersurface, as depicted in Fig.~\ref{fig:RvsXY}. The precise projection of the observer's past light cone onto this early-time surface depends on the details of the cosmology, and is discussed below. However, in the limit where the observed curvature is small, the causally-relevant portion of the collision will be a small region around the origin. As expected, $\mathcal{R}$ has an approximate planar symmetry (lines of constant $\mathcal{R}$ are very nearly lines of constant $X$) on scales much smaller than the curvature radius in the open FRW.

We expect the comoving perturbation $\mathcal{R}$ to align with the symmetries of the simulation. That is, we expect that at fixed $\tau$, $\mathcal{R}$ is a function of $\xi$ only.
We can explore the perturbation from the vantage of any observer by performing the mapping described in Sec.~\ref{sec:frames} (see in particular Fig.~\ref{fig:XYxirho}; lines of constant $\xi$ correspond to lines of constant $\mathcal{R}$). The benefit of working with $\xi$ is that, in the limit where $E$ is small,
going to the Observation Frame for an arbitrary observer amounts to a translation in $\xi$. 
Setting $E$ to zero, there is only a conformal perturbation to the spatial part of the metric, making the coordinate transformation from Cartesian to anisotropic hyperbolic coordinates trivial:
\begin{eqnarray}
 \left(1 - 2 \mathcal{R}(\vec{X}) \right) a^2(\tau) \frac{\delta_{ij} dX^i dX^j}{\left( 1 - \frac{R^2}{4} \right)^2 }\rightarrow \left(1 - 2 \mathcal{R}(\xi) \right) a^2(\tau) \left[ d\xi^2 + \cosh^2 \xi \ dH_2^2 \right], \nonumber
\end{eqnarray}
where $dH_2^2 = d\rho^2 + \sinh^2 \rho \ d\varphi^2$, and we restored the factor of $\left( 1 - \frac{R^2}{4} \right)^{-2}$ (the appropriate non-linear completion of $D^\curv$).  Moving from $\xi_{\rm obs} = 0$ to another Observation Frame with $\xi_{\rm obs}$ then amounts to taking  
\begin{equation}
 \left(1 - 2 \mathcal{R}(\xi) \right) a^2(\tau) \left[ d\xi^2 + \cosh^2 \xi \ dH_2^2 \right] \rightarrow  \left(1 - 2 \mathcal{R}(\xi', \rho')  \right) a^2(\tau) \left[ d\xi'^2 + \cosh^2 \xi' \ dH_2'^2 \right]. \nonumber
\end{equation}
That is, we simply change coordinates in $\mathcal{R}$. Lines of constant $\mathcal{R}$ in the Observation Frame will appear much as in Fig.~\ref{fig:XYxirho}. If we are interested in a region close to the observer's origin, we can approximate 
\begin{equation}
\mathcal{R}(\xi', \rho') \simeq \mathcal{R}(\xi + \xi_{\rm obs}), \nonumber
\end{equation}
which is a simple translation in $\xi$.

 If $E=0$ exactly, we can also re-cast the perturbed metric in terms of variations in $N$. In this case, $g_{\X\X} = g_{\Y\Y} = g_{\Z\Z}$, and, at ${\varphi} = 0$:
\begin{equation}
\label{eq:sqrt_gZZ}
\sqrt{g_{\Z\Z}} = \sinh N \frac{1+ \cosh\xi_{\rm obs}\cosh\xi\cosh\rho  - \sinh\xi_{\rm obs}\sinh\xi}{2\cosh\xi} 
\end{equation}
or, for $\rho=0$,
\begin{equation}
\sqrt{g_{\Z\Z}} = \sinh N \frac{1+\cosh(\xi-\xi_{\rm obs})}{2\cosh\xi}.
\end{equation}
Therefore, if we consider the spatial variation in $N$ as a perturbation $\delta N$ on top of some background mean value $N_0$, for large $N_0$ we have 
\begin{gather}\label{eq:deltaN}
-2D^\syn = \delta g_{\Z\Z} = \frac{ g_{\Z\Z}^\coll -  g_{\Z\Z}^\nocoll}{ g_{\Z\Z}^\nocoll} = \frac{\Delta \sinh^2 N}{\sinh^2 N_0} \\
\longrightarrow \mathcal{R} \simeq - \delta N + H \frac{\delta \phi}{\partial_\tau \phi_0},
\label{eq:Rofxi}
\end{gather}
and the entire perturbation is a function of $\xi, \tau$ and $\Delta x_{\rm sep}$ only. 
This is also the behavior seen in Fig.~\ref{fig:RvsXY}: lines of constant $\mathcal{R}$ are very nearly exactly lines of constant $\xi$. Note that the coordinate $N$ roughly measures the number of $e$-folds, so Eq.~\ref{eq:deltaN} is reminiscent of the $\delta N$ formalism~\cite{Sasaki:1995aw}. 

\begin{figure}[t] 
   \centering
   \includegraphics[width=6in]{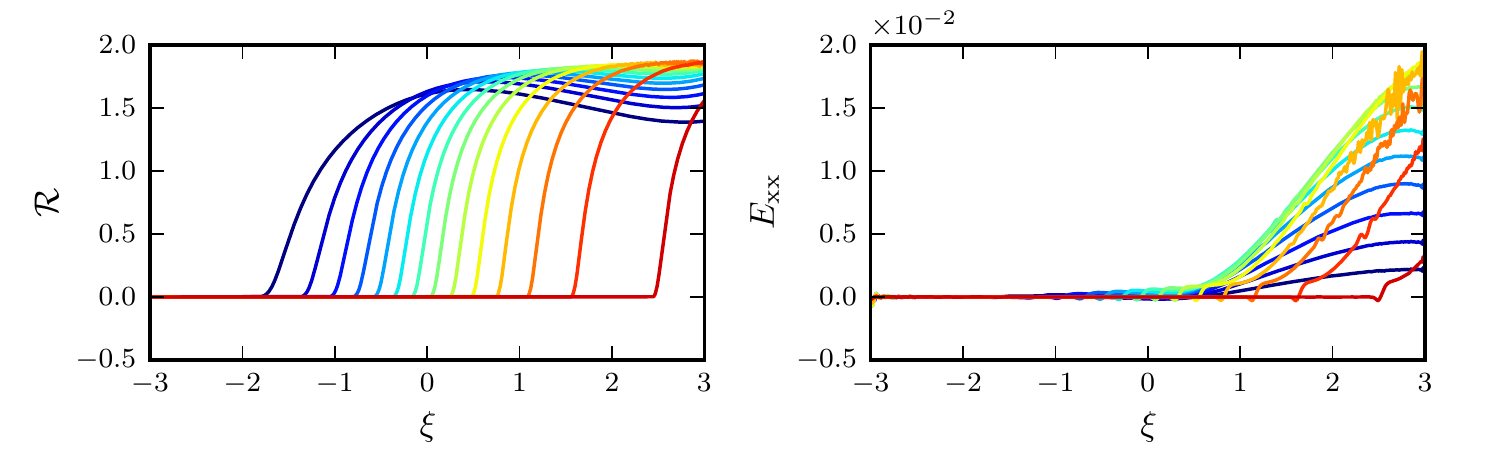} 
   \caption{Perturbations as a function of $\xi$ as measured by local observers. For each plotted value of $\xi$, we use Eq.~\ref{eq:comoving_perturb_full} to calculate $\mathcal{R}$ at the origin of an observer's coordinate system defined by $\xi_{\rm obs} = \xi$.
    Each line shows the perturbations at $\tau=60$ for a different value of the kinematic separation $\Delta x_{\rm sep}$, ranging from $\Delta x_{\rm sep} = 0.4$ to $\Delta x_{\rm sep} = 3.0$. As can be seen in Fig.~\ref{fig:position_kinematics}, the position of the collision boundary depends on the kinematics. We also calculate the perturbation $E_{\X\X}$ (right plot), finding $E_{\X\X} \ll \mathcal{R}$ everywhere.
   }
   \label{fig:Rvsxi}
\end{figure}

Fig.~\ref{fig:Rvsxi} shows the behavior of the perturbation 
for varying $\Delta x_{\rm sep}$. When the initial separation between bubbles is smaller, the bubbles collide at earlier times and the collision region penetrates farther into the observation bubble (bluer lines have smaller $\Delta x_{\rm sep}$). The amount of energy in the bubble walls is also smaller since they have had less time to accelerate, so the perturbation is not as steep in the immediate vicinity of the collision boundary. Far inside the collision region, the perturbation is roughly the same for different kinematics. However, far inside the collision region, linear perturbation theory is not reliable, so $\mathcal{R}$ does not have a clear physical meaning.

We find that the metric perturbation is almost entirely proportional to the identity matrix. That is, $D^\syn \gg E_{ij}$. Fig.~\ref{fig:Rvsxi} shows that $E_{\X\X}$ contributes no more than 0.25\% to $\mathcal{R}$ for $\xi < 3$, and much less than that for $\xi < 1$. Interestingly, $E_{\X\X}$ seems to be roughly proportional to $\partial (x^\coll-x^\nocoll)/\partial\xi$ (see Fig.~\ref{fig:geodesics}). At the observer's origin, $E_{\rm \Y\Y} = E_{\rm \Z\Z} = -\frac{1}{2} E_{\rm \X\X}$ exactly, and off-diagonal components disappear due to symmetry in $Y$ and $Z$. This is no longer true for $Y,Z \neq 0$, but we find that all components of $E_{ij}$ remain small at all coordinates (with $R \lesssim 0.3$) for all observers (boosted by $\xi_{\rm obs} \lesssim 3$). 

As discussed above, when $E$ can be safely neglected the perturbation from the perspective of different observers can be constructed by a straightforward transformation in the underlying coordinates (see Sec.~\ref{sec:frames}). For $Y=Z=0$ (or equivalently, $\rho=\rho'=0$), this corresponds to a simple translation in $\xi$. In the sections that follow, we will generally assume that $\mathcal{R}$ is a function of $\xi$ only. That is, we will assume that $E_{ij}=0$ exactly and that the perturbation is frozen out so that it has no time dependence. Under these assumptions, the left panel in Fig.~\ref{fig:Rvsxi} shows the complete comoving perturbation for \emph{all} observers (with $\xi_{\rm obs} \lesssim 3$): different observers just transform $\xi$ differently to their local Cartesian coordinates. For $R \ll 1$, the transformation is $X \approx \xi - \xi_{\rm obs}$, effectively translating each $\mathcal{R}(\xi)$ curve by a fixed amount.

\subsection{Fitting the comoving curvature perturbation}
\label{sec:perturbation_fits}

 \begin{figure}[t] 
   \centering
   \includegraphics[width=6in]{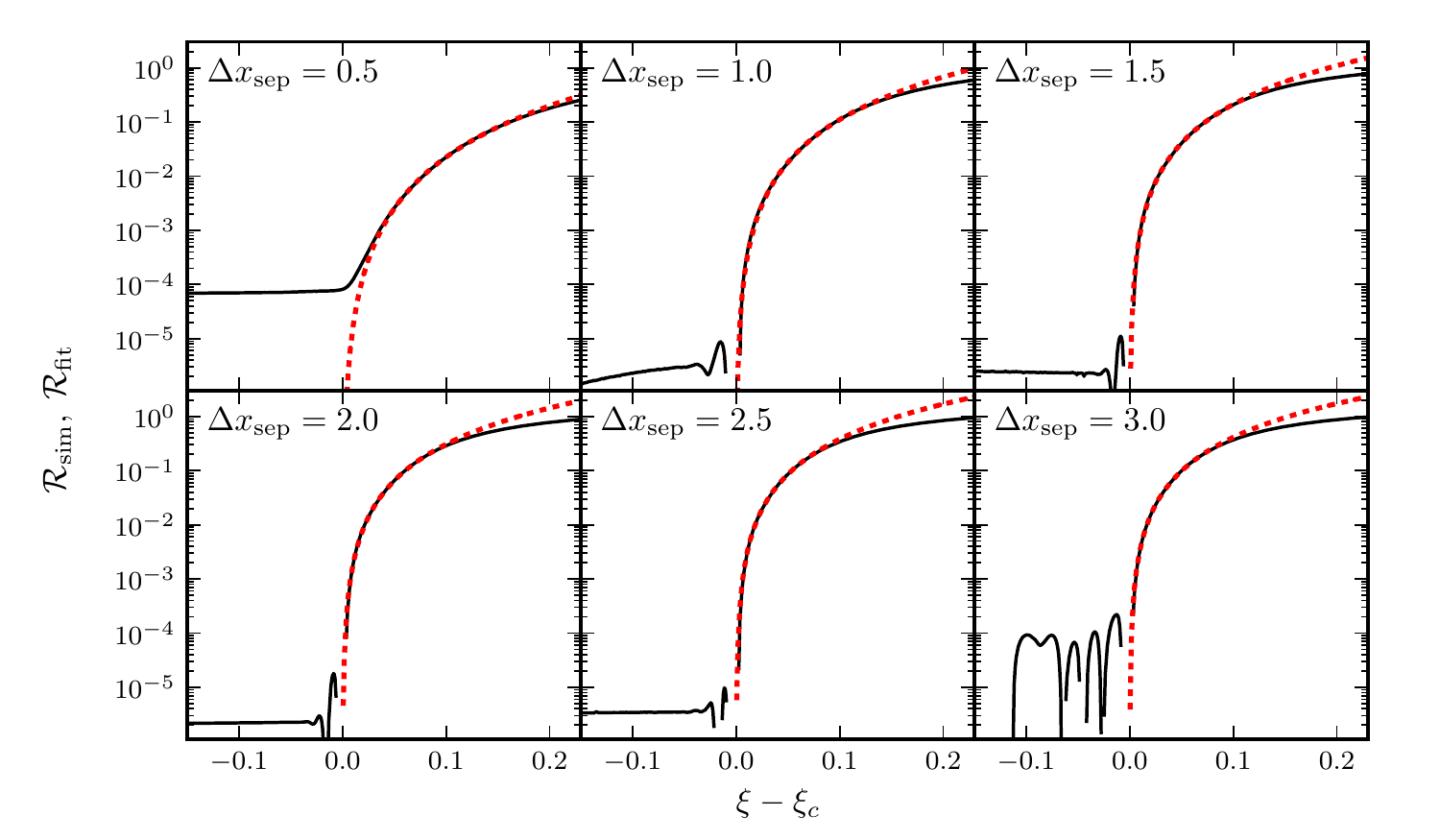} 
   \caption{The comoving perturbation as calculated by Eq.~\ref{eq:deltaN} (solid black lines) and the best fit power law from Eq.~\ref{eq:Rtemplate} (dotted red lines). The oscillations in $\mathcal{R}$ are numerical noise, showing the limits of our procedure with the default integration and output resolution.
   }
   \label{fig:example_fits}
\end{figure}
 
\begin{figure}[t]
   \centering
   \includegraphics[width=6in]{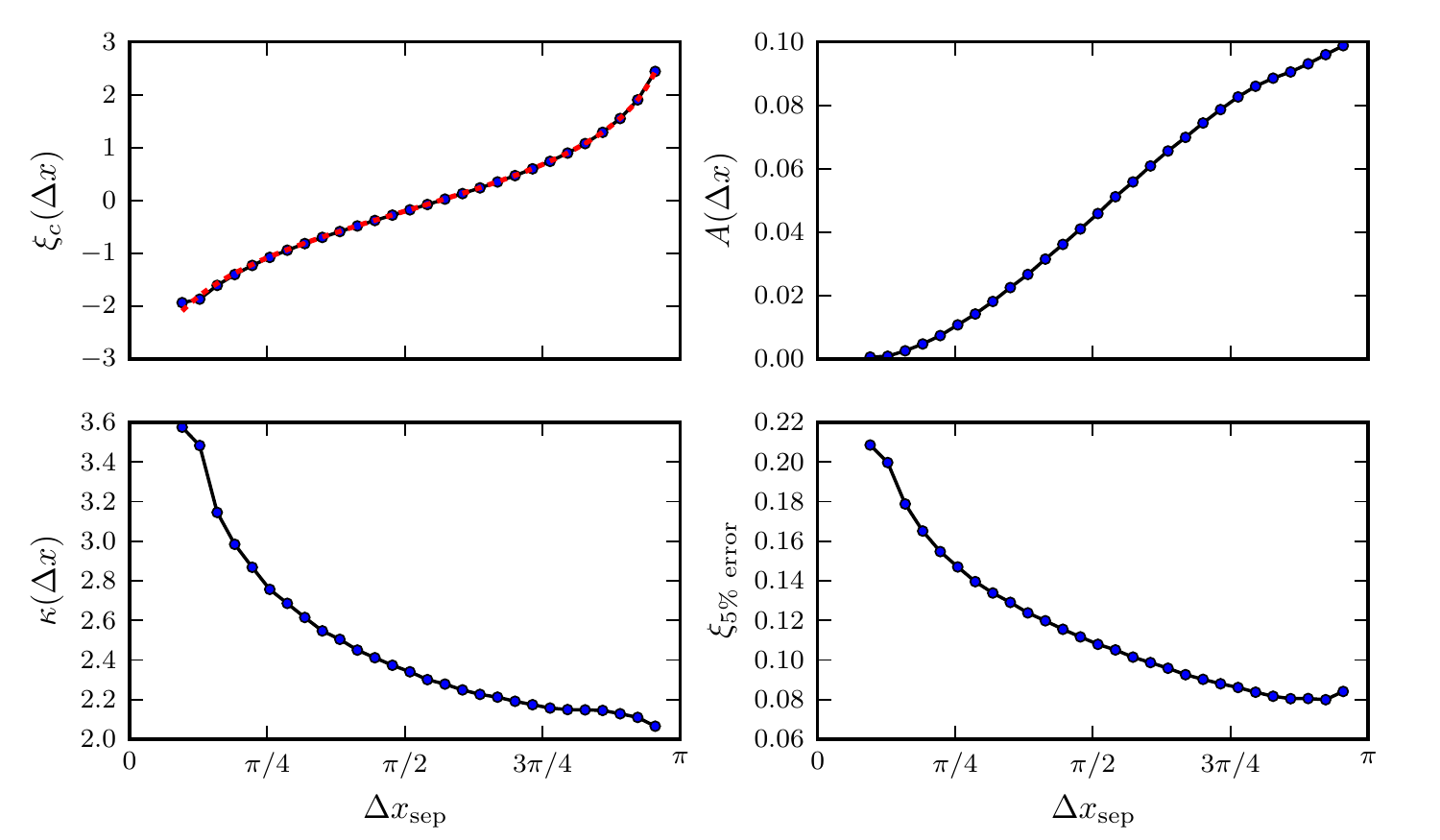} 
   \caption{The different fit parameters (defined in Eq.~\ref{eq:Rtemplate}) as a function of the bubbles' kinematic separation $\Delta x_{\rm sep}$. The dotted red line in the top-left plot, which almost exactly overlaps the best fit, shows the expected theoretical value of $\xi_c$ (Eq.~\ref{eq:xi_coll_theory}). The lower right plot shows how far into the collision region one can travel before the error between the simulated and fitted perturbations is always greater than 5\%.
   }
   \label{fig:template_plots}
\end{figure}

To facilitate comparison with observation, we define a template for the shape of the comoving curvature perturbation. We then fit for these coefficients as a function of the initial bubble separation. Very near the causal boundary of the collision, we find that the perturbation is well described by a power law,
\begin{equation}\label{eq:Rtemplate}
\mathcal{R} (\xi, \Delta x) = A (\Delta x) \left( \frac{\xi - \xi_c (\Delta x) }{\xi_0} \right)^{\kappa(\Delta x)}\Theta(\xi - \xi_c(\Delta x)).
\end{equation}
The coefficients $A(\Delta x)$, $\xi_c(\Delta x)$, and $\kappa (\Delta x)$ are all functions of the initial bubble separation; $\xi_0 = 0.05$ is a constant characteristic of the collision regions' width and the region over which the linear approximation holds.

We find the coefficients by minimizing $\sum w_i \left[\mathcal{R}_{\rm fit}(\xi_i) - \mathcal{R}_i\right]^2$ in some small region surrounding the collision boundary, where the weights $w_i$ are chosen to increase the fit's accuracy very close to the boundary where $\mathcal{R}$ is small. We use a Nelder-Mead downhill simplex algorithm as implemented in the {\tt scipy.optimize} package to perform the minimization. Fig.~\ref{fig:example_fits} shows some representative fits for different values of $\Delta x$, and Fig.~\ref{fig:template_plots} shows the fit parameters as a function of $\Delta x$ as well as the range in $\xi$ over which the power law solution is a good approximation. 

When the initial separation between the bubbles is very small, the collision region does not have a clear boundary. There is always a well-defined region that is in the causal future of the \emph{center} of the colliding bubble. However, the exterior of the bubble wall will never actually be in this region, and the wall itself does not have a sharp edge (see Fig.~\ref{fig:instanton}). The bubble wall length-contracts as the bubble grows, so the wall sharpens as it asymptotically approaches the well-defined bubble interior. However, for small $\Delta x_{\rm sep}$, the bubbles have not had much time to grow before they collide, and the bubble walls extend significantly beyond the bubble interiors. Because the collision does not have a sharp boundary, a power-law fit cannot, in principle, accurately reproduce the perturbation arbitrarily close to the collision region boundary. This is seen in Fig.~\ref{fig:example_fits} at $\Delta x_{\rm sep} = 0.5$, where the perturbation does not go to zero at what one would naively guess to be the collision boundary. The power-law fit does a much better job at larger kinematic separations: there, the non-zero perturbation outside the collision region is primarily due to (small) numerical integration errors.
 
 Most previous work has assumed that the curvature perturbation in the collision region could be considered piecewise continuous, and linear near the boundary.  This does not appear to be the case.  Near the boundary, we find that a good description of the perturbation requires continuous first derivatives; a power law with an index between 2 and 3 is a good fit.  Moreover, the region over which this power law describes the perturbation well is more than an order of magnitude larger than the width of the collision shock as seen in the simulations (making the appropriate transformation from $x$ to $\xi$).  Thus in the thin-wall limit, even if the function could be considered piecewise continuous, it is not clear that it would be linear in $\xi$ (and indeed, as the perturbation in this case would not be an analytic function at the junction point, there is no reason it need be described by a Taylor expansion there.)
 
In this section, we have outlined a mapping procedure between the metric extracted from our simulated bubble collisions (Sec.~\ref{sec:cosmo_metric}) and the scalar metric fluctuations in comoving gauge. We only considered the case where the causal boundary of the collision is very near the position of a fiducial observer at the origin of coordinates, and focussed on a window of radius $R \ll 1$ around the origin. Within this small window, we find that the collision is nearly planar, which allows us to calculate $E^{(\rm syn)}$, a necessary ingredient in the transformation from the synchronous to the comoving gauge. Empirically, we find $E^{(\rm syn)}$ to be small, in which case we can express the perturbation entirely in terms of the comoving curvature perturbation $\mathcal{R}$. In this limit, we can easily translate $\mathcal{R}$ to explore the perturbation from different vantage points. (Note that this is a convenience, and not in general necessary.) The comoving curvature perturbation $\mathcal{R}$ is well described by the power law (Eq.~\ref{eq:Rtemplate}, with coefficients shown in Fig.~\ref{fig:template_plots}) in the vicinity of the causal boundary of the collision. This represents a complete set of predictions given an underlying scalar field Lagrangian, and plays a key role in the following discussion of CMB observables.   
 
\section{CMB Signature}
\label{sec:cmb_signatures}

In this section, we compute the observational signatures of bubble collisions in the temperature anisotropies of the CMB. As we saw in the previous section, the comoving curvature perturbation is time-independent in the spatial region over which perturbation theory is valid. For the single-field models we are studying, only adiabatic fluctuations are excited by the collision, and the relevant part of the comoving curvature perturbation will remain time-independent throughout the duration of inflation. Once reheating occurs, and the standard Big Bang cosmology begins, the fine-grained shape of the template will be processed by the transfer function. The observed temperature anisotropy in the CMB will depend on both the underlying curvature perturbation and the position of the observer. 

Centering the collision on the north celestial pole, the temperature anisotropy will depend only on the polar angle $\theta$, and as depicted in Fig.~\ref{fig:RvsXY} we have
\begin{equation}
R_{\rm ls} \cos \theta = X \simeq \xi-\xi_{\rm obs} .
\end{equation}
Here, $X$ is the position in Cartesian Observer Coordinates of a point on the observer's lightcone, which has radius $R_{\rm ls}$. The approximation $X \simeq \xi-\xi_{\rm obs}$ holds in the small-$R_{\rm ls}$ approximation, which applies when the observed curvature is small (as quantified below), and in which  lines of constant $\xi$ correspond to lines of constant $X$.   The location of the causal boundary of the collision, $\theta_c$, is then given by
\begin{equation}\label{eq:costhetac}
\cos \theta_{c} \simeq \frac{\xi_c - \xi_{\rm obs}}{R_{\rm ls}} ,
\end{equation}
where $\xi_c$ is the location of the causal boundary in the collision frame and $\xi_{\rm obs}$ is the position of the observer. 

The comoving curvature perturbation (see Eq.~\ref{eq:Rtemplate}) is determined entirely by the kinematics, $\Delta x_{\rm sep}$, while the projection is determined entirely by the observer position, $\xi_{\rm obs}$, and the size of the observer's past light cone on the surface of last-scattering, $R_{\rm ls}$. This is the minimal set of variables determining the temperature anisotropy caused by a collision. A set of variables which will be more closely tied to the observed signature would include the parameters in our fit for $\mathcal{R}$ (as a proxy for $\Delta x_{\rm sep}$), $\theta_c$ (as a proxy for $\xi_{\rm obs}$), and $R_{\rm ls}$. Since all the parameters in $\mathcal{R}$ are a function of $\Delta x_{\rm sep}$, and $\kappa$ is monotonic in $\Delta x_{\rm sep}$, we are free to choose $\kappa$ as a proxy for the template, which then uniquely specifies the other fitting parameters. With this, we have
\begin{equation}
\frac{\delta T}{T} = \frac{\delta T}{T} (\Delta x_{\rm sep}, \xi_{\rm obs}, R_{\rm ls}) = \frac{\delta T}{T} (\kappa, \theta_c, R_{\rm ls}) .
\end{equation}
The kinematics and observer position are independent of the details of the inner-bubble cosmology; $R_{\rm ls}$ is determined entirely by the cosmology inside an undisturbed bubble. Returning to Fig.~\ref{fig:summary}, the details of $\mathcal{R}$ are fixed by region 1 of the potential and the kinematics, while $R_{\rm ls}$ is determined by region 3. For a $\Lambda$CDM cosmology~\cite{Aguirre:2009ug}, we can identify
\begin{equation}\label{eq:omegatoRls}
\Omega_k \simeq \left( \frac{R_{\rm ls}}{2} \right)^2 .
\end{equation}
In the limit of instantaneous reheating, the size of the observer's past light cone is directly related to the number of inflationary $e$-folds: $R_{\rm ls} \propto e^{-N_e}$. 

To get an idea of the shape of the observable signature in the CMB, we can compute the temperature anisotropy in the Sachs-Wolfe~\cite{Sachs:1967er} limit, 
\begin{equation}\label{eq:sw}
\frac{\delta T}{T} = \frac{\mathcal{R} (\vec{X}_{\rm ls})}{5} .
\end{equation}
This gives an accurate form for the temperature anisotropies on the largest scales. 

Using the template for the curvature perturbation, Eq.~\ref{eq:Rtemplate}, and the Sachs-Wolfe formula, Eq.~\ref{eq:sw}, we find
\begin{equation}\label{eq:deltat_template}
\frac{\delta T}{T} = R_{\rm ls}^{\kappa} \frac{A(\kappa)}{5 \ \xi_0^{\kappa}} \left( \cos \theta - \cos \theta_c \right)^{\kappa} \Theta(\theta_c - \theta).
\end{equation}
The central amplitude of the template (using the terminology of Refs.~\cite{Feeney_etal:2010dd,Feeney_etal:2010jj,Feeney:2012hj}) is given by
\begin{equation}\label{eq:z0}
z_0 = \frac{A(\kappa)}{5} \left( \frac{R_{\rm ls}}{\xi_0} \right)^{\kappa} \left( 1 - \cos \theta_c \right)^{\kappa} .
\end{equation}

 \begin{figure}[t] 
   \centering
   \includegraphics[width=6in]{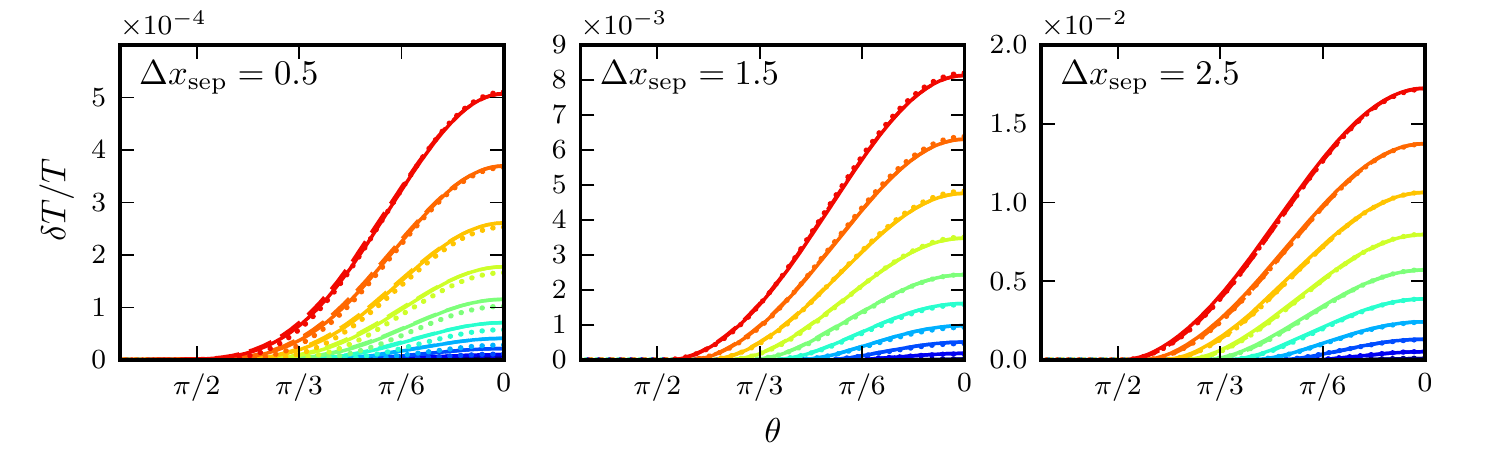} 
   \caption{The full-sky temperature perturbation $\delta T/T$ as seen by observers with $R_{\rm ls}=0.05$ using: $\mathcal{R}(X,Y)$ (solid lines); $\mathcal{R}(\xi)$, approximating $E=0$ (dashed lines, which are nearly coincident with the solid lines); and $\mathcal{R}_{\rm fit}(\xi)$ (dotted lines). We approximate $\xi \approx \xi_{\rm obs} + R_{\rm ls}\cos\theta$, whereas we use the precise coordinate transformations for $X=R_{\rm ls}\cos\theta$ and $Y=R_{\rm ls}\sin\theta$. Different lines show different observer positions $\xi_{\rm obs}$, or equivalently different bubble sizes $\theta_c$. For $\Delta x_{\rm sep} = 0.5$, we subtract off a constant perturbation such that $\mathcal{R}(\theta=\pi)=0$. 
   }
   \label{fig:projectedR}
\end{figure}

 \begin{figure}[h] 
   \centering
   \includegraphics[width=6in]{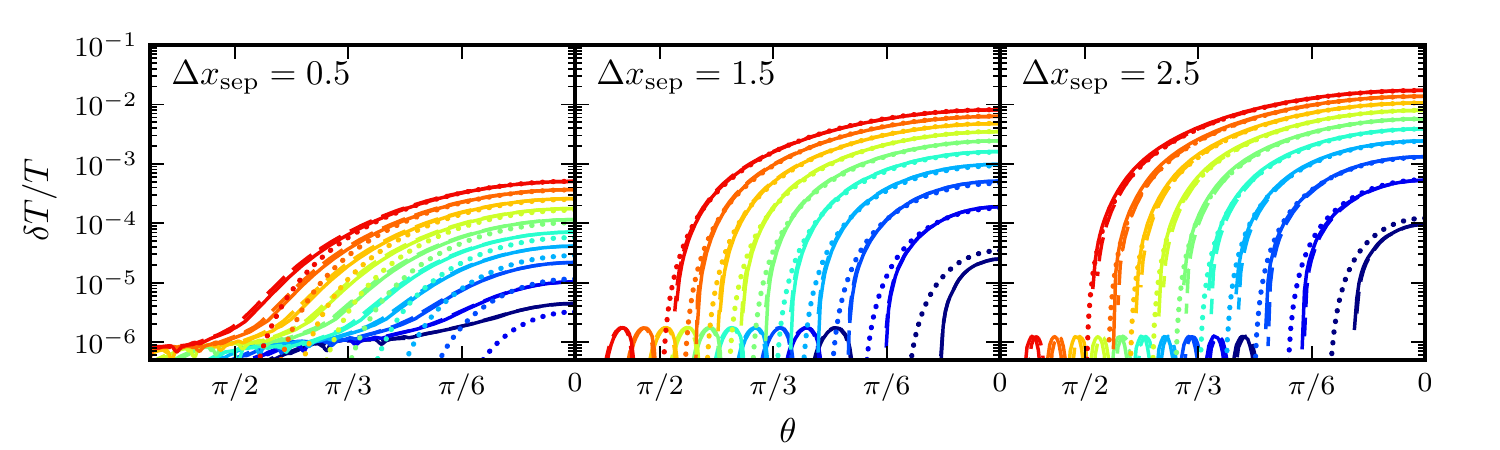} 
   \caption{The same plots as seen in Fig.~\ref{fig:projectedR}, but with a logarithmic $y$-axis. 
   The power-law is not a perfect fit all the way down to the collision boundary, and errors in finding $\xi_c$ appear larger when mapped to $\theta$ for $\theta \ll 1$ (since $d\theta/d\xi$ diverges at $\theta=0$). 
   }
   \label{fig:projectedR_log}
\end{figure}

A representative set of profiles from the simulation are shown in Figs.~\ref{fig:projectedR} and~\ref{fig:projectedR_log}. In these figures, we compare the result for the temperature profile using the full simulation data for $\mathcal{R}$ (with and without neglecting the contribution from $E$) with the fitting function Eq.~\ref{eq:Rtemplate}. The fitting function works very well for collisions with large $\theta_c$, and for large kinematics $\Delta x$. This reflects the increased accuracy of the fit for $\mathcal{R}$ for larger values of $\Delta x$.   
 
For the model we have chosen, the signature has a number of noteworthy properties:
\begin{itemize}
\item For this model, the collision is always a hot spot ($\delta T/ T > 0$).
\item Fixing $\theta_c$ and $R_{\rm ls}$, the central amplitude increases with increasing $\Delta x_{\rm sep}$. This makes intuitive sense: larger initial bubble separations give rise to higher energy collisions, yielding a stronger signal.
\item Fixing $\kappa$ and $R_{\rm ls}$, the central amplitude and size are not independent parameters. Larger collisions (in terms of $\theta_c$) generally have a larger amplitudes.
\item The central amplitude is large in magnitude, and therefore phenomenologically not viable, unless $R_{\rm ls}$ is quite small. Considering large-size collisions ($\theta_c \sim \pi/2$), for $\delta T/ T < 10^{-5}$ (a rough limit from observation), we require $R_{\rm ls} < \xi_0 (5 \times 10^{-5} / A)^{1/\kappa}$.  Kinematics with the largest $\Delta x_{\rm sep}$ provide the most stringent constraint, yielding $R_{\rm ls} < 10^{-3}$ for the potential we have studied. The  energy density in curvature at last-scattering is related to $R_{\rm ls}$ by $\Omega_k = \left( \frac{R_{\rm ls}}{2} \right)^2$, indicating a limit $\Omega_k < 2 \times 10^{-7}$ on the curvature in this model. The (in-principle) detectable level of curvature is $\Omega_k \sim 10^{-5}$. Therefore, in this case, detectable curvature and detectable (but not-yet-detected) bubble collisions are incompatible. This statement may vary for other models.
\end{itemize}

Since we have analyzed only one model, it is far from clear that the template parameters found above are representative of a generic single-field potential. Because the form of the curvature perturbation is different from that found in previous studies of the bubble collision signature, our template for the CMB temperature anisotropy is likewise somewhat different. The first theoretical calculation of the CMB signature was performed in Ref.~\cite{Chang_Kleban_Levi:2009}, where the collision was treated in the thin-wall limit, and the backreaction of the perturbed inflaton on the geometry was neglected. These authors found, for a particular set of kinematics, a template identical to Eq.~\ref{eq:deltat_template} with $\kappa = 1$. 
In Ref.~\cite{Gobbetti_Kleban:2012}, a power series ansatz was made for the comoving curvature perturbation. With sufficient inflation, and assuming a thin bubble wall, the leading term with $\kappa = 1$ was found to dominate. Incorporating a finite wall thickness, it was suggested that the template can change. A mapping from the numerical simulation onto the initial conditions in the open FRW universe would allow a direct comparison between the analytic model of Ref.~\cite{Gobbetti_Kleban:2012} and the output of the numerical simulations. We leave this to a more systematic study of different models to be performed in future work. However, in Sec.~\ref{sec:perturbation_fits}, we have shown that wall thickness alone cannot account for the shape of the template.
These changes have important implications for the detailed polarization signature discussed in Refs.~\cite{Kleban_Levi_Sigurdson:2011,Czech:2010rg}; we will return to predictions for polarization in future work. What is clear from our analysis is that the power law index $\kappa$ can vary significantly with kinematics, meaning there is definitely no universal index. 
\section{Collision Prior}
\label{sec:prior}

What are the detailed predictions of theories with bubble collisions? In the previous section, we saw that any individual collision can be described by the minimal set of parameters $\theta_c$, $\kappa$, and $R_{\rm ls}$. However, since bubble nucleation is a random stochastic process, a model of eternal inflation can only make probabilistic predictions for how many collisions, $N_{\rm coll}$, should appear in the CMB, and for what values the template parameters should take. In addition, there is uncertainty in the model itself, which has implications for observable signatures. 

Considering the predictions of all models is beyond the scope of this paper. However, the properties of $\mathcal{R}$ are insensitive to: (1) an overall re-scaling $\beta$ of the scalar field potential in the simulation; and (2) changes in the total number of post-nucleation $e$-folds of inflation $N_e$. Returning to Fig.~\ref{fig:summary}, this corresponds to considering variations in the overall scale of the potential in regions 1 and 2 with independent variations of the shape of the potential in region 3. Rescaling the potential leads, as we will see below, to changes in the prediction for the total number of observable collisions. 
 Changing the total number of $e$-folds leads to different values of $R_{\rm ls}$. We therefore have a restricted class of models in which four fundamental parameters can vary: $\Delta x_{\rm sep}$, $\xi_{\rm obs}$, $N_e$, and $\beta$. We have four observable parameters for which we want predictions: $\theta_c$, $\kappa$, $R_{\rm ls}$, and $N_{\rm coll}$. We hold the details of reheating fixed; allowing this to vary would provide additional sources of variation in $R_{\rm ls}$ and $N_{\rm coll}$.

Given a set of prior probabilities over the fundamental parameters, the goal of this section is to implement a change of variables,
\begin{equation}\label{eq:prior_general}
{ \rm Pr} (N_{\rm coll}, R_{\rm ls}, \kappa, \theta_c)  =  {\rm Pr} (\beta, R_{\rm ls}, \Delta x_{\rm sep}, \xi_{\rm obs}) \ \left| \frac{dN_{\rm coll} \ d R_{\rm ls} \ d\kappa \ d\theta_c}{d\beta \ dR_{\rm ls} \ d\Delta x_{\rm sep} \ d\xi_{\rm obs} } \right|^{-1} .
\end{equation}
As discussed above, we consider an ensemble of models in which $R_{\rm ls}$ and $\beta$ are produced by independent variations of the underlying scalar field potential. The kinematics and observer position are independent variables. We assume that these parameters are also independent of the variations in the potential (a good assumption as long as the initial radius of colliding bubbles is a small fraction of the false vacuum cosmological horizon size). 
The prior over fundamental parameters is therefore separable,
\begin{equation}
{\rm Pr} (\beta, R_{\rm ls}, \Delta x_{\rm sep}, \xi_{\rm obs}) =  {\rm Pr} (\beta) {\rm Pr} (R_{\rm ls})  {\rm Pr} (\Delta x_{\rm sep}) {\rm Pr}( \xi_{\rm obs}) \, ,
\end{equation}
allowing us to discuss each piece separately before finally assembling the desired distribution in Eq.~\ref{eq:prior_general}.

\subsection{Prior over potential re-scalings}

First, we discuss the prior over overall re-scalings of the potential, and its relation to the predicted number of collisions in the CMB. The number of collisions whose causal boundary intersects the observable portion of the surface of last-scattering is given by~\cite{Aguirre:2009ug,Freivogel_etal:2009it}
\begin{equation}
N_{\rm coll} \simeq \frac{16 \pi}{3} \left( \lambda H_F^{-4} \right) \left( \frac{H_F}{H_I} \right)^2 \frac{R_{\rm ls}}{2} ,
\end{equation}
where $\lambda$ is the probability of bubble nucleation per unit four-volume, $H_F$ is the Hubble scale in the false vacuum, and $H_I$ is the Hubble scale during inflation. This expression is valid in the limit where the collision is a small perturbation on the background cosmology which, as we have shown above, is a valid assumption over some range of parameters. The nucleation probability can be written as
\begin{equation}
\lambda H_F^{-4} = (A H_F^{-4} ) e^{-B/\beta} .
\end{equation}
In the thin-wall limit, and neglecting gravitational effects, the tunneling exponent is given by
\begin{equation}\label{eq:thinwallexponent}
B = \frac{27 \pi^2 \tilde{\sigma}^4}{2 (v(x_F) - v(x_T))^3 } \, ,
\end{equation}
where we have parameterized the potential, field, and wall tension by the scales
\begin{equation}
V (\phi) = \beta M^4 v(x), \ \  \phi = M x, \ \ \sigma = \beta^{1/2} M^3 \tilde{\sigma} .
\end{equation}
The total number of predicted collisions depends both on the detailed shape of the barrier and the details of the cosmology inside an undisturbed bubble. With this scaling, we also have
\begin{equation}
H_F^2 \rightarrow \beta H_F^2, \ \ H_I^2 \rightarrow \beta H_I^2 .
\end{equation}
Note that we have further assumed that the combination $A H_F^{-4}$ does not change under rescalings. 
The predicted number of collisions is therefore
\begin{equation}\label{eq:Ncolldef}
N_{\rm coll} \simeq N_0 e^{-B/\beta} R_{\rm ls}\, ,
\end{equation}
where we have defined
\begin{equation}
N_0 =  \frac{8 \pi}{3} \left( A H_F^{-4} \right) \left( \frac{H_F}{H_I} \right)^2 ,
\end{equation}
which does not change under re-scalings by $\beta$. For the models studied in this paper, we have $H_I \sim H_F$, although a large hierarchy between these scales may be essential for obtaining a significant probability for observable bubble collisions. 

It is not clear what prior probability we should assign to the overall scale of the potential. We can assume a flat prior for illustration, in which case
\begin{equation}
{\rm Pr} (\beta) = \frac{1}{\Delta \beta} , 
\end{equation}
where $\Delta \beta$ is the range in $\beta$ spanned by our ensemble of models. We consider the fiducial set of parameters: $B = 100$, $10 < \beta < 100$, and $N_0 = 10^5$ in the following. 

\subsection{Prior over $R_{\rm ls}$}

The prior over $R_{\rm ls}$ is essentially a prior over the number $e$-folds of inflation (for fixed reheating temperature), 
since $R_{\rm ls} \propto e^{-N_e}$. A number of proposals have been made for the prior over the number of inflationary $e$-folds~\cite{DeSimone:2009dq,Freivogel:2005vv,BlancoPillado:2012cb,Guth:2012ww,Gibbons:2006pa,Yang:2012jf,Marsh:2013qca}. The results are highly measure- and model-dependent, but most proposals are in agreement that obtaining a large number of $e$-folds is suppressed. As an example, consider a power law distribution for the number of $e$-folds ${\rm Pr} (N_e) \propto N_e^{-m}$ (such as was found by Refs.~\cite{Freivogel:2005vv,BlancoPillado:2012cb}). Changing variables, we find
\begin{equation}
{\rm Pr} (R_{\rm ls}) = {\rm Pr} (N_e) \frac{dN_e}{dR_{\rm ls}} \propto \frac{1}{R_{\rm ls} (-\log R_{\rm ls})^m} \, .
\end{equation}
Having too few $e$-folds of inflation is not consistent with observation, providing a model-independent upper bound of roughly $R_{\rm ls}^{\rm max} \sim 10^{-2}$. The minimal value of $R_{\rm ls}^{\rm min}$ is a property of the ensemble of models we are considering. For concreteness, let us impose a bound of $R_{\rm ls}^{\rm min} = 10^{-20}$. With these choices, for $m=4$ there is a $\sim 50\%$ probability ($35\%$ for $m=3$) that $R_{\rm ls}$ lies in the range producing observable curvature $\Omega_k > 10^{-5}$.

\subsection{Prior for kinematics and observer position}

\begin{figure}
\begin{center}
\includegraphics[width=8 cm]{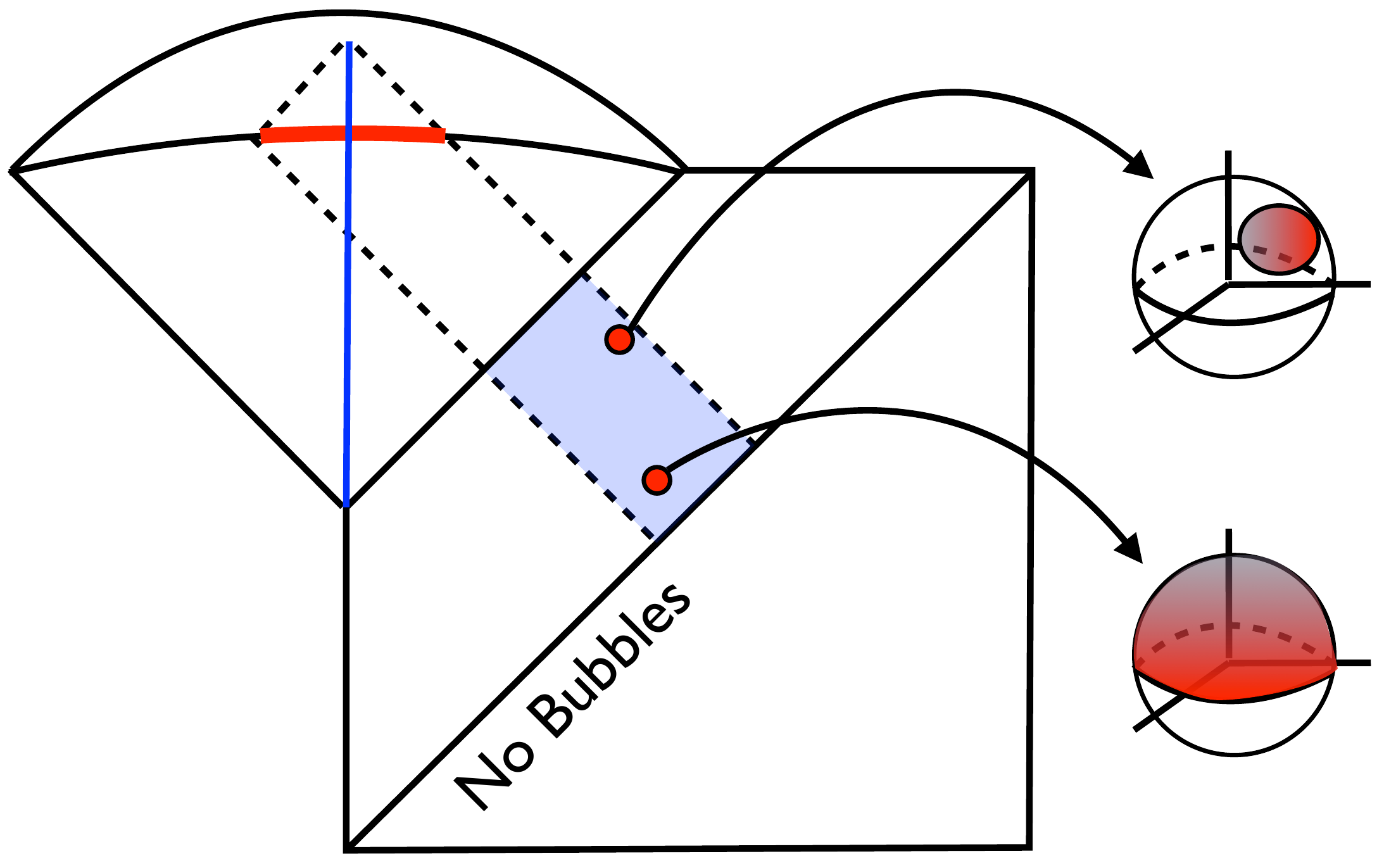}
\end{center}
\caption{A conformal diagram depicting the prior calculation. An observer inside the observation bubble has causal access to a portion of the surface of last-scattering (red line). Bubbles nucleating in the shaded region will produce a collision whose causal boundary projects onto various angular sizes $\theta_c$ on the CMB sky of the observer. The probability for bubble nucleation is uniform in each parcel of four-volume in the false vacuum.
  \label{fig:obsframe}
}
\end{figure}

The prior over $\Delta x_{\rm sep}$ and $\xi_{\rm obs}$ is most easily computed in the Observation Frame, where these variables map onto a nucleation site in the false vacuum. This is depicted in Fig.~\ref{fig:obsframe}; a detailed treatment of this calculation can be found in Ref.~\cite{Aguirre:2009ug}. A particularly convenient foliation of the false vacuum exterior to the observation bubble is given by the metric
\begin{equation}\label{eq:outsidehypercoords}
ds^2 = \frac{1}{H_F^2 \cosh^2 \Upsilon} \left( -d \Psi^2 + d\Upsilon^2 + \cosh^2 \Psi d\Omega_2^2 \right) ,
\end{equation}
where $- \infty < \Upsilon < \infty$ and $-\infty < \Psi < \infty$. The bubble wall is located at $\Upsilon \rightarrow -\infty$ in the limit where the wall sits on the light cone, and at finite $\Upsilon$ in the realistic case where the bubble has finite size. 

In the Observation Frame (unprimed) a position $\{\Psi, \Upsilon \}$ in the exterior is mapped to a position $\{\Psi' = 0, \Upsilon' = \Upsilon \}$ in the Collision Frame (primed). 
This will translate the observer in the open FRW universe along $x=0$ to the position
\begin{equation}
\xi' = - \Psi .
\end{equation}
Therefore, we see that the $\Upsilon$ coordinate maps onto the kinematics in the Collision Frame, and the $\Psi$ coordinate maps onto the observer's position in the Collision Frame at fixed kinematics. Note that, as described in Sec.~\ref{sec:frames}, points not along the $x-$axis will transform in a non-trivial way under a transformation between the Collision and Observation Frames. The kinematics in the simulation are determined by $\Delta x_{\rm sep}$, which is related to $\Upsilon$ by
\begin{equation}
\Delta x_{\rm sep} = 2 \arctan \left[ \tanh \left( \frac{\Upsilon}{2} \right) \right] + \frac{\pi}{2} .
\end{equation}

The differential number of bubbles coming from each volume element in the Observation Frame is given by
\begin{equation}
dN = \lambda H_F^{-4} \frac{\cosh^2 \Psi}{\cosh^4 \Upsilon} \sin\theta_0 d\Psi d\Upsilon d\theta_0 d\phi_0 .
\end{equation}
We can therefore write the probability density over nucleation sites as
\begin{equation}
{\rm Pr}(\Psi, \Upsilon, \theta_0, \phi_0) = \frac{ \lambda H_F^{-4}}{N_{\rm coll}}  \frac{\cosh^2 \Psi}{\cosh^4 \Upsilon} \sin\theta_0 .
\end{equation} 
As a first step, we make the transformation of variables from $\Psi$ to $\xi_{\rm obs}$, the position of the observer in the Collision Frame, via $\xi_{\rm obs} = - \Psi$. We then integrate over the angular position of the collision center, $\{ \theta_0, \phi_0 \}$. Making the coordinate transformation from $\Upsilon$ to $\Delta x_{\rm sep}$ yields
\begin{eqnarray}
 {\rm Pr}(\xi_{\rm obs}, \Delta x_{\rm sep}) &=& 4 \pi {\rm Pr}(\Psi, \Upsilon) \left| \frac{d\xi_{\rm obs} d\Delta x_{\rm sep}}{d\Psi d\Upsilon} \right|^{-1} \nonumber \\
&=& \frac{4 \pi \lambda H_F^{-4} }{N_{\rm coll}}\cosh^2 (\xi_{\rm obs}) \sin^3 (\Delta x_{\rm sep}) \\
&=& \frac{3}{2 R_{\rm ls}} \left( \frac{H_I}{H_F} \right)^2 \cosh^2 (\xi_{\rm obs}) \sin^3 (\Delta x_{\rm sep}) .
\end{eqnarray}

The range in $\Delta x_{\rm sep}$ has a lower bound determined by the size of the colliding bubbles (for our model, about $\Delta x_{\rm sep} \sim 0.3$) and an upper bound of $\pi$. The range in $\xi_{\rm obs}$ is determined by
\begin{equation}
\xi_{\rm c} - \xi_{\rm ls} < \xi_{\rm obs} < 2 \xi_{\rm c} \, .
\end{equation}
From Eq.~\ref{eq:costhetac}, this corresponds to collisions that encompass zero angular scale all the way to the full sky.\footnote{For collisions with $ \xi_{\rm obs} > \xi_c$ (larger than half-sky), it is not clear that the Sachs-Wolfe formula Eq.~\ref{eq:sw} for the CMB temperature captures all of the important contributions to the observable signal~\cite{Aguirre:2009ug,Kozaczuk:2012sx}. In particular, the Doppler component associated with motion in the gravitational potential can contribute significantly to the observed temperature anisotropy (as in the ``tilted universe" scenario~\cite{Turner:1991di,Erickcek:2008jp}).}
The position of the collision is a function of $\Delta x_{\rm sep}$ only. We can estimate it in the thin wall limit as
\begin{equation}
\label{eq:xi_coll_theory}
\xi_{\rm c} = \log \left[ \frac{H_I}{H_F} \tan \left( \frac{\Delta {x}}{2}\right) \right] .
\end{equation}
This quantity in practice is directly extracted from simulations: see Fig.~\ref{fig:template_plots}.

\subsection{Assembling the prior}

Dropping constant factors, the full prior over the parameters describing the ensemble of models and nucleation sites is given by
\begin{equation}
{\rm Pr} (\beta, R_{\rm ls}, \Delta x_{\rm sep}, \xi_{\rm obs}) \propto \frac{1}{R_{\rm ls}^2 (-\log R_{\rm ls})^m} \cosh^2 (\xi_{\rm obs}) \sin^3 (\Delta x_{\rm sep}) \, .
\end{equation}
Our final task is to change variables to find the prior over observables, Eq.~\ref{eq:prior_general}. We first evaluate the Jacobian in this expression. From the functional dependence of the observables ($N_{\rm coll}(\beta, R_{\rm ls})$, $\kappa(\Delta x_{\rm sep})$, and $\theta_c(\Delta x_{\rm sep}, \xi_{\rm obs}, R_{\rm ls})$) the Jacobian is
\begin{eqnarray}
\left| \frac{dN_{\rm coll} \ d R_{\rm ls} \ d\kappa \ d\theta_c}{d\beta \ dR_{\rm ls} \ d\Delta x_{\rm sep} \ d\xi_{\rm obs} } \right| &=& \frac{\partial N_{\rm coll}}{\partial \beta} \frac{\partial \kappa }{\partial \Delta x_{\rm sep}} \frac{\partial \theta_c}{\partial \xi_{\rm obs}} \\
&=& N_{\rm coll } \log\left[ \frac{N_{\rm coll}}{N_0 R_{\rm ls}} \right]^2 \frac{1}{R_{\rm ls} \sin \theta_c} \frac{\partial \kappa }{\partial \Delta x_{\rm sep}} \, ,
\end{eqnarray}
where we have dropped constant factors. With this, we can evaluate the prior over observables, 
\begin{eqnarray}\label{eq:finalprior}
{ \rm Pr} (N_{\rm coll}, R_{\rm ls}, \kappa, \theta_c)  
&=& \frac{\sin \theta_c }{N_{\rm coll } \log\left[ \frac{N_{\rm coll}}{N_0 R_{\rm ls}} \right]^2}  \left(\frac{\partial \kappa }{\partial \Delta x_{\rm sep}} \right)^{-1} \nonumber \\ 
& \times& \frac{ \cosh^2 \left( \xi_{\rm c}(\kappa) - R_{\rm ls} \cos \theta_c \right) \sin^3 (\Delta x_{\rm sep} [\kappa] ) }{R_{\rm ls} (-\log R_{\rm ls})^m} \, .
\end{eqnarray}
In this expression, $\kappa(\Delta x_{\rm sep})$ and $\xi_c (\Delta x_{\rm sep})$ are obtained directly from the simulation (see Fig.~\ref{fig:template_plots}), and then inverted to find $\xi_c (\kappa)$ and $\Delta x_{\rm sep}(\kappa)$. From Eq.~\ref{eq:Ncolldef}, the possible values for $N_{\rm coll}$ are constrained by the constants $N_0$, $B$, and $\beta$ as well as the value of $R_{\rm ls}$. 

 \begin{figure}[t] 
   \centering
   \includegraphics[width=2.7in]{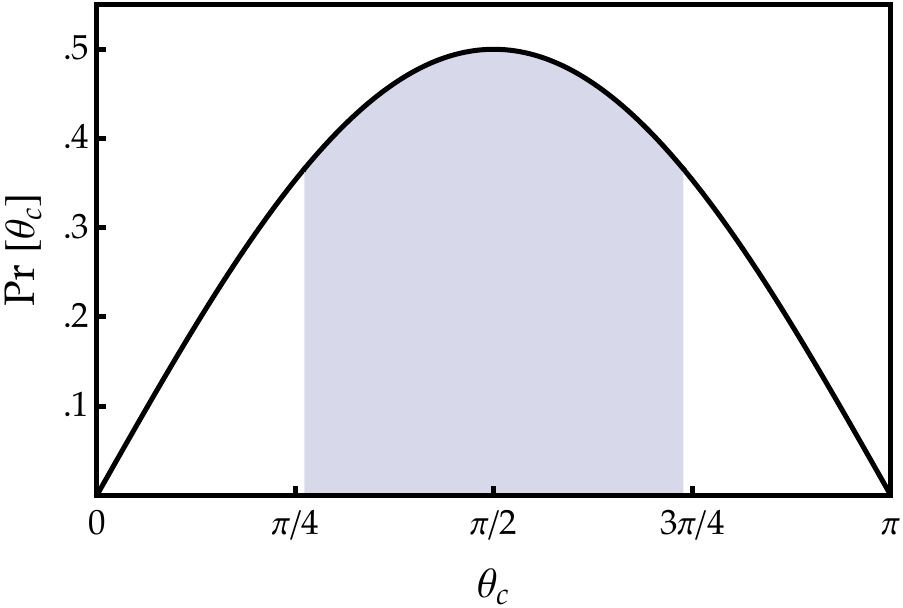} 
      \includegraphics[width=2.75in]{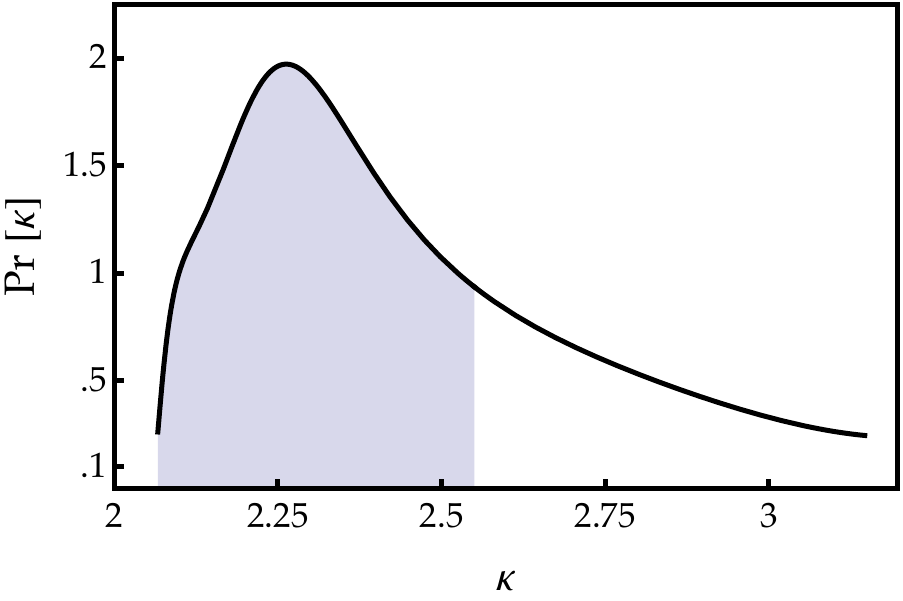} 
           \includegraphics[width=2.8in]{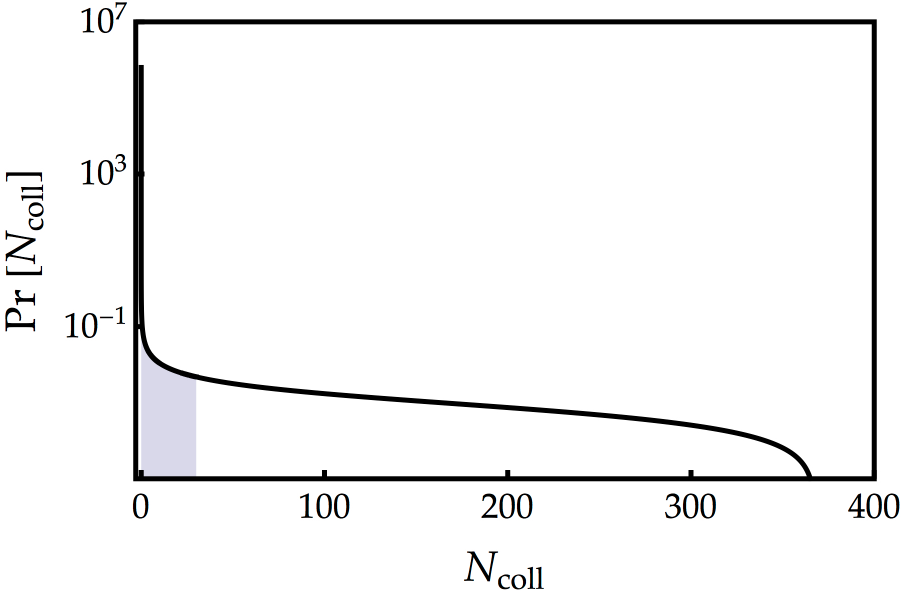} 
                \includegraphics[width=2.75in]{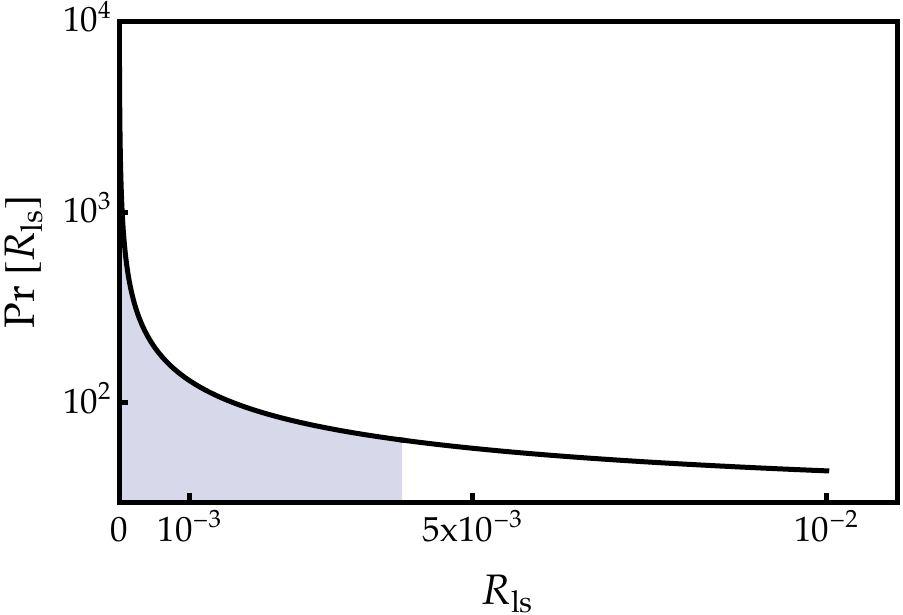} 
   \caption{The marginalized distribution function Eq.~\ref{eq:finalprior} over the four phenomenological parameters for a model of bubble collisions $\{\theta_c, \kappa, N_{\rm coll}, R_{\rm ls} \}$ for a set of models that has $B=100$, $10 < \beta < 100$, $N_0 = 10^5$, and $m=3$. Regions containing $68 \%$ of the probability surrounding the maxima are shaded. 
   }
   \label{fig:distribution}
\end{figure}

In Fig.~\ref{fig:distribution} we show the marginalized distribution function Eq.~\ref{eq:finalprior} for a fiducial set of models with $B=100$, $10 < \beta < 100$, $N_0 = 10^5$, and $m=3$. There are a number notable properties of the marginalized probability distributions. First, looking at the properties of the individual collisions, the distribution over angular scales is very close to ${\rm Pr} \sim \sin \theta_c$, preferring large-scale collisions on the sky. The power-law index in Eq.~\ref{eq:deltat_template} is preferentially near $\kappa = 2.25$. Moving to the distribution over $N_{\rm coll}$, we see that there is significant probability for $N_{\rm coll} >1$. The distribution is weakly weighted towards fewer total collisions in the range $N_{\rm coll} > 1$. For the fiducial model we study, there is equal marginalized probability for the no collision ($N_{\rm coll} < 1$), few collisions ($1 < N_{\rm coll} < 25$), and many collisions ($N_{\rm coll} > 25$) cases. Lowering the lower bound on $\beta$ gives higher weight to the no-collision case ($N_{\rm coll} < 1$),
and maintains the relative weights between the few- and many-collision cases. Increasing $N_0$ gives higher weight to the many-collision case, preserving the relative weight of the no- and few-collision cases. Therefore, the allowed ranges for these parameters can play an important role in the nature of the prediction. Finally, the marginalized distribution over $R_{\rm ls}$ gives significant weight to relatively large values of $R_{\rm ls}$. There is roughly equal weight for $R_{\rm ls}$ in the range that produces amplitudes compatible with the observed CMB ($R_{\rm ls}< 10^{-3}$) and incompatible with the observed CMB. The observable range might span roughly an order of magnitude in $R_{\rm ls}$; there is roughly a $20 \%$ chance to find $R_{\rm ls}$ in this interesting region between $10^{-4} < R_{\rm ls} < 10^{-3}$.

The prior probability distributions relevant for observational searches must take into account some of the properties of the experiment itself. For example, a more relevant prior is over the expected number of observable collisions as opposed to the expected total number (see Refs.~\cite{Feeney_etal:2010dd,Feeney_etal:2010jj,Feeney:2012hj}). This is a conditional probability, determined by making a cut in the size $\theta_c$ and amplitude $z_0$ (see Eq.~\ref{eq:z0}) of the signature. These cuts will affect the shape of the marginalized distributions shown in Fig.~\ref{fig:distribution}; we leave a detailed discussion of this point for future work.
\section{Conclusions}
\label{sec:collisions}

This paper has provided, for the first time, a direct quantitative set of predictions for the observable signatures of bubble collisions from an ensemble of scalar-field models that give rise to eternal inflation. This is an important proof-of-principle --- we have shown that it is feasible to directly compute the detailed predictions of eternal inflation, linking cosmological observations (such as that of the CMB) to fundamental theory. For the ensemble of models we have studied, we have rigorously shown that bubble collisions can be consistent with a mildly inhomogeneous cosmology, and produce interesting, possibly observable signatures. We have also identified a minimal set of phenomenological parameters, and computed the prior probability distribution over them. This constitutes a complete set of predictions for the ensemble of models we have considered.  

An important product of this work is a computer code that allowed us to simulate bubble collisions and the entire post-collision inflationary cosmology inside a bubble. We have extracted a map from the simulation to the standard variable for cosmological perturbation theory, the comoving curvature perturbation $\mathcal{R}$, which allows us to make direct contact with observables. For a comprehensive roadmap of the paper, and an overview of how this was accomplished, see Sec.~\ref{sec:summary}. 

Our results corroborate most of the qualitative features of previous analytic, numerical, and semi-analytic studies of bubble collisions. There are nevertheless some quantitative differences. The detailed form of the curvature perturbation depends on the kinematics of the collision, and we have been able to simulate the entire kinematic range. The curvature perturbation is well described by a power law (of index between 2 and 3) in the region to the future of the collision for the ensemble of models we have studied. This leads to an azimuthally-symmetric CMB temperature anisotropy that falls off with angular distance from the center. In the Sachs-Wolfe limit, the temperature anisotropy follows a power-law in the cosine of the angle from the center. This fall-off is faster than was found in previous studies, which assumed that the curvature perturbation was linear in the region to the future of the collision. 

We have identified four phenomenological parameters describing bubble collisions in this model: the total number of predicted collisions ($N_{\rm coll}$), the radius of our past light-cone at last-scattering ($R_{\rm ls}$), and the angular size ($\theta_c$) and detailed shape (specified by $\kappa$) of each collision. We derived the prior probability over these phenomenological parameters for an ensemble of models by  considering the distribution of allowed kinematics and observer positions, and by varying the overall scale of the underlying scalar-field potential as well as the number of $e$-folds of inflation inside the bubble. This prior probability prefers large-scale collisions, a power-law index of around $\kappa = 2.25$, and with some reasonable assumptions about measure, appreciable weight for $N_{\rm coll} > 1$ and observationally interesting values of $R_{\rm ls}$. 

The methods developed in this paper can be used to study the detailed phenomenology of bubble collisions in many models of eternal inflation. The model studied in this paper is simple, but it is finely tuned and features a somewhat unnatural hierarchy of scales between the inflationary and barrier regions of the potential. There is an extremely large landscape of more complicated, and perhaps more realistic, inflationary models, featuring for example multiple scalar fields, branes, fluxes, and extra dimensions. It should be straightforward to extend our program to cover some of these cases. In particular, the code can simulate multiple scalar fields with virtually no modification. More studies will be necessary to determine whether the results presented here for the template and prior are universal across different models.

The results of this paper will be instrumental in obtaining the best limits possible from searches for bubble collisions in data from CMB experiments and other cosmological probes. A previous Bayesian analysis~\cite{Feeney:2012hj,McEwen:2012uk,Feeney_etal:2010dd,Feeney_etal:2010jj} of WMAP 7-year data relied on assumptions about the prior. The results of this paper support some of the assumptions made about the form of the template and prior, but invalidate others. Future work on the universality of our findings will lead to a clear picture for the template and prior, allowing for the most stringent possible tests based on data from the Planck satellite.

The enormity of the implications for detecting a bubble collision signature as a direct experimental test of eternal inflation is obvious; even just the formulation of an observational test of the resulting ÒMultiverseÓ is a breakthrough. Feedback between theory, simulations, and observational tests will allow us to determine how the fundamental theory is constrained, or if a detection is made, to understand what has been discovered.

\acknowledgments 

We would like to thank Jim Bardeen, Eichiro Komatsu, Karim Malik, Kendrick Smith, and Kris Sigurdson for helpful conversations on constructing the comoving gauge. This work was partially supported by a New Frontiers in Astronomy and Cosmology grant \#37426, and by a grant from the Foundational Questions Institute (FQXi) Fund, a donor advised fund of the Silicon Valley Community Foundation on the basis of proposal FQXi-RFP3-1015 to the Foundational Questions Institute. Research at Perimeter Institute is supported by the Government of Canada through Industry Canada and by the Province of Ontario through the Ministry of Research and Innovation. MCJ is supported by the National Science and Engineering Research Council through a Discovery grant. HVP is supported by STFC, the Leverhulme Trust, and the European Research Council under the European Community's Seventh Framework Programme (FP7/2007-2013) / ERC grant agreement no 306478-CosmicDawn. AA was supported in part by a time release grant from the Foundational Questions Institute (FQXi), of which he is Associate Director. LL is supported by NSERC and CIFAR. SLL is supported by the NSF (PHY-0969827 \& PHY-1308621) and NASA (NNX13AH01G). 

\bibliographystyle{JHEP}
\bibliography{NRbubbles}

\end{document}